\documentclass[11pt]{article}

\usepackage[final]{acl}

\usepackage{times}
\usepackage{latexsym}
\usepackage[T1]{fontenc}
\usepackage[utf8]{inputenc}
\usepackage{microtype}
\usepackage{inconsolata}
\usepackage{graphicx}

\usepackage{amsmath}
\usepackage{amssymb}
\usepackage{mathtools}
\usepackage{amsthm}
\usepackage{booktabs}
\usepackage{multirow}
\usepackage{subcaption}      
\usepackage{wrapfig}
\usepackage{nicefrac}
\usepackage[table]{xcolor}
\usepackage{colortbl}
\usepackage{tabularx}
\usepackage{float}
\usepackage{placeins} 
\usepackage{algorithm}
\usepackage{algpseudocode}
\usepackage{tikz}
\usetikzlibrary{positioning, arrows.meta}
\definecolor{Gray}{gray}{0.9}

\newcommand{\R}{\mathbb{R}}
\newcommand{\Prob}{\mathbb{P}}
\newcommand{\E}{\mathbb{E}}
\newcommand{\calA}{\mathcal{A}}
\newcommand{\calB}{\mathcal{B}}
\newcommand{\calD}{\mathcal{D}}
\newcommand{\calO}{\mathcal{O}}
\newcommand{\calU}{\mathcal{U}}
\newcommand{\calT}{\mathcal{T}}

\newcommand{\method}{Col-Bandit}

\newcommand{\LCB}{\mathrm{LCB}}
\newcommand{\UCB}{\mathrm{UCB}}

\newtheoremstyle{compactplain}{4pt}{4pt}{\itshape}{}{\bfseries}{.}{.5em}{}
\newtheoremstyle{compactdef}{4pt}{4pt}{}{}{\bfseries}{.}{.5em}{}
\newtheoremstyle{compactremark}{4pt}{4pt}{}{}{\itshape}{.}{.5em}{}
\theoremstyle{compactplain}
\newtheorem{theorem}{Theorem}

\theoremstyle{compactdef}

\theoremstyle{compactremark}
\newtheorem{remark}[theorem]{Remark}

\newif\ifcameraready
\camerareadytrue
\ifcameraready
  \newcommand{\ar}[1]{}\newcommand{\on}[1]{}\newcommand{\ub}[1]{}\newcommand{\rp}[1]{}
  \newcommand{\cdel}[1]{}\newcommand{\cnote}[1]{}
\else
  \newcommand{\ar}[1]{\textcolor{orange}{[AR: #1]}}
  \newcommand{\on}[1]{\textcolor{teal}{[ON: #1]}}
  \newcommand{\ub}[1]{\textcolor{blue}{[UB: #1]}}
  \newcommand{\rp}[1]{\textcolor{violet}{[RP: #1]}}
  \newcommand{\cdel}[1]{\textcolor{red}{\textbf{[DEL]}\,#1}}

  \newcommand{\cnote}[1]{\textcolor{orange}{\textit{[note: #1]}}}
\fi

\usepackage[capitalize,noabbrev]{cleveref}

\title{\method{}: Query-Time Top-$K$ Estimation for Late-Interaction Retrieval}

  \author{Roi Pony\thanks{Corresponding author: \texttt{roi.pony@ibm.com}.} \quad Adi Raz Goldfarb \quad Oshri Naparstek \quad Idan Friedman \quad Udi Barzelay \quad Eli Schwartz  \\
    IBM Research Israel}

\begin{document}
\maketitle

\begin{abstract}
Multi-vector late-interaction retrievers such as ColBERT achieve state-of-the-art quality, but their query-time cost is dominated by exhaustively computing token-level MaxSim interactions for every candidate document.
The MaxSim scores of $N$ candidates against $T$ query tokens form an $N\times T$ matrix whose row-sums are the late-interaction scores, and identifying the top-$K$ rarely requires every entry.
We introduce \method{},\footnote{Code will be released as open-source software upon publication.} a query-time estimator of the exhaustive-MaxSim top-$K$: it reveals matrix entries in batches, maintains a finite-population Bernstein--Serfling confidence interval on each candidate's score, and permanently drops any document whose upper bound falls below the $K$-th largest lower bound, computing only the cells needed to separate the top-$K$.
A single relaxation knob $\alpha_{\mathrm{ef}}\in(0,1]$ tunes the compute--fidelity trade-off. We deploy $\alpha_{\mathrm{ef}}{=}0.2$, while $\alpha_{\mathrm{ef}}{=}1$ admits a $\delta$-PAC guarantee under a simplified radius.
On BEIR and REAL-MM-RAG, \method{} preserves $\geq 90\%$ fidelity to the exhaustive top-$5$ on every corpus while cutting MaxSim FLOPs by up to ${\sim}8\times$, for up to ${\sim}13\times$ single-thread CPU speedups across x86 and ARM.
A drop-in reranking layer, it needs no retraining or index changes. Released
open-source at \url{https://github.com/roipony/ColBandit}.
\end{abstract}

\section{Introduction}
\label{sec:introduction}

\begin{figure*}[t]
\centering
\includegraphics[width=0.85\textwidth,trim=0 6pt 0 12pt,clip]{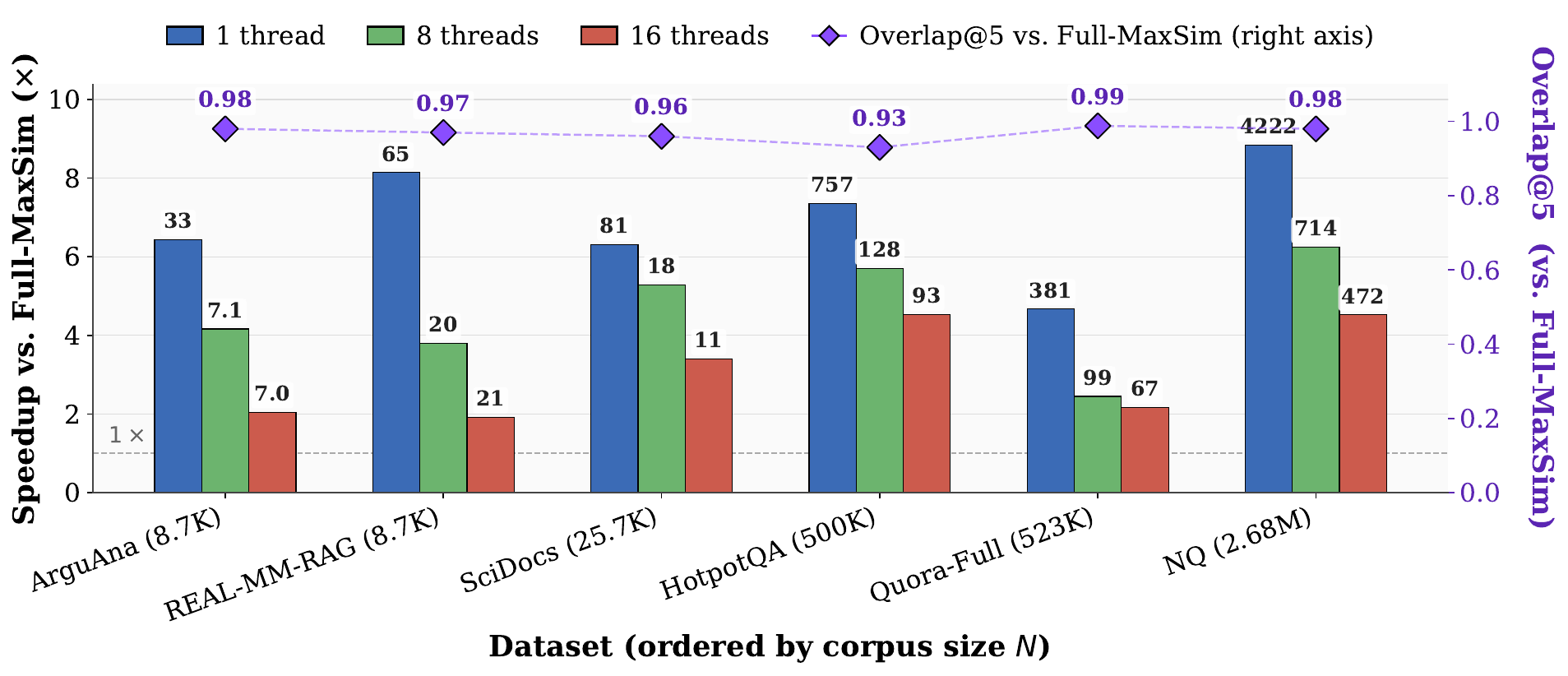}
\caption{\textbf{Wall-clock speedup of \method{} vs.\ Full-MaxSim across BEIR and REAL-MM-RAG} on \textsc{Cpu-S} at the deployed knob ($\alpha_{\mathrm{ef}}{=}0.2$, $M{=}5$, $\delta{=}0.01$, $K{=}5$). Bars are speedup factors at $1$/$8$/$16$ threads. Numbers above each bar are \method{} per-query latency in ms. Datasets are ordered by corpus size $N$. Right-axis purple diamonds: Overlap@$5$ vs.\ Full-MaxSim\textquotesingle s exhaustive top-$K$. Fidelity stays $\geq 0.93$ on $\textsc{Cpu-S}$ for every corpus (HotpotQA-$500$\,K is the floor). See Table~\ref{tab:m1_neon} for the cross-platform breakdown. REAL-MM-RAG is the four multimodal corpora merged into one collection. Extended sweeps at other $\alpha_{\mathrm{ef}}$ and $K$ values in Appendix~\ref{sec:app_headline_variants}.}
\label{fig:headline_speedup}
\end{figure*}

Multi-vector late-interaction retrievers, such as ColBERT~\citep{khattab2020colbert}, have emerged as a powerful alternative to single-vector dense retrieval. By representing each query and document as a \emph{set} of token embeddings, these models capture fine-grained semantic matches that single-vector representations miss \cite{wang2023reproducibility,formal2021white}. This paradigm has been widely adopted in recent text and multimodal systems \cite{faysse2024colpali,granite_vision_embedding,warner2025smarter_modern_bert, nomicembedmultimodal2025,xu2025llama_nemo,gunther2025jina_v4}, becoming a standard foundation for high-accuracy neural retrieval.
However, this granularity comes with a cost. Unlike single-vector retrieval, where scoring is a cheap dot product, exact late interaction requires evaluating a grid of token-level operations (MaxSim) for every document. Consequently, this computation often becomes the bottleneck in modern pipelines, motivating methods that reduce these operations without sacrificing ranking fidelity~\citep{plaid,dessert}.

\textbf{The ``Hiring'' Analogy.} Consider a manager hiring the top-$K$ candidates from $N$ applicants, where each takes $T$ independent tests and the final score is the sum. An efficient manager proceeds in rounds: each round administers a small batch of new tests to the surviving applicants, then drops those who cannot reach the top-$K$ even if they ace every untaken test. Standard late-interaction retrieval skips the rounds: it scores every token interaction on every document, even those already ruled out.

\textbf{Our Approach: \method{} (CB).}
We view this as \emph{progressive matrix completion}: the token-level scores are values in a table revealed on demand, and our objective is to estimate the exhaustive scorer's Top-$K$ identity from a partially revealed matrix, minimizing computation while maintaining a user-defined level of statistical reliability (Figure~\ref{fig:algo}).
To this end, we introduce \method{} (CB), a \emph{purely query-time} algorithm that operates directly on \textit{vanilla ColBERT}. \method{} exploits a fundamental asymmetry that index-time methods cannot use: \textbf{the query is only known at query time}, so only at query time can the system decide which document tokens matter for \emph{this} query, and the relevant cells of the MaxSim matrix are unknowable a priori. We refer to the algorithm as CB and to its concrete kernel implementation built on the \textsc{numkong}~\citep{Vardanian_NumKong_2000_Mixed} C extension as CB-NK. Unlike prior acceleration methods that \emph{lossily} compress or distill document representations, \method{} targets the exact-MaxSim top-$K$ (surviving candidates are always rescored exactly on all $T$ tokens, and the rule that eliminates the rest is provably correct at $\alpha_{\mathrm{ef}}{=}1$) and requires \textbf{no retraining and no changes to the model or retrieval index}. Its fast kernel uses only a one-time offline repack of the candidate embeddings into a cache-friendly layout (Appendix~\ref{app:implementation}). It composes with retrieval-side systems such as PLAID and MUVERA rather than competing with them.
We formulate the task as a \textit{finite-population Top-$K$ identification problem}. By exploiting the fact that document token sequences are finite, we utilize the empirical Bernstein--Serfling inequality~\citep{bardenet2015concentration} to construct tighter confidence intervals than standard bandit approaches.
\paragraph{Contributions.}
\begin{itemize}
\setlength{\itemsep}{2pt}
\setlength{\topsep}{2pt}
\item \textbf{Formulation.} We cast late-interaction reranking as a finite-population Top-$K$ identification problem using a progressive scoring framework, exploiting the query at query time to decide which MaxSim cells to compute, a signal that index-time accelerators cannot use.
\item \textbf{Algorithm.} We introduce \method{}, a progressive multi-round elimination algorithm that leverages the empirical Bernstein--Serfling inequality~\citep{bardenet2015concentration} for tight per-document confidence bounds, with a tunable relaxation parameter $\alpha_{\mathrm{ef}}$ that provides a $\delta$-PAC certificate at $\alpha_{\mathrm{ef}}{=}1$ (under the simplified radius of Eq.~\ref{eq:effective_radius}, see Remark~\ref{rem:pac_scope}). The deployed default $\alpha_{\mathrm{ef}}{=}0.2$ is a calibrated relaxation that retains the Pareto-dominant cost--fidelity profile we report in \S\ref{sec:experiments}.
\item \textbf{Drop-in Acceleration.} We demonstrate a $\mathbf{7.0\times}$/$\mathbf{4.6\times}$ ($1$/$8$ thread) wall-clock speedup on a server CPU (AMD EPYC 7763, AVX2) and $\mathbf{12.9\times}$/$\mathbf{10.3\times}$ ($1$/$8$ thread, $K{=}5$) on Apple M1 Max (NEON, four-corpus subset), across BEIR and REAL-MM-RAG (nine corpora, up to $2.68$\,M documents, Figure~\ref{fig:headline_speedup}), preserving Overlap@$5\geq 0.90$ on every corpus tested ($\geq 0.96$ on \textsc{Cpu-S} for every corpus but HotpotQA-$500$\,K), with no index modifications or retraining. Latency scales \emph{sub-linearly} in $K$.
\end{itemize}

\section{Background and Related Work}
\label{sec:background}

\subsection{Preliminaries: Late Interaction Retrieval}
\label{sec:col_retrieval_background}

\paragraph{ColBERT Late Interaction Scoring.}
Consider a query $Q$ and a document $d$ from a collection $\mathcal{D}$ of size $N$. ColBERT represents both as sets of token embeddings:
\[
    Q=\{\mathbf{q}_{t}\}_{t=1}^{T}, \quad \mathbf{E}(d)=\{\mathbf{e}_{d,j}\}_{j=1}^{L_d},
\]
where $\mathbf{q}_t, \mathbf{e}_{d,j} \in \mathbb{R}^{l}$, $l$ is the embedding dimension, $T$ the query length, and $L_d$ the document length. The score is computed via \emph{late interaction}: for each query token $t \in [T]$, ColBERT identifies the most similar document token (MaxSim) and sums:
\begin{equation}
    \label{eq:colbert_score_maxsim}
    h(d,t) \triangleq \max_{j\in[L_d]} \mathrm{sim}( \mathbf{e}_{d,j},\mathbf{q}_t),
\end{equation}
\begin{equation}
    \label{eq:colbert_score}
    S(d;Q) \triangleq \sum_{t=1}^{T} h(d,t),
\end{equation}
where $\mathrm{sim}(\cdot, \cdot)$ is a similarity function (typically cosine).\footnote{More generally, we assume $\mathrm{sim}$ is bounded in a known interval $[a,b]$ (e.g., $[-1,1]$ for cosine on normalized vectors), hence each $h(d,t)$ is also bounded.}

\paragraph{Top-$K$ Ranking.}
The ranking objective is to identify the $K$ documents from a search set $\mathcal{D}$ (the full corpus, or a candidate set produced by an upstream stage) with the highest MaxSim scores:
\begin{equation}
\mathcal{T}^\star_K \triangleq \operatorname*{arg\,topK}_{d \in \mathcal{D}} S(d; Q).
\label{eq:topk_intro}
\end{equation}

\paragraph{Index-Time vs.\ Query-Time.}
Retrieval systems separate \textit{index-time} (offline) representation/index construction from \textit{query-time} (online) scoring. In late-interaction systems the latter is typically a \emph{reranking} stage. \textbf{This separation matters for pruning:} index-time methods must commit before seeing the query and prune conservatively, while query-time methods can use the actual query to decide what to compute. \method{} is, to our knowledge, the first method to exploit this signal at the atomic MaxSim-cell level. Standard reranking evaluates all $N \times T$ MaxSim values $h(d,t)$ (Eq.~\ref{eq:colbert_score}), which dominates query-time cost even after candidate retrieval.

\subsection{Related Work}
\label{sec:related_work}
We categorize related work by \emph{when} and \emph{what} they prune (visual taxonomy in Appendix~\ref{app:taxonomy}). To our knowledge, \method{} is the first method to \emph{adaptively use the query} to prune at the \emph{atomic} MaxSim-cell level during query-time scoring.

\paragraph{Index-time accelerators.}
Centroid compression and fixed-dimensional encodings, including PLAID \citep{plaid}, ColBERTv2 \citep{colbertv2}, MUVERA \citep{muvera}, LEMUR \citep{jaasaari2026lemur}, and WARP \citep{scheerer2025warp}, accelerate retrieval by quantizing or projecting document representations before the query arrives. Offline token pruning \citep{lassance2021study,tonellotto2021query,clavie2024reducing,macavaney2025efficient} reduces the index size $N$ or query length $T$ by permanently discarding low-utility tokens. Both families must commit at index time, before seeing the query. \method{} is \emph{orthogonal} and stacks on top, since any such system can produce the candidate set $\mathcal{D}$ that \method{} reranks.

\paragraph{Bound-based skipping.}
In sparse retrieval, WAND \citep{broder2003efficient_wand} and BMW~\citep{ding2011faster_bmw} use per-term upper bounds to skip low-scoring documents, and DESSERT \citep{dessert} applies similar bound-based reasoning to approximate set search. \method{} brings this idea to \emph{dense late-interaction}, pruning atomic MaxSim cells $h(d,t)$ with statistical (rather than term-level) bounds.

\paragraph{Top-$K$ arm identification.}
\method{} extends fixed-confidence Top-$K$ arm identification \citep{kalyanakrishnan2012pac,chen2014combinatorial,audibert2010best} from stochastic arms to a \emph{finite population sampled without replacement} (each MaxSim row is a fixed set of $T$ deterministic values). We therefore use the empirical Bernstein--Serfling concentration of \citet{bardenet2015concentration} in place of the standard sub-Gaussian radius (\S\ref{sec:bounds}).

\section{Problem Formulation}
\label{sec:formulation}

\begin{figure*}[t]
\centering
\includegraphics[width=0.85\textwidth, trim=1cm 5.5cm 3cm 0.5cm, clip]{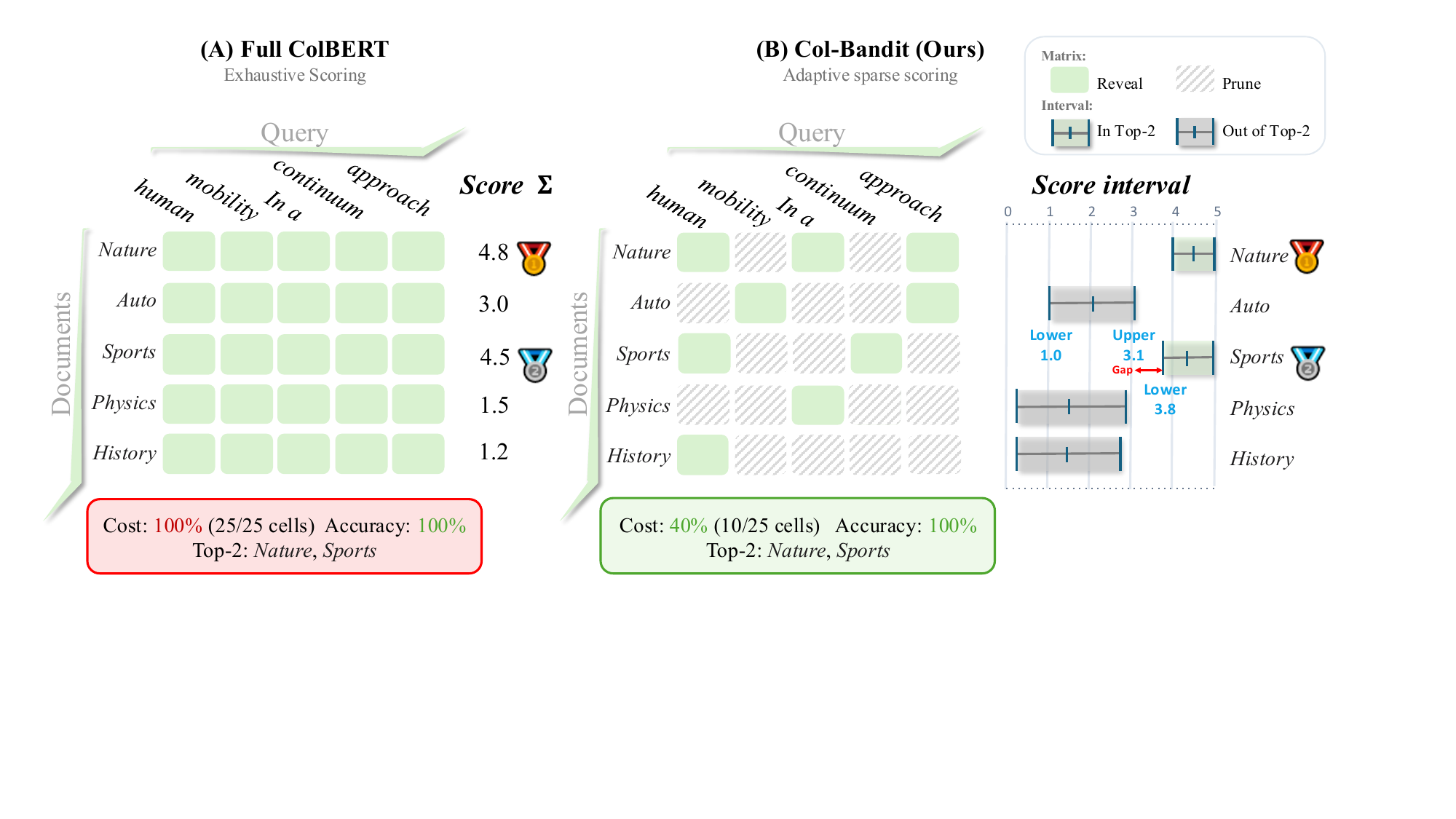}
\caption{\textbf{Intuition: the top-$K$ is identifiable from a partial view of the MaxSim matrix $H$.} Given a query and a candidate set, the goal is to identify the top-$K$ documents under exhaustive MaxSim. \textbf{(A)} Full ColBERT scores every cell of the $N\times T$ matrix. \textbf{(B)} \method{} reveals only a subset of cells (green) and skips the rest (hatched), maintaining a confidence interval $[\LCB_i, \UCB_i]$ on each document's total score. Once a positive \emph{separation gap} opens between the weakest winner's $\LCB$ and the strongest loser's $\UCB$, the top-$K$ is determined at a fraction of the compute. This panel illustrates \emph{why} partial observation suffices. The deployed reveal schedule (shared random token batches with whole-document pruning) is given in Algorithm~\ref{alg:progressive}.}
\label{fig:algo}
\end{figure*}

We construct \method{} as a statistical estimator of $\mathcal{T}^\star_K$ defined in Eq.~\eqref{eq:topk_intro}. Given a query $Q$, it outputs $\hat{\mathcal{T}}_K \subset \mathcal{D}$. Under confidence parameter $\delta$ and relaxation $\alpha_{\mathrm{ef}}$, it recovers $\mathcal{T}^\star_K$ with probability at least $1-\delta$ when $\alpha_{\mathrm{ef}}=1$. We cast this as a fixed-confidence Multi-Armed Bandit (MAB) problem over the sparsely observed, finite-population MaxSim matrix.

\subsection{The MaxSim Matrix and Observation Model}
\label{sec:sparse_matrix}
Consider a query $Q$ with $T$ tokens and a search set of $N$ documents, $\calD = \{d_1, \dots, d_N\}$.
We define the implicit \emph{MaxSim Matrix}, $H \in \R^{N \times T}$, where each entry corresponds to the maximum similarity (Eq.~\ref{eq:colbert_score}) of a query token with a document's tokens:
\begin{equation}
    H_{i,t} \triangleq h(d_i, t) = \max_{j \in [L_{d_i}]} \mathrm{sim}(\mathbf{e}_{d_i,j}, \mathbf{q}_t).
    \label{eq:def_H_matrix}
\end{equation}
The total late-interaction score for document $i$ is the row-sum:
\begin{equation}
S_i \triangleq \sum_{t=1}^T H_{i,t}.
\label{eq:def_row_sum}
\end{equation}
Our objective is to identify the set of indices $\calT^\star_K$ corresponding to the $K$ documents with the highest scores $S_i$.\\
At any step, the algorithm holds an observed set $\Omega \subseteq [N] \times [T]$ with per-document indices $\calO_i \triangleq \{t : (i,t) \in \Omega\}$ and $\calU_i \triangleq [T] \setminus \calO_i$. Revealing $(i,t) \notin \Omega$ incurs unit cost and returns $H_{i,t}$.
We measure computational cost via \emph{coverage}, defined as the fraction of the matrix revealed. At any time step of our algorithm the cost is:
\begin{equation}
    \gamma(\Omega) \triangleq \frac{|\Omega|}{N \times T} \;=\; \frac{1}{NT}\sum_{i=1}^N |\calO_i|.
    \label{eq:coverage_gamma}
\end{equation}


\section{Method: \method{}}
\label{sec:method}

\paragraph{Overview.} Figure~\ref{fig:algo} previews \method{}: from a partially revealed MaxSim matrix we maintain per-document lower/upper bounds (\S\ref{sec:bounds}) and identify the top-$K$ once the weakest winner's lower bound exceeds the strongest loser's upper bound (a positive \emph{separation gap}), and \S\ref{sec:alg_progressive} gives the round-by-round procedure that drives this gap.

\subsection{Decision Bounds}
\label{sec:bounds}

Let $n_i = |\calO_i|$ be the number of revealed query-token positions (cells) for document $i$, $\widehat{\mu}_i = \tfrac{1}{n_i} \sum_{t\in\calO_i} H_{i,t}$ the empirical mean, and $\widehat{S}_i \triangleq T\widehat{\mu}_i$ the partial-sum estimator used to order candidates.
Using the known global support $[a,b]$ of unrevealed entries (\textit{e.g.}, $[-1, 1]$ for cosine similarity), the deterministic hard bounds are $LB^{\mathrm{hard}}_i = \sum_{t\in\calO_i} H_{i,t} + (T-n_i)\,a$ and $UB^{\mathrm{hard}}_i = \sum_{t\in\calO_i} H_{i,t} + (T-n_i)\,b$. Tighter per-cell bounds (\textit{e.g.}, per-document token-norm or centroid-based upper/lower bounds) would tighten both $LB^{\mathrm{hard}}_i$ and $UB^{\mathrm{hard}}_i$. We leave them for future work and use the simpler global support throughout. Correctness depends only on $H_{i,t}\!\in\![a,b]$ holding. If $\mathrm{sim}$ is not strictly bounded (\textit{e.g.}, un-normalized embeddings), widening $[a,b]$ keeps the bounds valid and only loosens the radii, never breaking elimination safety.
We combine these with an empirical Bernstein--Serfling style decision radius~\citep{bardenet2015concentration},
\begin{equation}
    r^{\mathrm{eff}}_i \triangleq \alpha_{\mathrm{ef}} \cdot T \widehat{\sigma}_i \sqrt{\frac{2 \log(c N T / \delta)}{n_i}} \cdot \sqrt{\rho_{n_i}},
    \label{eq:effective_radius}
\end{equation}
where $\widehat{\sigma}_i$ is the empirical standard deviation over revealed entries and $\rho_{n_i}$ is a finite-population correction with $\rho_{n_i}\!\to\!0$ as $n_i\!\to\!T$ (Appendix~\ref{app:variance_details}).
The relaxation $\alpha_{\mathrm{ef}}\in(0,1]$ controls conservativeness: $\alpha_{\mathrm{ef}}{=}1$ recovers the unshrunk empirical Bernstein--Serfling form, while $\alpha_{\mathrm{ef}}{<}1$ tightens the radius.
The hybrid decision interval is
\begin{equation}
\begin{aligned}
\LCB_i &= \max(LB^{\mathrm{hard}}_i,\, \widehat{S}_i - r^{\mathrm{eff}}_i), \\
\UCB_i &= \min(UB^{\mathrm{hard}}_i,\, \widehat{S}_i + r^{\mathrm{eff}}_i).
\end{aligned}
\label{eq:LCB_UCB}
\end{equation}

\subsection{Batched Progressive Elimination}
\label{sec:alg_progressive}

\paragraph{From rule to schedule.} A per-cell loop would be dominated by dispatch overhead (a MaxSim cell is only a few FLOPs), so we run \method{} in \emph{rounds} over an \emph{active set} $\calA_r\subseteq[N]$ ($\calA_1=[N]$). Each round reveals the next $B$ cells for every $i\in\calA_r$ in one vectorized pass, refreshes the bounds, and eliminates any document whose $\UCB_i$ falls below the $K$-th largest $\LCB$ over $\calA_r$. Per-round bookkeeping amortizes across $B|\calA_r|$ cells.

\paragraph{Pay-to-prune.} Each round costs $B|\calA_r|$ evaluations, but every eliminated document saves its remaining $T-rB$ cells, so the active set shrinks monotonically and the cumulative budget is $\sum_r B|\calA_r|\ll NT$ in practice (\S\ref{sec:experiments}).

We instantiate this as Algorithm~\ref{alg:progressive} (fully annotated in Appendix~\ref{app:algo_detailed}). The reveal schedule is a single per-query permutation $\pi$ of $[T]$, drawn uniformly with a fixed RNG seed and shared across surviving documents (the uniform-without-replacement structure is what makes Theorem~\ref{thm:pac} applicable, see Appendix~\ref{app:uniform_pac}). A safety margin $M$ ($M{=}5$ by default) preserves up to $K{+}M$ borderline survivors, which are then rescored on all $T$ query tokens via the same fused MaxSim kernel as Full-MaxSim, so the final survivor scores are \emph{bit-identical} to the exhaustive baseline. Ties at the $K$-th boundary are broken by document index and deferred to the exact rescore. We use $B{=}4$ throughout, matching the SIMD register tile of our fused C kernel (Appendix~\ref{app:implementation}). The Bernstein--Serfling pre-factor $c$ in Eq.~\ref{eq:effective_radius} and the $n_i\!\le\!1$ case are detailed in Appendix~\ref{app:variance_details}.

\begin{algorithm}[t]
\caption{\method{}: batched progressive elimination for top-$K$.}
\label{alg:progressive}
\small
\begin{algorithmic}[1]
\Require Query $Q$ ($T$ tokens), candidates $\calD$ ($N$ docs); $K, M, B, \alpha_{\mathrm{ef}}, \delta$.
\Ensure Estimated top-$K$ set $\widehat{\calT}_K$.
\State $r \gets 1$;\quad $\calA_1 \gets [N]$;\quad $\Omega \gets \emptyset$;\quad draw a uniform permutation $\pi$ of $[T]$.
\While{$|\calA_r| > K + M$ \textbf{and} $(r{-}1)B < T$}
    \State Reveal the next $B$ cells of $\pi$ for every $i \in \calA_r$ (one SIMD pass); add them to $\Omega$ and refresh $[\LCB_i, \UCB_i]$ via Eq.~\ref{eq:LCB_UCB}.
    \State $\tau_r \gets$ $K$-th largest $\LCB_i$ over $i \in \calA_r$.
    \State $\calA_{r+1} \gets \{i \in \calA_r : \UCB_i \ge \tau_r\}$;\quad $r \gets r+1$.
\EndWhile
\State Rescore each survivor: $S_i \gets \sum_{t=1}^{T} H_{i,t}$ via the fused MaxSim kernel.
\State \Return $\widehat{\calT}_K \gets \arg\mathrm{topK}_{i \in \calA_r} S_i$.
\end{algorithmic}
\end{algorithm}

\paragraph{Certified corner and deployed relaxation.} Under uniform-without-replacement reveals, $\alpha_{\mathrm{ef}}{=}1$ admits a $\delta$-PAC guarantee (under the simplified radius of Eq.~\ref{eq:effective_radius}, see Remark~\ref{rem:pac_scope}): $\Prob(\widehat{\calT}_K{=}\calT^\star_K)\!\ge\!1{-}\delta$ (Theorem~\ref{thm:pac} and proof in Appendix~\ref{app:uniform_pac}). \emph{Sketch:} the empirical Bernstein--Serfling interval holds per row at a fixed reveal count. Setting the per-cell budget to $\delta/(NT)$ and union-bounding over all $N$ documents and $T$ reveal counts (the $\log(cNT/\delta)$ term in Eq.~\ref{eq:effective_radius}) makes every $[\LCB_i,\UCB_i]$ valid \emph{simultaneously}, hence also at the data-dependent stopping time, side-stepping optional-stopping issues. A document is eliminated only once its $\UCB_i$ provably falls below the $K$-th largest $\LCB$, so no true top-$K$ member is dropped on this $1{-}\delta$ event. The constant $c$ is the Bernstein--Serfling pre-factor, set to $1$ and absorbed into $\alpha_{\mathrm{ef}}$ (Appendix~\ref{app:variance_details}). For $\alpha_{\mathrm{ef}}{<}1$ the radii shrink and elimination is more aggressive, trading the certificate for coverage. The deployed default $\alpha_{\mathrm{ef}}{=}0.2$ is this calibrated relaxation, with no formal certificate but $\geq 0.90$ overlap with the exhaustive top-$K$ on every corpus tested (Overlap@$K$ defined in \S\ref{sec:experiments}).

\section{Experiments}
\label{sec:experiments}
\begin{table*}[t]
\centering
\caption{\textbf{Universal Efficiency Analysis: Overlap@$K$ and nDCG@$K$ at $K\!\in\!\{5,50\}$.} Each cell: \textbf{mean coverage \% (std across corpora)} to recover \textbf{95\% (near-lossless)} of Full-MaxSim's own Overlap@$K$ / nDCG@$K$ per dataset (lower is better). The 90\% threshold and per-dataset breakdowns are in Appendix~\ref{sec:app_per_dataset_eff}. Quora's nDCG/Recall/MRR are undefined under our harness (a qrels/doc-id mapping issue; Table~\ref{tab:dataset_stats}) and do not contribute to the nDCG averages here, which match the Quora-excluded per-corpus means (Appendix~\ref{sec:app_per_dataset_eff}); Overlap@$K$ is unaffected.}
\label{tab:main_benchmark_universal_eff_combined}
\label{tab:main_benchmark_text_top1_top5_compact}
\label{tab:main_benchmark_ndcg_compact}
\small
\setlength{\tabcolsep}{8pt}
\renewcommand{\arraystretch}{1.15}
\resizebox{0.65\textwidth}{!}{%
\begin{tabular}{l|cc||cc}
\toprule
\textbf{Method} &
\textbf{Overlap@5} & \textbf{Overlap@50} &
\textbf{nDCG@5} & \textbf{nDCG@50} \\
\midrule

\multicolumn{5}{l}{\textit{\textbf{ColBERTv2 (BEIR)}}} \\
Doc-Uniform
    & 98\% (3.8)
    & 100\% (0.0)
    & 57\% (10.3)
    & 38\% (11.7) \\
Ball-carving
    & 46\% (13.3)
    & 43\% (12.2)
    & 29\% (11.1)
    & 23\% (4.3) \\
\textbf{\method{} (Ours)}
    & \textbf{14\% (1.7)}
    & \textbf{18\% (5.1)}
    & \textbf{13\% (0.8)}
    & \textbf{13\% (0.6)} \\

\midrule
\multicolumn{5}{l}{\textit{\textbf{Jina-ColBERTv2 (BEIR)}}} \\
Doc-Uniform
    & 100\% (0.0)
    & 100\% (0.0)
    & 59\% (30.7)
    & 57\% (27.7) \\
Ball-carving
    & 69\% (32.6)
    & 88\% (21.3)
    & 20\% (6.3)
    & 19\% (5.6) \\
\textbf{\method{} (Ours)}
    & \textbf{26\% (16.9)}
    & \textbf{39\% (17.8)}
    & \textbf{9\% (3.5)}
    & \textbf{14\% (7.5)} \\

\midrule
\multicolumn{5}{l}{\textit{\textbf{Granite-Vision-Embedding (REAL-MM-RAG)}}} \\
Doc-Uniform
    & 100\% (0.0)
    & 98\% (4.1)
    & 54\% (18.6)
    & 49\% (16.0) \\
Ball-carving
    & 76\% (1.7)
    & 76\% (1.7)
    & 28\% (9.8)
    & 27\% (10.9) \\
\textbf{\method{} (Ours)}
    & \textbf{22\% (1.9)}
    & \textbf{41\% (3.1)}
    & \textbf{15\% (5.7)}
    & \textbf{15\% (2.7)} \\

\bottomrule
\end{tabular}%
}
\end{table*}

\subsection{Experimental Setup}
\label{sec:exp_setup}
\noindent
\textbf{Corpora.} Five BEIR language corpora (ArguAna, SciDocs, NQ-$2.68$M, HotpotQA-$500$K, Quora-Full~\citep{thakur2021beir}) and four REAL-MM-RAG~\citep{wasserman2025real_mm} multimodal corpora (FinSlides, FinReport, TechSlides, TechReport), plus a merged \textsc{MM} set (the four REAL-MM-RAG corpora pooled) for the CPU wall-clock benchmarks ($\textsc{Cpu-S}$ and $\textsc{Cpu-M1}$). Per-corpus details in Appendix~\ref{app:datasets_models} (Table~\ref{tab:dataset_stats}). $N$ is the full corpus rerank size (e.g., $N{=}2.68$M for NQ).\\
\textbf{Encoders.} ColBERTv2~\citep{colbertv2} on all text corpora, Jina-ColBERTv2~\citep{jha2024jina} ($d{\in}\{128,64\}$ Matryoshka) on four BEIR corpora, and Granite Vision Embedding 3.2~\citep{granite_vision_embedding} for REAL-MM-RAG.\\
\textbf{Hardware.} $\textsc{Cpu-S}$ = AMD EPYC 7763 (AVX2 server), $\textsc{Cpu-M1}$ = Apple M1 Max (NEON laptop), and $\textsc{Gpu}$ = NVIDIA A100 $80$\,GB. A single \textsc{numkong}~\citep{Vardanian_NumKong_2000_Mixed} C extension drives both CPUs via SIMD-backend swap (Appendix~\ref{app:impl_details}).\\
\textbf{Baselines.}\label{sec:exp_baselines}~Full-MaxSim (exhaustive oracle), Doc-Uniform (non-adaptive random-cell reveal at coverage $\gamma$, Appendix~\ref{app:bl_doc_uniform}), and Ball carving~\citep{muvera} (Stage-2 query-time pruning peer, Appendix~\ref{app:bl_ball_carving}). All methods are built on the same \textsc{numkong} kernel and differ in what they reveal or compress per query (Appendix~\ref{app:experiments}).\\
\textbf{Metrics.}\label{sec:exp_metrics}~Our metric measures how \method{} reproduces Full-MaxSim's top-$K$ set:
\begin{equation}
\mathrm{Overlap}@K \;\triangleq\; \frac{\bigl|\,\mathcal{T}^\star_K \cap \hat{\mathcal{T}}_K\,\bigr|}{K}\, ,
\label{eq:overlap_def}
\end{equation}
where $\mathcal{T}^\star_K$ is the exhaustive top-$K$ and $\hat{\mathcal{T}}_K$ is \method{}'s estimate, with $\mathrm{Overlap}@K\!=\!1$ iff the two sets coincide. We report $K\!\in\!\{5,50,100\}$, coverage $\gamma$ (Eq.~\ref{eq:coverage_gamma}), and per-query wall-clock latency. Task-level metrics (nDCG, Recall, MRR) are in Appendix~\ref{sec:app_effectiveness_k1}--\ref{sec:app_effectiveness_k5_realmm}.\\
\textbf{CB knobs.} Deployed: $\alpha_{\mathrm{ef}}{=}0.2$, $M{=}5$ (rescore margin), $\delta{=}0.01$. Round size $B{=}4$ matches the SIMD register tile of the fused C kernel. Pareto sweeps span $\alpha_{\mathrm{ef}}\in(0,1]$ (Appendix~\ref{app:parameter_selection}). The $\alpha_{\mathrm{ef}}{=}1$ corner satisfies Theorem~\ref{thm:pac}.

\subsection{Main Results: Coverage Savings Translate to Wall-Clock Latency}
\label{sec:main_results}
\method{} reaches the exhaustive late-interaction top-$K$ at a small fraction of the MaxSim cost, portably across server (AVX2) and edge (ARM NEON) CPUs.

\begin{figure}[t]
\centering
\includegraphics[width=\columnwidth]{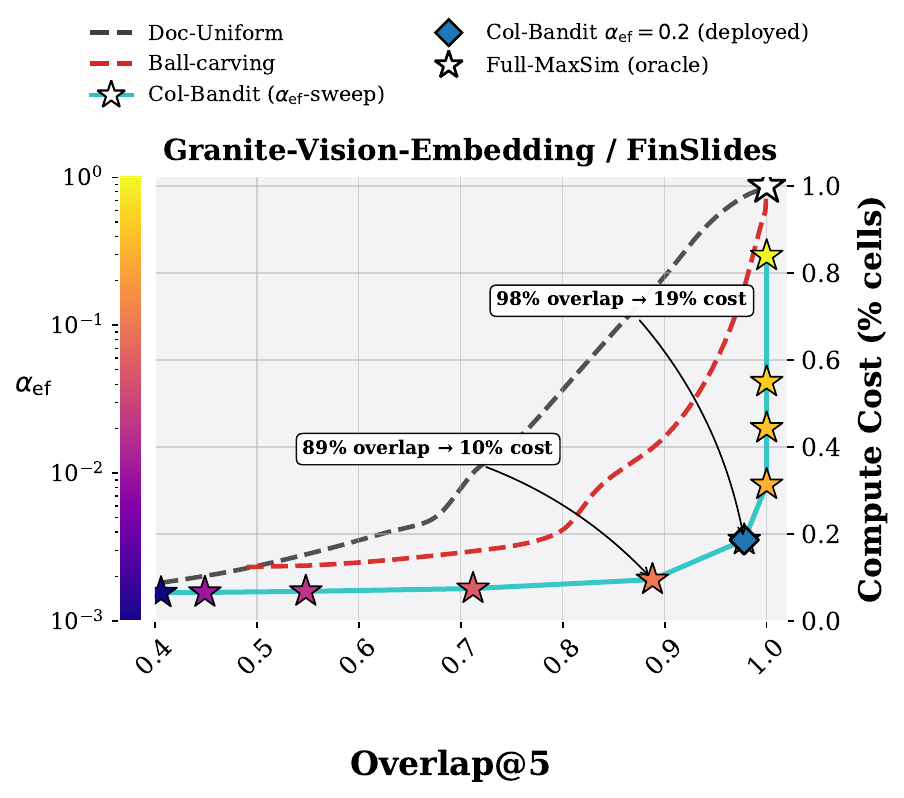}
\caption{\textbf{Quality--coverage Pareto frontier ($K{=}5$), GVE/FinSlides.} \method{} (sweeping $\alpha_{\mathrm{ef}}$) vs.\ Doc-Uniform and Ball-carving (MUVERA App.~C.3). \method{} dominates both baselines, reaching high fidelity at $\leq 20\%$ of the cell budget. The deployed knob $\alpha_{\mathrm{ef}}{=}0.2$ marked, Full-MaxSim oracle at top-right. Additional comparisons in Appendix Figure~\ref{fig:results_grid_}.}
\label{fig:pareto_finslides}
\end{figure}

\paragraph{Coverage savings.}
Table~\ref{tab:main_benchmark_universal_eff_combined} reports the mean coverage budget needed to recover $95\%$ (near-lossless) of Full-MaxSim's own Overlap@$K$ and nDCG@$K$ at $K\!\in\!\{5,50\}$ across the BEIR (language) and REAL-MM-RAG (multimodal) suites. \method{} dominates both non-adaptive baselines on every dataset. Figure~\ref{fig:pareto_finslides} confirms this for GVE/FinSlides: the $\alpha_{\mathrm{ef}}$-swept frontier (compute cost vs.\ Overlap@$5$) lies strictly below Doc-Uniform and Ball-carving. The deployed $\alpha_{\mathrm{ef}}{=}0.2$ recovers ${\approx}98\%$ Overlap@$5$ at $19\%$ cost. \textbf{Headline:} at the $95\%$ near-lossless threshold, \method{} reaches Overlap@$5$ at $\mathbf{14}$\%/$\mathbf{26}$\%/$\mathbf{22}$\% coverage on ColBERTv2/Jina/GVE respectively ($\mathbf{4.6}$--$\mathbf{7.0\times}$ savings), versus $98$--$100\%$ for Doc-Uniform and $46$--$76\%$ for Ball-carving, a $2$--$5\times$ gap that holds on every corpus and at $K{=}50$ as well. Doc-Uniform's near-total coverage makes it a redundancy \emph{upper bound}, so Ball-carving is the substantive baseline. The savings widen for ranking quality: nDCG@$5$ recovers at just $\mathbf{9\%}$ coverage on Jina ($\mathbf{10.9\times}$) and $\mathbf{13\%}$ on ColBERTv2 ($\mathbf{7.6\times}$). Recall@$5$ and MRR@$5$ follow the same coverage-savings pattern (\method{} reaches $95\%$ retention at $13$--$26\%$ coverage on every corpus, with per-corpus tables in Appendix~\ref{sec:app_per_dataset_eff}). The looser $90\%$ threshold is in Appendix~\ref{sec:app_universal_eff_90pct}.

\paragraph{Wall-clock translation.}
\label{sec:exp_m1}
Coverage savings translate to wall-clock speedups via a first-order model: every method here shares the same fused C MaxSim kernel (\textsc{numkong}, Appendix~\ref{app:impl_details}) and differs mainly in which $(d, t)$ cells it touches (\method{} uses an int8 elimination pass and an exact fp32 rescore, bit-identical to Full-MaxSim), so per-query latency is dominated by kernel work. The bandit bookkeeping (elimination decisions + the $K{+}M$ rescore) stays below $4\%$ of total runtime across the eight corpora profiled in Appendix~\ref{app:overhead_breakdown}, giving the heuristic $\mathrm{Speedup}\!\sim\!100/\mathrm{Cov}\%$. Per-query wall-clock latency on $\textsc{Cpu-S}$ is reported in Table~\ref{tab:wallclock_cpu} (visualized as Figure~\ref{fig:headline_speedup}, with extended sweeps at other $\alpha_{\mathrm{ef}}$ and $K$ values in Appendix~\ref{sec:app_headline_variants}), and on $\textsc{Cpu-M1}$ in Table~\ref{tab:m1_neon}. At the deployed knob ($\alpha_{\mathrm{ef}}{=}0.2$, $M{=}5$, $\delta{=}0.01$), \method{} preserves Overlap@$K \geq 0.90$ on every corpus and delivers a $\mathbf{7.0\times}$ mean single-thread speedup on $\textsc{Cpu-S}$ ($K{=}5$, falling to $5.3\times$ at $K{=}100$) and $\mathbf{12.9\times}$/$\mathbf{10.3\times}$ ($1$/$8$ thread, $K{=}5$) on $\textsc{Cpu-M1}$ ($9.0\times$/$7.0\times$ at $K{=}100$). Against \emph{maxsim-cpu}~\citep{maxsimcpu2025mxbai}, the published Rust SIMD MaxSim baseline, \method{} is $\mathbf{4.8\times}$ faster on average at $1$ thread on $\textsc{Cpu-M1}$ ($K{=}5$) at comparable fidelity (Table~\ref{tab:m1_neon}). Absolute end-task quality tracks the oracle: at ${\sim}20\%$ coverage \method{} retains $\geq 98.7\%$ of Full ColBERT's Recall@$1$/nDCG@$1$/MRR@$1$ (vs.\ $81\text{--}82\%$ Ball-carving, $56\%$ Doc-Uniform, Appendix~\ref{sec:app_effectiveness_k1}).

\begin{table*}[t]
\centering
\caption{\textbf{Wall-clock CPU benchmarks on \textsc{Cpu-S}} (AMD EPYC 7763, AVX2) at the deployed knob ($\alpha_{\mathrm{ef}}{=}0.2$, $M{=}5$, $\delta{=}0.01$), \textbf{single thread}. Each cell shows $K{=}5$\,/\,$K{=}100$. Ov@$K$ is overlap with Full-MaxSim's exhaustive top-$K$ (Eq.~\ref{eq:overlap_def}); Cov is the fraction of $(d,t)$ MaxSim cells revealed. \method{} delivers a $\mathbf{7.0\times}$\,/\,$\mathbf{5.3\times}$ ($K{=}5$\,/\,$K{=}100$) mean single-thread speedup across the six \textsc{Cpu-S} settings (five BEIR corpora and the REAL-MM-RAG aggregate) while preserving Ov@$5\!\geq\!0.93$ on every corpus. Full-MaxSim latency is $K$-invariant up to measurement noise. Threads parallelise across queries, so multi-thread figures are throughput (per-query latency under concurrent load), not single-query speedup. Multi-thread ($1$t\,/\,$8$t) breakdown in Appendix Table~\ref{tab:wallclock_cpu_threads}.}
\label{tab:wallclock_cpu}
\scriptsize
\setlength{\tabcolsep}{5pt}
\renewcommand{\arraystretch}{1.1}
\resizebox{0.8\textwidth}{!}{%
\begin{tabular}{lr|ccccc}
\toprule
\textbf{Dataset} & \textbf{$N$} & \textbf{Ov@$K$} & \textbf{Cov} & \textbf{Full (ms)} & \textbf{CB (ms)} & \textbf{Sp.} \\
\midrule
ArguAna  & $8.7$\,K   & 0.98 / 0.97 & 14\% / 20\% & 211 / 210           & \textbf{33 / 50}          & 6.4$\times$ / 4.2$\times$ \\
SciDocs  & $25.7$\,K  & 0.96 / 0.93 & 14\% / 18\% & 513 / 515           & \textbf{81 / 108}         & 6.3$\times$ / 4.8$\times$ \\
HotpotQA & $500$\,K   & 0.93 / 0.93 & 13\% / 14\% & 5{,}568 / 5{,}153   & \textbf{757 / 825}        & 7.4$\times$ / 6.2$\times$ \\
Quora    & $522.9$\,K & 0.99 / 0.97 & 13\% / 13\% & 1{,}778 / 1{,}806   & \textbf{381 / 393}        & 4.7$\times$ / 4.6$\times$ \\
NQ       & $2.68$\,M  & 0.98 / 0.97 & 13\% / 13\% & 37{,}283 / 37{,}224 & \textbf{4{,}222 / 4{,}384} & 8.8$\times$ / 8.5$\times$ \\
MM       & $8.6$\,K   & 0.97 / 0.95 & 21\% / 47\% & 527 / 523           & \textbf{65 / 164}         & 8.1$\times$ / 3.2$\times$ \\
\midrule
\rowcolor{gray!15}\textbf{Mean} & --- & --- & --- & --- & --- & \textbf{7.0$\times$ / 5.3$\times$} \\
\bottomrule
\end{tabular}%
}
\end{table*}

\begin{table*}[t]
\centering
\caption{\textbf{Cross-platform benchmark on \textsc{Cpu-M1}} (Apple M1 Max, NEON laptop) at the deployed knob, \textbf{single thread}. Each cell shows $K{=}5$\,/\,$K{=}100$; ColBERTv2 on text, GVE on MM. \method{} delivers a $\mathbf{12.9\times}$\,/\,$\mathbf{9.0\times}$ ($K{=}5$\,/\,$K{=}100$) mean single-thread speedup vs.\ Full-MaxSim and $\mathbf{4.8\times}$\,/\,$\mathbf{3.5\times}$ vs.\ \emph{m-cpu}~\citep{maxsimcpu2025mxbai} (Rust SIMD baseline). Overlap@$K$ can differ slightly from \textsc{Cpu-S} because int8 dot products round differently across SIMD back-ends (Appendix~\ref{app:impl_details}). Threads parallelise across queries, so multi-thread figures are throughput, not single-query latency. Multi-thread ($1$t\,/\,$8$t) breakdown in Appendix Table~\ref{tab:m1_neon_threads}.}
\label{tab:m1_neon}
\scriptsize
\setlength{\tabcolsep}{4pt}
\renewcommand{\arraystretch}{1.1}
\resizebox{0.8\textwidth}{!}{%
\begin{tabular}{lr|cccccc}
\toprule
\textbf{Dataset} & \textbf{$N$} & \textbf{Ov@$K$} & \textbf{Cov} & \textbf{Full (ms)} & \textbf{CB (ms)} & \textbf{Sp.\ vs Full} & \textbf{vs m-cpu} \\
\midrule
ArguAna  & $8.7$\,K   & 0.95 / 0.96 & 14\% / 20\% & 284 / 289           & \textbf{23 / 36}    & 12.3$\times$ / 8.0$\times$  & 4.6$\times$ / 2.8$\times$ \\
SciDocs  & $25.7$\,K  & 0.92 / 0.93 & 14\% / 18\% & 821 / 832           & \textbf{68 / 93}    & 12.1$\times$ / 8.9$\times$  & 6.0$\times$ / 4.4$\times$ \\
MM       & $8.6$\,K   & 0.99 / 0.97 & 21\% / 47\% & 1{,}636 / 1{,}657   & \textbf{103 / 170}  & 15.9$\times$ / 9.7$\times$  & 4.0$\times$ / 2.4$\times$ \\
HotpotQA & $500$\,K   & 0.90 / 0.94 & 13\% / 14\% & 7{,}394 / 7{,}215   & \textbf{660 / 766}  & 11.2$\times$ / 9.4$\times$  & 4.7$\times$ / 4.3$\times$ \\
\midrule
\rowcolor{gray!15}\textbf{Mean} & --- & --- & --- & --- & --- & \textbf{12.9$\times$ / 9.0$\times$} & \textbf{4.8$\times$ / 3.5$\times$} \\
\bottomrule
\end{tabular}%
}
\end{table*}

\paragraph{Commodity CPU vs.\ GPU at the deployed knob.}
Table~\ref{tab:cpu_vs_a100} pairs $16$-thread \method{} on $\textsc{Cpu-S}$ against $\textsc{Gpu}$ PyTorch dense Full-MaxSim ($K{=}5$). This is an \emph{architecture-level} pairing, not an apples-to-apples algorithm comparison. On the two text corpora with $N \geq 500$\,K ($\alpha_{\mathrm{ef}}{=}0.2$, Ov@5\,$\geq 0.92$), $\textsc{Cpu-S}$ lands within $1.06$--$1.41\times$ $\textsc{Gpu}$ latency. NQ-$2.68$M overflows the $80$\,GB $\textsc{Gpu}$ budget ($\sim\!140$\,GB at fp16) yet $\textsc{Cpu-S}$ reranks it in $472$\,ms. On smaller corpora ($N \leq 26$\,K) $\textsc{Gpu}$ is $2$--$4\times$ faster, as expected.

\subsection{Sensitivity, Composability, and Calibration}
\label{sec:exp_ksensitivity}
\paragraph{Sub-linear scaling in $K$.}
Does the algorithm collapse outside the small-$K$ rerank regime? Sweeping $K\!\in\!\{5,100,500,1000\}$ at the deployed knob (Table~\ref{tab:k_sensitivity}), \method{} latency grows far slower than $K$ (sub-linearly) while Full-MaxSim is essentially flat: a $200\times$ $K$-jump ($5\!\to\!1000$) costs only $1.3\times$ on HotpotQA and $2.7\times$ on SciDocs, though $4.5\times$ on the small ArguAna corpus. This reflects the adaptivity of the elimination phase rather than a property of $K$ per se: with the $K$-margin rescore fixed at $M{=}5$, larger $K$ only forces the elimination phase to retain a bigger active set at termination, which it absorbs cheaply on large corpora (per-document decisions amortize over many documents) but less so on small ones like ArguAna. The headline-speedup chart at $K{=}100$ is shown in Appendix~\ref{sec:app_headline_variants} (Figure~\ref{fig:headline_K100}).

\paragraph{Rescore margin $M$.}
We sweep $M\in\{0,5,10\}$ on SciDocs and HotpotQA-$500$\,K at $K{=}5$, $\alpha_{\mathrm{ef}}{=}0.2$, $\delta{=}0.01$, $1$ thread, $50$ queries per corpus. Without the rescore ($M{=}0$), borderline-eliminated documents leak through and Ov@$5$ drops to $0.93\text{--}0.94$. Enabling $M{=}5$ lifts Ov@$5$ to $\geq 0.98$ at a small latency cost ($+6\%$/$+21\%$ on HotpotQA/SciDocs). $M{=}10$ matches $M{=}5$ on fidelity: five slots already absorb the eliminator's borderline mistakes (Appendix~\ref{app:parameter_selection}, Table~\ref{tab:ablations_M}).

\paragraph{Cross-encoder: trends are not encoder-specific.}\label{sec:exp_crossencoder}
Are the savings an artifact of ColBERTv2's embedding geometry? We re-run the universal-efficiency analysis with Jina-ColBERTv2~\citep{jha2024jina} ($L_d{\le}256$, $d{=}128$) and its $d{=}64$ Matryoshka projection on four BEIR corpora (Table~\ref{tab:main_benchmark_universal_eff_combined}, middle block, where Jina-NQ embeddings were not built and are marked ``n/a'' in Appendix~\ref{sec:app_per_dataset_eff}). At $95\%$ Overlap@$5$, \method{} costs $\mathbf{26\%}$ coverage on Jina (vs.\ $100\%$ Doc-Uniform, $69\%$ Ball-carving). On nDCG@$5$ the saving widens to $\mathbf{9\%}$ coverage ($\mathbf{10.9\times}$). The same conclusion holds at $d{=}64$, confirming the savings behave as a property of multi-vector representations, not of a single encoder.

\paragraph{Compression composability.}
\method{} composes orthogonally with two complementary axes: \emph{dimension reduction} (Matryoshka) and \emph{token-axis pooling} (Ward~\citep{clavie2024reducing}). Under Matryoshka $d{=}128\!\to\!64$, Ov@$5$ retention stays $\geq 0.95$ on every Jina-ColBERTv2 BEIR corpus and coverage is essentially invariant to the projection (Appendix~\ref{app:matryoshka}). Under Ward $k\times$ token pooling ($k\!\in\!\{2,4,8\}$, per-doc proportional), the end-to-end Ov@$5$ loss decomposes empirically into a pooling-induced term ($9$--$41$\,pp, median $20$) and \method{}'s elimination cost ($0.8$--$4.0$\,pp, median $2.0$). Pooling dominates by an order of magnitude across the six (corpus, $k$) cells we measured (Appendix~\ref{app:ward}, single-seed measurement). The two axes stack with \method{}'s cell-skipping multiplicatively: \emph{fewer dimensions $\times$ fewer tokens $\times$ fewer cells}.

\paragraph{Choosing the calibration knob $\alpha_{\mathrm{ef}}$.}\label{sec:exp_alpha_choice}
$\alpha_{\mathrm{ef}}\in(0,1]$ is the single continuous Pareto knob: smaller values tighten the decision radius and cut more aggressively. Setting $\alpha_{\mathrm{ef}}{=}1$ recovers the unshrunk empirical Bernstein--Serfling radius (the $\delta$-PAC corner of Theorem~\ref{thm:pac}). Table~\ref{tab:alpha_guidance} gives operating-point ranges across the four BEIR text corpora at $K{=}5$: the aggressive $\alpha_{\mathrm{ef}}{=}0.1$ holds ${\sim}13\%$ coverage but drops Overlap@$5$ to $0.76$ on the toughest corpus, the deployed $\alpha_{\mathrm{ef}}{=}0.2$ keeps $13$--$14\%$ coverage at Overlap@$5\geq 0.92$, and the $\alpha_{\mathrm{ef}}{=}1$ setting reaches Overlap@$5{=}1.00$ at $28$--$59\%$ coverage. Practitioners pick along this Pareto: $0.2$ when speed dominates, $0.3$ for fidelity headroom, $1.0$ for the PAC guarantee. The headline-speedup chart at each $\alpha_{\mathrm{ef}}\!\in\!\{0.1, 0.3, 1.0\}$ is in Appendix~\ref{sec:app_headline_variants} (Figures~\ref{fig:headline_K5_a01}, \ref{fig:headline_K5_a03}, \ref{fig:headline_K5_a1}).
\newblock
\section{Conclusion}
\label{sec:conclusion}

We presented \method{}, an adaptive framework for accelerating late-interaction reranking at query time by selectively revealing MaxSim cells until the top-$K$ set stabilizes.
Across BEIR and REAL-MM-RAG, \method{} consistently exposes substantial redundancy in dense late-interaction scoring, reducing MaxSim FLOPs by up to $\mathbf{\sim 8\times}$ and translating these savings to a $\mathbf{7.0\times}/\mathbf{4.6\times}$ ($1$/$8$ thread) wall-clock speedup on a server CPU and $\mathbf{12.9\times}/\mathbf{10.3\times}$ ($1$/$8$ thread) on Apple M1 Max (four-corpus subset), while preserving high overlap with exhaustive reranking (Ov@$5\geq 0.90$ on every corpus tested, and $\geq 0.96$ on \textsc{Cpu-S} for every corpus but HotpotQA-$500$\,K).
A single calibration knob, $\alpha_{\mathrm{ef}}$ (Eq.~\ref{eq:effective_radius}), provides a practical control over the quality--compute trade-off and yields strong Pareto frontiers against non-adaptive baselines. The $\alpha_{\mathrm{ef}}{=}1$ corner recovers a $\delta$-PAC guarantee (Theorem~\ref{thm:pac}, scoped in Remark~\ref{rem:pac_scope}).
\method{} is a drop-in reranking layer requiring no retraining or index changes, making it easy to deploy on top of standard search pipelines.

\newpage
\section*{Limitations}
\method{} has three main limitations. \textbf{(i) Hardware scope.} Wall-clock evaluation is restricted to two CPU classes (an AVX2 server and a NEON laptop). We have not built a GPU-native fused kernel, so the GPU comparison uses an unfused PyTorch baseline. \textbf{(ii) Fidelity ceiling.} \method{} estimates the exhaustive MaxSim top-$K$. It cannot exceed the quality of the underlying scorer, only recover it at lower cost. \textbf{(iii) Theory vs.\ deployment.} The $\delta$-PAC guarantee (Theorem~\ref{thm:pac}) holds at the $\delta$-PAC corner $\alpha_{\mathrm{ef}}{=}1$ with the full Bernstein--Serfling radius. The deployed $\alpha_{\mathrm{ef}}{=}0.2$ uses a calibrated, simplified radius (pre-factor $c{=}1$, lower-order term omitted) and makes elimination decisions on int8-quantised estimates, so it carries no formal certificate, trading it for the cost--fidelity operating point we report. Separately, Quora's end-task metrics (nDCG/Recall/MRR) are unavailable in our harness (a qrels/doc-id mapping issue), so its task-metric averages are excluded throughout. Overlap@$K$, our primary fidelity metric, is unaffected.

\section*{Ethics Statement}
This work accelerates an existing retrieval primitive (MaxSim) on public
benchmarks (BEIR, REAL-MM-RAG). It introduces no new datasets, models, or
human-subjects data. The only dual-use consideration is the generic one
shared by any retrieval-efficiency method (faster search of an already
indexed corpus). No new model capability is introduced.


\bibliography{custom}

\appendix
\onecolumn

\setlength{\intextsep}{3pt plus 1pt minus 1pt}
\setlength{\textfloatsep}{4pt plus 1pt minus 1pt}
\setlength{\floatsep}{3pt plus 1pt minus 1pt}
\setlength{\abovecaptionskip}{3pt}
\setlength{\belowcaptionskip}{2pt}

\section{Details of Variance-Adaptive Radius}
\label{app:variance_details}
\paragraph{Empirical Standard Deviation.}
The empirical standard deviation $\widehat{\sigma}_i$ used in the standard variance bound is calculated over the set of observed tokens $\mathcal{O}_i$:
\begin{equation}
\widehat{\sigma}_i^2
=
\frac{1}{n_i-1}
\sum_{t\in\mathcal{O}_i}
\left(H_{i,t}-\widehat{\mu}_i\right)^2.
\end{equation}
In the edge case where $n_i \le 1$, the variance is undefined; we strictly set $r^{\mathrm{eff}}_i = +\infty$ and rely solely on the deterministic hard bounds.

\paragraph{Finite Population Correction ($\rho_n$).}
The term $\rho_{n_i}$ in Eq.~\eqref{eq:effective_radius} accounts for sampling without replacement from a finite set of size $T$. It is defined piecewise as:
\begin{equation}
\rho_n \triangleq
\begin{cases}
1-\dfrac{n-1}{T}, & n \le T/2,\\[8pt]
\left(1-\dfrac{n}{T}\right)\left(1+\dfrac{1}{n}\right), & n > T/2.
\end{cases}
\end{equation}
This formulation ensures that the confidence interval shrinks faster than standard Bernstein bounds as $n \to T$. Specifically, when $n=T$, the term $(1 - n/T)$ becomes zero, collapsing the radius entirely as required for a fully observed document.

\paragraph{Bernstein--Serfling Constant $c$ and $O(1/n)$ Bias Term.}
The constant $c$ in Eq.~\eqref{eq:effective_radius} is the standard absolute Bernstein--Serfling pre-factor of \citet[Theorem~4.3]{bardenet2015concentration}. We treat it as $c{=}1$ in the implementation, since any difference (along with the $O(1/n)$ bias term standardly omitted from the simplified radius) is absorbed by the calibration factor $\alpha_{\mathrm{ef}}$. Moreover, the operative quantity is the \emph{hybrid} interval $[\LCB_i,\UCB_i]$ (Eq.~\ref{eq:LCB_UCB}), which intersects the empirical radius with the deterministic support bounds $[LB^{\mathrm{hard}}_i,UB^{\mathrm{hard}}_i]$; the latter tend to be the binding constraint in the small-$n$ regime where the omitted $O(1/n)$ term is largest, so in practice they cushion much of its effect. While the stopping time is adaptive, the procedure requires full separation of the top-$K$ set, making it substantially less sensitive to optional-stopping risks than classical sequential hypothesis tests.

\subsection{Reranking Setting}
\label{app:two_stage_ret_ann}
Our evaluation operates in the standard \emph{full-corpus reranking} regime: given a query $Q=\{\mathbf{q}_1,\dots,\mathbf{q}_T\}$ and a candidate set $\mathcal{D}=\{d_1,\dots,d_N\}$, the goal is to return the top-$K$ documents according to exact ColBERT-style late-interaction scoring (Eq.~\ref{eq:def_row_sum}). The candidate set is the entire corpus throughout the main results (e.g.\ $N{=}2.68$M for NQ); first-stage retrievers (PLAID inverted index, MUVERA-FDE) are an orthogonal axis treated as black-box upstream components, so $\mathcal{D}$ may be replaced by their output without changing the algorithm. \method{} acts on the $N\times T$ MaxSim matrix and returns the same top-$K$ that exhaustive Full-MaxSim would produce on the same $\mathcal{D}$, with $\delta$-PAC certification at $\alpha_{\mathrm{ef}}{=}1$ (Theorem~\ref{thm:pac}).

\subsection{Datasets and Models}
\label{app:datasets_models}
We evaluate \method{} on five widely used text retrieval datasets from the BEIR benchmark~\citep{thakur2021beir}: \textbf{ArguAna}~\citep{wachsmuth2018retrieval_arguana}, \textbf{Quora}~\citep{thakur2021beir}, \textbf{SciDocs}~\citep{cohan2020specter_scidocs}, \textbf{NQ}~\citep{kwiatkowski2019natural_NQ}, and \textbf{HotPotQA}~\citep{yang2018hotpotqa}. We use two state-of-the-art late-interaction text embedding models: \textbf{ColBERTv2}\footnote{\url{https://huggingface.co/lightonai/colbertv2.0}}~\citep{colbertv2} and \textbf{Jina-ColBERT-v2}\footnote{\url{https://huggingface.co/jinaai/jina-colbert-v2}}~\citep{jha2024jina}. Both models produce token embeddings of dimension $d=128$ and use a fixed query token length of $T=32$. In addition, we evaluate \method{} on a visual document retrieval task using the \textbf{REAL-MM-RAG}~\citep{wasserman2025real_mm} benchmark which include 4 subsets: FinReports, FinSlides, TechReports and TechSlides. In this setting, we employ the \textbf{Granite Vision Embedding 3.2}\footnote{\url{https://huggingface.co/ibm-granite/granite-vision-3.3-2b-embedding}}~\citep{granite_vision_embedding} model, a vision-language embedding model that produces $d=128$-dimensional token embeddings, with variable-length query representations and 729 document tokens per image. 
Table~\ref{tab:dataset_stats} summarizes the key statistics of all evaluation datasets.
\begin{table}[t]
\centering
\footnotesize
\setlength{\tabcolsep}{4pt}
\renewcommand{\arraystretch}{1.1}
\caption{Evaluation datasets statistics. $T_q$: query token count; $L_d$: document token count (mean for text, fixed for vision). Quora end-task metrics (nDCG/Recall/MRR) are zero across \emph{all} methods due to a known qrels/doc-id mapping issue in our harness; Overlap@$K$ (the primary fidelity metric) is unaffected.}
\label{tab:dataset_stats}
\begin{tabular}{l r r c c l c}
\toprule
\textbf{Dataset} & \textbf{Corpus $N$} & \textbf{Queries} & \textbf{$T_q$} & \textbf{$L_d$} & \textbf{Encoder} & \textbf{Modality} \\
\midrule
\multicolumn{7}{l}{\textit{BEIR (language)}} \\
ArguAna       & 8{,}674       & 1.4K  & 32 & 192 & ColBERTv2, Jina-ColBERT-v2 & Text \\
SciDocs       & 25{,}657      & 1K    & 32 & 187 & ColBERTv2, Jina-ColBERT-v2 & Text \\
HotpotQA-500K & 500{,}000     & 1K    & 32 & 68  & ColBERTv2, Jina-ColBERT-v2 & Text \\
Quora         & 522{,}931     & 5K    & 32 & 32  & ColBERTv2, Jina-ColBERT-v2 & Text \\
NQ-2.68M      & 2{,}681{,}468 & 3.5K  & 32 & 68  & ColBERTv2 & Text \\
\midrule
\multicolumn{7}{l}{\textit{REAL-MM-RAG (multimodal)}} \\
FinSlides     & 2{,}280       & 1K    & 32 & 729 & Granite Vision Embedding 3.2 & Image+Text \\
FinReport     & 2{,}687       & 853   & 32 & 729 & Granite Vision Embedding 3.2 & Image+Text \\
TechSlides    & 1{,}963       & 1.4K  & 32 & 729 & Granite Vision Embedding 3.2 & Image+Text \\
TechReport    & 1{,}674       & 1.3K  & 32 & 729 & Granite Vision Embedding 3.2 & Image+Text \\
\bottomrule
\end{tabular}
\end{table}

\paragraph{Licensing and intended use.}
All evaluation artifacts are publicly released research resources, used here in accordance with their intended use. The BEIR corpora~\citep{thakur2021beir} and the REAL-MM-RAG benchmark~\citep{wasserman2025real_mm} are distributed for research under their respective licenses, and the pretrained encoders (ColBERTv2~\citep{colbertv2}, Jina-ColBERT-v2~\citep{jha2024jina}, and Granite Vision Embedding~3.2~\citep{granite_vision_embedding}) are released via Hugging Face under their respective model licenses. \method{} is a purely inference-time reranking layer over these artifacts: it introduces no new datasets or model weights and redistributes none of the underlying data. The one artifact we contribute, the CB-NK kernel, will be released as open-source software for research and general use (Appendix~\ref{app:impl_details}).

\subsection{Compared Methods}
\label{app:experiments}
All compared methods operate, like \method{}, on the Stage-2 candidate set (Appendix~\ref{app:two_stage_ret_ann}) and target the same exhaustive late-interaction top-$K$ on that set. We compare against two zero-shot query-time pruning baselines that consume the same $(N, L_d, d)$ corpus tensor as \method{}:

\paragraph{Doc-Uniform ($\gamma$-sweep).}
\label{app:bl_doc_uniform}
Doc-Uniform is a non-adaptive ``blind random scoring budget'': for each query, sample $\lceil\gamma T\rceil$ cells uniformly at random per document (${\approx}\gamma\cdot N\cdot T$ cells in total, matching Algorithm~\ref{alg:doc_random_uniform}), score them, and predict the top-$K$ by the sum of the revealed cells per document. We sweep $\gamma\in\{0.05, 0.10, 0.15, \ldots, 0.50, 0.70, 0.80, 0.90, 1.00\}$ and report the resulting Overlap@$K$ vs.\ coverage curve. This baseline is the \emph{redundancy upper bound}: if \method{} cannot beat Doc-Uniform on the quality--coverage Pareto, then nothing query-adaptive is happening and we are merely paying for some random subset of cells. This is exactly the unstructured-budget reference we use in the §4/§5 Pareto plots.

\paragraph{Ball-carving ($\tau$-sweep).}
\label{app:bl_ball_carving}
Ball-carving is the procedure described in MUVERA, Appendix~C.3~\citep{muvera}, and is the strongest published query-time pruning baseline that operates on the same $(N, L_d, d)$ corpus tensor we score. In contrast to Doc-Uniform, Ball-carving is a \emph{query-side} compression: it greedily clusters the $T$ query tokens into $k$ groups using a cosine-similarity threshold $\tau$, replaces each cluster $\mathcal{C}_i$ with a centroid $c_i = \sum_{q\in\mathcal{C}_i} q$, and rescores every document via Chamfer/MaxSim over the $k$ centroids: $\mathrm{Score}(d) = \sum_{i=1}^{k}\max_{j\le L_d} \langle c_i, e^{(d)}_j\rangle$. Coverage is reported as $k/T$: smaller $\tau$ collapses more tokens into fewer, larger clusters (lower coverage), while $\tau=1$ keeps every token as its own cluster and recovers exhaustive Chamfer. We sweep $\tau\in\{0.40, 0.50, 0.60, 0.70, 0.80, 0.90\}$ and average $k/T$ across queries to place the operating point on the coverage axis. Our implementation matches MUVERA \S3 / App.~C.3 exactly.

\noindent
\begin{minipage}[t]{0.48\linewidth}
    \begin{algorithm}[H]
        \centering
        \caption{Doc-Uniform (Static Random Reveal)}
        \label{alg:doc_random_uniform}
        \footnotesize
        \begin{algorithmic}[1]
            \Require Docs $\mathcal{D}$, Query $Q$, $K$, $\gamma \in [0,1]$
            \State $N \gets |\mathcal{D}|$, $B \gets \lceil \gamma T \rceil$ \Comment{Cells per row}
            \State $\Omega \gets \emptyset$, $H \in \mathbb{R}^{N\times T}$
            \For{$i = 1$ \textbf{to} $N$}
                \State Sample $\mathcal{R}_i \subseteq [T]$ uniformly \Comment{w/o replacement}
                \State \textbf{s.t.} $|\mathcal{R}_i|=B$
                \For{\textbf{each} $t \in \mathcal{R}_i$}
                    \State $H_{i,t} \gets h(d_i,t)$ \Comment{Reveal MaxSim}
                    \State $\Omega \gets \Omega \cup \{(i,t)\}$
                \EndFor
                \State $\widetilde{S}_i \gets \sum_{t \in \mathcal{R}_i} H_{i,t}$ \Comment{Static score}
            \EndFor
            \State \Return $\arg\text{topK}_{i\in[N]} \widetilde{S}_i$
        \end{algorithmic}
    \end{algorithm}
\end{minipage}\hfill
\begin{minipage}[t]{0.48\linewidth}
    \begin{algorithm}[H]
        \centering
        \caption{Ball-carving query compression (MUVERA Appendix~C.3)}
        \label{alg:ball_carving}
        \footnotesize
        \begin{algorithmic}[1]
            \Require Query $Q=(q_1,\ldots,q_T)\in\mathbb{R}^{T\times d}$, threshold $\tau\in(0,1]$
            \Require Docs $\mathcal{D}$ with embeddings $(e^{(d)}_1,\ldots,e^{(d)}_{L_d})$, $K$
            \State \textit{// Step 1: greedy query-side clustering}
            \State $\hat{q}_i \gets q_i / \lVert q_i \rVert$ \textbf{ for all } $i\in[T]$
            \State $\mathcal{R} \gets \{1,\ldots,T\}$, \quad $k \gets 0$
            \While{$\mathcal{R} \neq \emptyset$}
                \State $k \gets k+1$, \quad $p \gets \min \mathcal{R}$ \Comment{pivot}
                \State $\mathcal{C}_k \gets \{ i\in\mathcal{R} : \langle \hat{q}_p, \hat{q}_i\rangle \geq \tau \}$
                \State $c_k \gets \sum_{i\in\mathcal{C}_k} q_i$
                \State $\mathcal{R} \gets \mathcal{R} \setminus \mathcal{C}_k$
            \EndWhile
            \State \textit{// Step 2: rescore with one MaxSim term per cluster}
            \For{\textbf{each} $d \in \mathcal{D}$}
                \State $\mathrm{Score}(d) \gets \sum_{i=1}^{k} \max_{j\le L_d} \langle c_i, e^{(d)}_j\rangle$
            \EndFor
            \State \Return $\arg\text{topK}_{d\in\mathcal{D}}\, \mathrm{Score}(d)$
        \end{algorithmic}
    \end{algorithm}
\end{minipage}

\FloatBarrier
\section{Extended Experimental Results}
\label{sec:app_extended_results}
\subsection{Universal Efficiency at the 90\% Threshold}
\label{sec:app_universal_eff_90pct}
The body table (Table~\ref{tab:main_benchmark_universal_eff_combined}) reports the stricter $95\%$ near-lossless threshold. For comparability with prior work that reports the looser $90\%$ retention threshold, we provide the companion table below; the layout is identical to the body table apart from the threshold.
\begin{table}[t]
\centering
\footnotesize
\renewcommand{\arraystretch}{1.1}
\setlength{\tabcolsep}{4pt}
\caption{\textbf{Universal Efficiency Analysis at the 90\% threshold (Appendix companion to Table~\ref{tab:main_benchmark_universal_eff_combined}).} Each cell reports \textbf{mean coverage \% (std) / savings $\times$} required to recover \textbf{90\%} of Full-MaxSim's own Overlap@$K$ / nDCG@$K$ per dataset; savings $= 100\%/\text{mean coverage}$. The body table reports the stricter $95\%$ near-lossless threshold; this appendix table provides the looser $90\%$ threshold for comparability with prior work.}
\label{tab:appendix_universal_eff_90pct}
\resizebox{\ifdim\width>\linewidth \linewidth\else \width\fi}{!}{%
\begin{tabular}{l|cc||cc}
\toprule
\textbf{Method} &
\textbf{Overlap@5} & \textbf{Overlap@50} &
\textbf{nDCG@5} & \textbf{nDCG@50} \\
\midrule

\multicolumn{5}{l}{\textit{\textbf{ColBERTv2 (BEIR)}}} \\
\hspace{3mm} Doc-Uniform
    & 94\% (7.5) / 1.1$\times$
    & 92\% (7.0) / 1.1$\times$
    & 38\% (11.7) / 2.7$\times$
    & 25\% (4.4) / 4.0$\times$ \\
\hspace{3mm} Ball-carving
    & 34\% (11.1) / 2.9$\times$
    & 32\% (8.9) / 3.2$\times$
    & 21\% (3.7) / 4.8$\times$
    & 19\% (5.9) / 5.2$\times$ \\
\hspace{3mm} \textbf{\method{} (Ours)}
    & \textbf{13\% (0.9) / 7.4$\times$}
    & \textbf{15\% (2.5) / 6.6$\times$}
    & \textbf{13\% (0.1) / 7.9$\times$}
    & \textbf{13\% (0.5) / 7.6$\times$} \\

\midrule
\multicolumn{5}{l}{\textit{\textbf{Jina-ColBERTv2 (BEIR)}}} \\
\hspace{3mm} Doc-Uniform
    & 93\% (7.8) / 1.1$\times$
    & 93\% (7.8) / 1.1$\times$
    & 52\% (26.6) / 1.9$\times$
    & 45\% (24.4) / 2.2$\times$ \\
\hspace{3mm} Ball-carving
    & 37\% (9.7) / 2.7$\times$
    & 34\% (6.2) / 2.9$\times$
    & 16\% (3.8) / 6.3$\times$
    & 16\% (3.8) / 6.3$\times$ \\
\hspace{3mm} \textbf{\method{} (Ours)}
    & \textbf{17\% (11.2) / 5.9$\times$}
    & \textbf{28\% (14.6) / 3.6$\times$}
    & \textbf{7\% (1.2) / 13.5$\times$}
    & \textbf{9\% (2.1) / 10.9$\times$} \\

\midrule
\multicolumn{5}{l}{\textit{\textbf{Granite-Vision-Embedding (REAL-MM-RAG)}}} \\
\hspace{3mm} Doc-Uniform
    & 83\% (4.3) / 1.2$\times$
    & 83\% (4.3) / 1.2$\times$
    & 34\% (11.7) / 2.9$\times$
    & 27\% (8.8) / 3.7$\times$ \\
\hspace{3mm} Ball-carving
    & 46\% (2.1) / 2.2$\times$
    & 39\% (5.3) / 2.6$\times$
    & 19\% (5.3) / 5.2$\times$
    & 19\% (5.3) / 5.2$\times$ \\
\hspace{3mm} \textbf{\method{} (Ours)}
    & \textbf{16\% (4.8) / 6.1$\times$}
    & \textbf{26\% (5.4) / 3.9$\times$}
    & \textbf{10\% (1.5) / 10.4$\times$}
    & \textbf{12\% (1.3) / 8.6$\times$} \\

\bottomrule
\end{tabular}%
}
\end{table}

\subsection{Detailed Efficiency Results per Dataset}
\label{sec:app_detailed_results}
\label{sec:app_per_dataset_eff}
In the main text (Table~\ref{tab:main_benchmark_universal_eff_combined}), we present efficiency metrics averaged across the BEIR and REAL-MM-RAG suites. The tables in this subsection break those averages down per dataset, organized by metric (Overlap, nDCG, Recall, MRR) and target rank ($K\in\{5, 10, 50\}$). Each cell reports the coverage budget (\%) at which \method{} reaches \textbf{90\%} (white) and \textbf{95\%} (gray) of Full-MaxSim's metric on that dataset. \emph{Note:} cells marked ``n/a'' indicate that the corresponding (encoder, dataset) sweep was not run; in particular, the Jina-ColBERTv2 sweep covers four of the five BEIR text corpora (NQ-2.68M not benchmarked).

This detailed view confirms that the efficiency gains of \method{} are robust across diverse data distributions: \method{} consistently outperforms the baselines on every individual dataset, with the exact magnitude varying with document length and query difficulty.

The eight tables below cover the four metrics (Overlap@$K$, nDCG@$K$, Recall@$K$, MRR@$K$) on the BEIR text suite (top half) and the REAL-MM-RAG multimodal suite (bottom half).

\begin{table}[t]
\centering
\footnotesize
\setlength{\tabcolsep}{4pt}
\renewcommand{\arraystretch}{1.1}
\caption{\textbf{Per-dataset Efficiency Analysis: Overlap@$K$ (BEIR).} We report the coverage budget required to achieve \textbf{90\%} (white) and \textbf{95\%} (gray) of Full-MaxSim\textquotesingle{}s \textbf{Overlap}@$K$ per corpus. Under \textbf{Average}, we report \textbf{mean coverage (std)} across datasets, and \textbf{Savings (vs.\ Full)} is the compute reduction factor relative to full reranking (i.e., $100\%/\textbf{Mean}$).}
\label{tab:per_dataset_eff_overlap_beir_combinedK}
\resizebox{\ifdim\width>\linewidth \linewidth\else \width\fi}{!}{%
\begin{tabular}{l|cc|cc|cc|cc|cc||cc|cc}
\toprule
\textbf{Task Domain} & \multicolumn{10}{c||}{\textbf{Text Retrieval Benchmarks (BEIR)}} & \multicolumn{4}{c}{\textbf{Average}} \\
\cmidrule(lr){1-1} \cmidrule(lr){2-11} \cmidrule(lr){12-15}
\textbf{Method} & \multicolumn{2}{c|}{\textbf{SciDocs}} & \multicolumn{2}{c|}{\textbf{Quora}} & \multicolumn{2}{c|}{\textbf{NQ}} & \multicolumn{2}{c|}{\textbf{HotpotQA}} & \multicolumn{2}{c|}{\textbf{ArguAna}} & \multicolumn{2}{c|}{\textit{Mean (std)}} & \multicolumn{2}{c}{\textit{Savings (vs.\ Full)}} \\
\midrule

\multicolumn{15}{c}{\textit{$K = 5$}} \\
\midrule
\multicolumn{15}{l}{\textit{\textbf{ColBERTv2}}} \\
\hspace{3mm} Doc-Uniform & 100\% & \cellcolor{gray!20}100\% & 81\% & \cellcolor{gray!20}100\% & 81\% & \cellcolor{gray!20}91\% & 100\% & \cellcolor{gray!20}100\% & 91\% & \cellcolor{gray!20}100\% & 91\% (8.4) & \cellcolor{gray!20}98\% (3.8) & 1.10$\times$ & \cellcolor{gray!20}1.02$\times$ \\
\hspace{3mm} Ball-carving & 31\% & \cellcolor{gray!20}36\% & 22\% & \cellcolor{gray!20}27\% & 21\% & \cellcolor{gray!20}28\% & 47\% & \cellcolor{gray!20}61\% & 44\% & \cellcolor{gray!20}60\% & 33\% (10.7) & \cellcolor{gray!20}42\% (14.9) & 3.02$\times$ & \cellcolor{gray!20}2.36$\times$ \\
\hspace{3mm} \textbf{\method{} (Ours)} & \textbf{14\%} & \cellcolor{gray!20}\textbf{17\%} & \textbf{13\%} & \cellcolor{gray!20}\textbf{13\%} & \textbf{13\%} & \cellcolor{gray!20}\textbf{13\%} & \textbf{13\%} & \cellcolor{gray!20}\textbf{13\%} & \textbf{15\%} & \cellcolor{gray!20}\textbf{15\%} & \textbf{13\% (0.9)} & \cellcolor{gray!20}\textbf{14\% (1.7)} & \textbf{7.42$\times$} & \cellcolor{gray!20}\textbf{7.06$\times$} \\
\midrule
\multicolumn{15}{l}{\textit{\textbf{Jina-ColBERT-V2}}} \\
\hspace{3mm} Doc-Uniform & 100\% & \cellcolor{gray!20}100\% & 91\% & \cellcolor{gray!20}100\% & \multicolumn{2}{c|}{\textit{n/a}} & 100\% & \cellcolor{gray!20}100\% & 81\% & \cellcolor{gray!20}100\% & 93\% (7.8) & \cellcolor{gray!20}100\% (0.0) & 1.08$\times$ & \cellcolor{gray!20}1.00$\times$ \\
\hspace{3mm} Ball-carving & 33\% & \cellcolor{gray!20}100\% & 24\% & \cellcolor{gray!20}24\% & \multicolumn{2}{c|}{\textit{n/a}} & 40\% & \cellcolor{gray!20}100\% & 51\% & \cellcolor{gray!20}51\% & 37\% (9.7) & \cellcolor{gray!20}69\% (32.6) & 2.69$\times$ & \cellcolor{gray!20}1.45$\times$ \\
\hspace{3mm} \textbf{\method{} (Ours)} & \textbf{19\%} & \cellcolor{gray!20}\textbf{35\%} & \textbf{7\%} & \cellcolor{gray!20}\textbf{11\%} & \multicolumn{2}{c|}{\textit{n/a}} & \textbf{34\%} & \cellcolor{gray!20}\textbf{49\%} & \textbf{7\%} & \cellcolor{gray!20}\textbf{8\%} & \textbf{17\% (11.2)} & \cellcolor{gray!20}\textbf{26\% (16.9)} & \textbf{5.95$\times$} & \cellcolor{gray!20}\textbf{3.92$\times$} \\
\midrule
\multicolumn{15}{c}{\textit{$K = 10$}} \\
\midrule
\multicolumn{15}{l}{\textit{\textbf{ColBERTv2}}} \\
\hspace{3mm} Doc-Uniform & 100\% & \cellcolor{gray!20}100\% & 91\% & \cellcolor{gray!20}100\% & 81\% & \cellcolor{gray!20}100\% & 100\% & \cellcolor{gray!20}100\% & 91\% & \cellcolor{gray!20}100\% & 92\% (7.0) & \cellcolor{gray!20}100\% (0.0) & 1.08$\times$ & \cellcolor{gray!20}1.00$\times$ \\
\hspace{3mm} Ball-carving & 29\% & \cellcolor{gray!20}36\% & 22\% & \cellcolor{gray!20}27\% & 21\% & \cellcolor{gray!20}28\% & 38\% & \cellcolor{gray!20}61\% & 50\% & \cellcolor{gray!20}60\% & 32\% (10.9) & \cellcolor{gray!20}42\% (14.9) & 3.10$\times$ & \cellcolor{gray!20}2.36$\times$ \\
\hspace{3mm} \textbf{\method{} (Ours)} & \textbf{15\%} & \cellcolor{gray!20}\textbf{15\%} & \textbf{13\%} & \cellcolor{gray!20}\textbf{13\%} & \textbf{13\%} & \cellcolor{gray!20}\textbf{13\%} & \textbf{13\%} & \cellcolor{gray!20}\textbf{15\%} & \textbf{15\%} & \cellcolor{gray!20}\textbf{19\%} & \textbf{14\% (1.2)} & \cellcolor{gray!20}\textbf{15\% (2.4)} & \textbf{7.28$\times$} & \cellcolor{gray!20}\textbf{6.68$\times$} \\
\midrule
\multicolumn{15}{l}{\textit{\textbf{Jina-ColBERT-V2}}} \\
\hspace{3mm} Doc-Uniform & 100\% & \cellcolor{gray!20}100\% & 91\% & \cellcolor{gray!20}100\% & \multicolumn{2}{c|}{\textit{n/a}} & 100\% & \cellcolor{gray!20}100\% & 70\% & \cellcolor{gray!20}100\% & 90\% (12.1) & \cellcolor{gray!20}100\% (0.0) & 1.11$\times$ & \cellcolor{gray!20}1.00$\times$ \\
\hspace{3mm} Ball-carving & 33\% & \cellcolor{gray!20}100\% & 24\% & \cellcolor{gray!20}100\% & \multicolumn{2}{c|}{\textit{n/a}} & 40\% & \cellcolor{gray!20}100\% & 39\% & \cellcolor{gray!20}51\% & 34\% (6.2) & \cellcolor{gray!20}88\% (21.3) & 2.94$\times$ & \cellcolor{gray!20}1.14$\times$ \\
\hspace{3mm} \textbf{\method{} (Ours)} & \textbf{60\%} & \cellcolor{gray!20}\textbf{60\%} & \textbf{16\%} & \cellcolor{gray!20}\textbf{16\%} & \multicolumn{2}{c|}{\textit{n/a}} & \textbf{34\%} & \cellcolor{gray!20}\textbf{69\%} & \textbf{11\%} & \cellcolor{gray!20}\textbf{11\%} & \textbf{30\% (19.0)} & \cellcolor{gray!20}\textbf{39\% (25.7)} & \textbf{3.31$\times$} & \cellcolor{gray!20}\textbf{2.57$\times$} \\
\midrule
\multicolumn{15}{c}{\textit{$K = 50$}} \\
\midrule
\multicolumn{15}{l}{\textit{\textbf{ColBERTv2}}} \\
\hspace{3mm} Doc-Uniform & 100\% & \cellcolor{gray!20}100\% & 91\% & \cellcolor{gray!20}100\% & 81\% & \cellcolor{gray!20}100\% & 100\% & \cellcolor{gray!20}100\% & 91\% & \cellcolor{gray!20}100\% & 92\% (7.0) & \cellcolor{gray!20}100\% (0.0) & 1.08$\times$ & \cellcolor{gray!20}1.00$\times$ \\
\hspace{3mm} Ball-carving & 29\% & \cellcolor{gray!20}36\% & 22\% & \cellcolor{gray!20}27\% & 21\% & \cellcolor{gray!20}23\% & 38\% & \cellcolor{gray!20}61\% & 44\% & \cellcolor{gray!20}60\% & 31\% (8.9) & \cellcolor{gray!20}41\% (15.9) & 3.23$\times$ & \cellcolor{gray!20}2.42$\times$ \\
\hspace{3mm} \textbf{\method{} (Ours)} & \textbf{17\%} & \cellcolor{gray!20}\textbf{22\%} & \textbf{14\%} & \cellcolor{gray!20}\textbf{15\%} & \textbf{13\%} & \cellcolor{gray!20}\textbf{13\%} & \textbf{14\%} & \cellcolor{gray!20}\textbf{16\%} & \textbf{19\%} & \cellcolor{gray!20}\textbf{26\%} & \textbf{15\% (2.4)} & \cellcolor{gray!20}\textbf{18\% (4.8)} & \textbf{6.54$\times$} & \cellcolor{gray!20}\textbf{5.46$\times$} \\
\midrule
\multicolumn{15}{l}{\textit{\textbf{Jina-ColBERT-V2}}} \\
\hspace{3mm} Doc-Uniform & 100\% & \cellcolor{gray!20}100\% & 91\% & \cellcolor{gray!20}100\% & \multicolumn{2}{c|}{\textit{n/a}} & 100\% & \cellcolor{gray!20}100\% & 81\% & \cellcolor{gray!20}100\% & 93\% (7.8) & \cellcolor{gray!20}100\% (0.0) & 1.08$\times$ & \cellcolor{gray!20}1.00$\times$ \\
\hspace{3mm} Ball-carving & 33\% & \cellcolor{gray!20}100\% & 24\% & \cellcolor{gray!20}100\% & \multicolumn{2}{c|}{\textit{n/a}} & 40\% & \cellcolor{gray!20}100\% & 39\% & \cellcolor{gray!20}51\% & 34\% (6.2) & \cellcolor{gray!20}88\% (21.3) & 2.94$\times$ & \cellcolor{gray!20}1.14$\times$ \\
\hspace{3mm} \textbf{\method{} (Ours)} & \textbf{39\%} & \cellcolor{gray!20}\textbf{52\%} & \textbf{15\%} & \cellcolor{gray!20}\textbf{29\%} & \multicolumn{2}{c|}{\textit{n/a}} & \textbf{45\%} & \cellcolor{gray!20}\textbf{61\%} & \textbf{12\%} & \cellcolor{gray!20}\textbf{16\%} & \textbf{28\% (14.6)} & \cellcolor{gray!20}\textbf{39\% (17.8)} & \textbf{3.58$\times$} & \cellcolor{gray!20}\textbf{2.55$\times$} \\
\bottomrule
\end{tabular}%
}
\end{table}

\begin{table}[t]
\centering
\footnotesize
\setlength{\tabcolsep}{4pt}
\renewcommand{\arraystretch}{1.1}
\caption{\textbf{Per-dataset Efficiency Analysis: Overlap@$K$ (REAL-MM-RAG).} Coverage budget required to achieve \textbf{90\%} (white) and \textbf{95\%} (gray) of Full-MaxSim\textquotesingle{}s \textbf{Overlap}@$K$ per corpus, on the Granite-Vision-Embedding (GVE) multimodal benchmarks. \textbf{Mean (std)} averages across the four corpora; \textbf{Savings} is $100\%/\textbf{Mean}$.}
\label{tab:per_dataset_eff_overlap_mm_combinedK}
\resizebox{\ifdim\width>\linewidth \linewidth\else \width\fi}{!}{%
\begin{tabular}{l|cc|cc|cc|cc||cc|cc}
\toprule
\textbf{Task Domain} & \multicolumn{8}{c||}{\textbf{Multimodal (REAL-MM-RAG, GVE)}} & \multicolumn{4}{c}{\textbf{Average}} \\
\cmidrule(lr){1-1} \cmidrule(lr){2-9} \cmidrule(lr){10-13}
\textbf{Method} & \multicolumn{2}{c|}{\textbf{FinSlides}} & \multicolumn{2}{c|}{\textbf{FinReport}} & \multicolumn{2}{c|}{\textbf{TechSlides}} & \multicolumn{2}{c|}{\textbf{TechReport}} & \multicolumn{2}{c|}{\textit{Mean (std)}} & \multicolumn{2}{c}{\textit{Savings (vs.\ Full)}} \\
\midrule

\multicolumn{13}{c}{\textit{$K = 5$}} \\
\midrule
\hspace{3mm} Doc-Uniform & 81\% & \cellcolor{gray!20}100\% & 91\% & \cellcolor{gray!20}100\% & 81\% & \cellcolor{gray!20}100\% & 81\% & \cellcolor{gray!20}100\% & 83\% (4.3) & \cellcolor{gray!20}100\% (0.0) & 1.20$\times$ & \cellcolor{gray!20}1.00$\times$ \\
\hspace{3mm} Ball-carving & 44\% & \cellcolor{gray!20}75\% & 44\% & \cellcolor{gray!20}74\% & 47\% & \cellcolor{gray!20}75\% & 49\% & \cellcolor{gray!20}79\% & 46\% (2.1) & \cellcolor{gray!20}76\% (1.7) & 2.18$\times$ & \cellcolor{gray!20}1.32$\times$ \\
\hspace{3mm} \textbf{\method{} (Ours)} & \textbf{19\%} & \cellcolor{gray!20}\textbf{19\%} & \textbf{12\%} & \cellcolor{gray!20}\textbf{23\%} & \textbf{12\%} & \cellcolor{gray!20}\textbf{22\%} & \textbf{23\%} & \cellcolor{gray!20}\textbf{23\%} & \textbf{16\% (4.8)} & \cellcolor{gray!20}\textbf{22\% (1.9)} & \textbf{6.11$\times$} & \cellcolor{gray!20}\textbf{4.59$\times$} \\
\midrule
\multicolumn{13}{c}{\textit{$K = 10$}} \\
\midrule
\hspace{3mm} Doc-Uniform & 91\% & \cellcolor{gray!20}100\% & 91\% & \cellcolor{gray!20}100\% & 91\% & \cellcolor{gray!20}100\% & 81\% & \cellcolor{gray!20}100\% & 88\% (4.3) & \cellcolor{gray!20}100\% (0.0) & 1.13$\times$ & \cellcolor{gray!20}1.00$\times$ \\
\hspace{3mm} Ball-carving & 44\% & \cellcolor{gray!20}75\% & 44\% & \cellcolor{gray!20}74\% & 33\% & \cellcolor{gray!20}47\% & 49\% & \cellcolor{gray!20}79\% & 43\% (5.9) & \cellcolor{gray!20}69\% (12.9) & 2.35$\times$ & \cellcolor{gray!20}1.46$\times$ \\
\hspace{3mm} \textbf{\method{} (Ours)} & \textbf{21\%} & \cellcolor{gray!20}\textbf{21\%} & \textbf{13\%} & \cellcolor{gray!20}\textbf{27\%} & \textbf{14\%} & \cellcolor{gray!20}\textbf{25\%} & \textbf{26\%} & \cellcolor{gray!20}\textbf{26\%} & \textbf{19\% (5.5)} & \cellcolor{gray!20}\textbf{25\% (2.3)} & \textbf{5.40$\times$} & \cellcolor{gray!20}\textbf{4.01$\times$} \\
\midrule
\multicolumn{13}{c}{\textit{$K = 50$}} \\
\midrule
\hspace{3mm} Doc-Uniform & 81\% & \cellcolor{gray!20}100\% & 91\% & \cellcolor{gray!20}100\% & 81\% & \cellcolor{gray!20}100\% & 81\% & \cellcolor{gray!20}91\% & 83\% (4.3) & \cellcolor{gray!20}98\% (4.1) & 1.20$\times$ & \cellcolor{gray!20}1.02$\times$ \\
\hspace{3mm} Ball-carving & 44\% & \cellcolor{gray!20}75\% & 44\% & \cellcolor{gray!20}74\% & 33\% & \cellcolor{gray!20}75\% & 34\% & \cellcolor{gray!20}79\% & 39\% (5.3) & \cellcolor{gray!20}76\% (1.7) & 2.58$\times$ & \cellcolor{gray!20}1.32$\times$ \\
\hspace{3mm} \textbf{\method{} (Ours)} & \textbf{35\%} & \cellcolor{gray!20}\textbf{35\%} & \textbf{21\%} & \cellcolor{gray!20}\textbf{43\%} & \textbf{23\%} & \cellcolor{gray!20}\textbf{42\%} & \textbf{24\%} & \cellcolor{gray!20}\textbf{43\%} & \textbf{26\% (5.4)} & \cellcolor{gray!20}\textbf{41\% (3.1)} & \textbf{3.85$\times$} & \cellcolor{gray!20}\textbf{2.46$\times$} \\
\bottomrule
\end{tabular}%
}
\end{table}

\begin{table}[t]
\centering
\footnotesize
\setlength{\tabcolsep}{4pt}
\renewcommand{\arraystretch}{1.1}
\caption{\textbf{Per-dataset Efficiency Analysis: nDCG@$K$ (BEIR).} We report the coverage budget required to achieve \textbf{90\%} (white) and \textbf{95\%} (gray) of Full-MaxSim\textquotesingle{}s \textbf{NDCG}@$K$ per corpus. Under \textbf{Average}, we report \textbf{mean coverage (std)} across datasets, and \textbf{Savings (vs.\ Full)} is the compute reduction factor relative to full reranking (i.e., $100\%/\textbf{Mean}$). Quora is shown as \textit{n/a} and excluded from the Average: its end-task metrics are zero across all methods under our harness (a qrels/doc-id mapping issue; Table~\ref{tab:dataset_stats}), so per-corpus task-metric budgets are undefined there. Quora retrieval fidelity is unaffected and reported via Overlap@$K$ (Table~\ref{tab:per_dataset_eff_overlap_beir_combinedK}).}
\label{tab:per_dataset_eff_ndcg_beir_combinedK}
\resizebox{\ifdim\width>\linewidth \linewidth\else \width\fi}{!}{%
\begin{tabular}{l|cc|cc|cc|cc|cc||cc|cc}
\toprule
\textbf{Task Domain} & \multicolumn{10}{c||}{\textbf{Text Retrieval Benchmarks (BEIR)}} & \multicolumn{4}{c}{\textbf{Average}} \\
\cmidrule(lr){1-1} \cmidrule(lr){2-11} \cmidrule(lr){12-15}
\textbf{Method} & \multicolumn{2}{c|}{\textbf{SciDocs}} & \multicolumn{2}{c|}{\textbf{Quora}} & \multicolumn{2}{c|}{\textbf{NQ}} & \multicolumn{2}{c|}{\textbf{HotpotQA}} & \multicolumn{2}{c|}{\textbf{ArguAna}} & \multicolumn{2}{c|}{\textit{Mean (std)}} & \multicolumn{2}{c}{\textit{Savings (vs.\ Full)}} \\
\midrule

\multicolumn{15}{c}{\textit{$K = 5$}} \\
\midrule
\multicolumn{15}{l}{\textit{\textbf{ColBERTv2}}} \\
\hspace{3mm} Doc-Uniform & 41\% & \cellcolor{gray!20}72\% & \multicolumn{2}{c|}{\textit{n/a}} & 22\% & \cellcolor{gray!20}50\% & 50\% & \cellcolor{gray!20}81\% & 50\% & \cellcolor{gray!20}50\% & 41\% (11.4) & \cellcolor{gray!20}63\% (13.6) & 2.45$\times$ & \cellcolor{gray!20}1.58$\times$ \\
\hspace{3mm} Ball-carving & 20\% & \cellcolor{gray!20}26\% & \multicolumn{2}{c|}{\textit{n/a}} & 17\% & \cellcolor{gray!20}17\% & 30\% & \cellcolor{gray!20}34\% & 26\% & \cellcolor{gray!20}44\% & 23\% (5.1) & \cellcolor{gray!20}30\% (10.0) & 4.30$\times$ & \cellcolor{gray!20}3.31$\times$ \\
\hspace{3mm} \textbf{\method{} (Ours)} & \textbf{13\%} & \cellcolor{gray!20}\textbf{14\%} & \multicolumn{2}{c|}{\textit{n/a}} & \textbf{13\%} & \cellcolor{gray!20}\textbf{13\%} & \textbf{13\%} & \cellcolor{gray!20}\textbf{13\%} & \textbf{13\%} & \cellcolor{gray!20}\textbf{13\%} & \textbf{13\% (0.0)} & \cellcolor{gray!20}\textbf{13\% (0.4)} & \textbf{7.69$\times$} & \cellcolor{gray!20}\textbf{7.55$\times$} \\
\midrule
\multicolumn{15}{l}{\textit{\textbf{Jina-ColBERT-V2}}} \\
\hspace{3mm} Doc-Uniform & 91\% & \cellcolor{gray!20}100\% & \multicolumn{2}{c|}{\textit{n/a}} & \multicolumn{2}{c|}{\textit{n/a}} & 50\% & \cellcolor{gray!20}70\% & 50\% & \cellcolor{gray!20}50\% & 64\% (19.3) & \cellcolor{gray!20}73\% (20.5) & 1.57$\times$ & \cellcolor{gray!20}1.36$\times$ \\
\hspace{3mm} Ball-carving & 15\% & \cellcolor{gray!20}18\% & \multicolumn{2}{c|}{\textit{n/a}} & \multicolumn{2}{c|}{\textit{n/a}} & 20\% & \cellcolor{gray!20}26\% & 19\% & \cellcolor{gray!20}24\% & 18\% (2.2) & \cellcolor{gray!20}23\% (3.4) & 5.56$\times$ & \cellcolor{gray!20}4.41$\times$ \\
\hspace{3mm} \textbf{\method{} (Ours)} & \textbf{10\%} & \cellcolor{gray!20}\textbf{15\%} & \multicolumn{2}{c|}{\textit{n/a}} & \multicolumn{2}{c|}{\textit{n/a}} & \textbf{7\%} & \cellcolor{gray!20}\textbf{7\%} & \textbf{7\%} & \cellcolor{gray!20}\textbf{7\%} & \textbf{8\% (1.4)} & \cellcolor{gray!20}\textbf{10\% (3.8)} & \textbf{12.50$\times$} & \cellcolor{gray!20}\textbf{10.34$\times$} \\
\midrule
\multicolumn{15}{c}{\textit{$K = 10$}} \\
\midrule
\multicolumn{15}{l}{\textit{\textbf{ColBERTv2}}} \\
\hspace{3mm} Doc-Uniform & 22\% & \cellcolor{gray!20}41\% & \multicolumn{2}{c|}{\textit{n/a}} & 22\% & \cellcolor{gray!20}50\% & 50\% & \cellcolor{gray!20}81\% & 41\% & \cellcolor{gray!20}72\% & 34\% (12.2) & \cellcolor{gray!20}61\% (16.1) & 2.96$\times$ & \cellcolor{gray!20}1.64$\times$ \\
\hspace{3mm} Ball-carving & 20\% & \cellcolor{gray!20}26\% & \multicolumn{2}{c|}{\textit{n/a}} & 12\% & \cellcolor{gray!20}17\% & 23\% & \cellcolor{gray!20}30\% & 26\% & \cellcolor{gray!20}36\% & 20\% (5.2) & \cellcolor{gray!20}27\% (6.9) & 4.94$\times$ & \cellcolor{gray!20}3.67$\times$ \\
\hspace{3mm} \textbf{\method{} (Ours)} & \textbf{13\%} & \cellcolor{gray!20}\textbf{15\%} & \multicolumn{2}{c|}{\textit{n/a}} & \textbf{13\%} & \cellcolor{gray!20}\textbf{13\%} & \textbf{13\%} & \cellcolor{gray!20}\textbf{13\%} & \textbf{13\%} & \cellcolor{gray!20}\textbf{13\%} & \textbf{13\% (0.0)} & \cellcolor{gray!20}\textbf{14\% (0.9)} & \textbf{7.69$\times$} & \cellcolor{gray!20}\textbf{7.41$\times$} \\
\midrule
\multicolumn{15}{l}{\textit{\textbf{Jina-ColBERT-V2}}} \\
\hspace{3mm} Doc-Uniform & 70\% & \cellcolor{gray!20}91\% & \multicolumn{2}{c|}{\textit{n/a}} & \multicolumn{2}{c|}{\textit{n/a}} & 50\% & \cellcolor{gray!20}70\% & 41\% & \cellcolor{gray!20}70\% & 54\% (12.1) & \cellcolor{gray!20}77\% (9.9) & 1.86$\times$ & \cellcolor{gray!20}1.30$\times$ \\
\hspace{3mm} Ball-carving & 15\% & \cellcolor{gray!20}18\% & \multicolumn{2}{c|}{\textit{n/a}} & \multicolumn{2}{c|}{\textit{n/a}} & 20\% & \cellcolor{gray!20}26\% & 19\% & \cellcolor{gray!20}30\% & 18\% (2.2) & \cellcolor{gray!20}25\% (5.0) & 5.56$\times$ & \cellcolor{gray!20}4.05$\times$ \\
\hspace{3mm} \textbf{\method{} (Ours)} & \textbf{22\%} & \cellcolor{gray!20}\textbf{60\%} & \multicolumn{2}{c|}{\textit{n/a}} & \multicolumn{2}{c|}{\textit{n/a}} & \textbf{7\%} & \cellcolor{gray!20}\textbf{34\%} & \textbf{7\%} & \cellcolor{gray!20}\textbf{7\%} & \textbf{12\% (7.1)} & \cellcolor{gray!20}\textbf{34\% (21.6)} & \textbf{8.33$\times$} & \cellcolor{gray!20}\textbf{2.97$\times$} \\
\midrule
\multicolumn{15}{c}{\textit{$K = 50$}} \\
\midrule
\multicolumn{15}{l}{\textit{\textbf{ColBERTv2}}} \\
\hspace{3mm} Doc-Uniform & 22\% & \cellcolor{gray!20}50\% & \multicolumn{2}{c|}{\textit{n/a}} & 22\% & \cellcolor{gray!20}22\% & 50\% & \cellcolor{gray!20}81\% & 31\% & \cellcolor{gray!20}41\% & 31\% (11.4) & \cellcolor{gray!20}48\% (21.3) & 3.20$\times$ & \cellcolor{gray!20}2.06$\times$ \\
\hspace{3mm} Ball-carving & 20\% & \cellcolor{gray!20}26\% & \multicolumn{2}{c|}{\textit{n/a}} & 12\% & \cellcolor{gray!20}17\% & 23\% & \cellcolor{gray!20}30\% & 26\% & \cellcolor{gray!20}26\% & 20\% (5.2) & \cellcolor{gray!20}25\% (4.8) & 4.94$\times$ & \cellcolor{gray!20}4.04$\times$ \\
\hspace{3mm} \textbf{\method{} (Ours)} & \textbf{14\%} & \cellcolor{gray!20}\textbf{14\%} & \multicolumn{2}{c|}{\textit{n/a}} & \textbf{13\%} & \cellcolor{gray!20}\textbf{13\%} & \textbf{13\%} & \cellcolor{gray!20}\textbf{14\%} & \textbf{13\%} & \cellcolor{gray!20}\textbf{14\%} & \textbf{13\% (0.4)} & \cellcolor{gray!20}\textbf{14\% (0.4)} & \textbf{7.55$\times$} & \cellcolor{gray!20}\textbf{7.27$\times$} \\
\midrule
\multicolumn{15}{l}{\textit{\textbf{Jina-ColBERT-V2}}} \\
\hspace{3mm} Doc-Uniform & 81\% & \cellcolor{gray!20}91\% & \multicolumn{2}{c|}{\textit{n/a}} & \multicolumn{2}{c|}{\textit{n/a}} & 50\% & \cellcolor{gray!20}50\% & 31\% & \cellcolor{gray!20}70\% & 54\% (20.6) & \cellcolor{gray!20}70\% (16.7) & 1.85$\times$ & \cellcolor{gray!20}1.42$\times$ \\
\hspace{3mm} Ball-carving & 15\% & \cellcolor{gray!20}23\% & \multicolumn{2}{c|}{\textit{n/a}} & \multicolumn{2}{c|}{\textit{n/a}} & 20\% & \cellcolor{gray!20}20\% & 19\% & \cellcolor{gray!20}24\% & 18\% (2.2) & \cellcolor{gray!20}22\% (1.7) & 5.56$\times$ & \cellcolor{gray!20}4.48$\times$ \\
\hspace{3mm} \textbf{\method{} (Ours)} & \textbf{13\%} & \cellcolor{gray!20}\textbf{27\%} & \multicolumn{2}{c|}{\textit{n/a}} & \multicolumn{2}{c|}{\textit{n/a}} & \textbf{10\%} & \cellcolor{gray!20}\textbf{12\%} & \textbf{8\%} & \cellcolor{gray!20}\textbf{9\%} & \textbf{10\% (2.1)} & \cellcolor{gray!20}\textbf{16\% (7.9)} & \textbf{9.68$\times$} & \cellcolor{gray!20}\textbf{6.25$\times$} \\
\bottomrule
\end{tabular}%
}
\end{table}

\begin{table}[t]
\centering
\footnotesize
\setlength{\tabcolsep}{4pt}
\renewcommand{\arraystretch}{1.1}
\caption{\textbf{Per-dataset Efficiency Analysis: nDCG@$K$ (REAL-MM-RAG).} Coverage budget required to achieve \textbf{90\%} (white) and \textbf{95\%} (gray) of Full-MaxSim\textquotesingle{}s \textbf{NDCG}@$K$ per corpus, on the Granite-Vision-Embedding (GVE) multimodal benchmarks. \textbf{Mean (std)} averages across the four corpora; \textbf{Savings} is $100\%/\textbf{Mean}$.}
\label{tab:per_dataset_eff_ndcg_mm_combinedK}
\resizebox{\ifdim\width>\linewidth \linewidth\else \width\fi}{!}{%
\begin{tabular}{l|cc|cc|cc|cc||cc|cc}
\toprule
\textbf{Task Domain} & \multicolumn{8}{c||}{\textbf{Multimodal (REAL-MM-RAG, GVE)}} & \multicolumn{4}{c}{\textbf{Average}} \\
\cmidrule(lr){1-1} \cmidrule(lr){2-9} \cmidrule(lr){10-13}
\textbf{Method} & \multicolumn{2}{c|}{\textbf{FinSlides}} & \multicolumn{2}{c|}{\textbf{FinReport}} & \multicolumn{2}{c|}{\textbf{TechSlides}} & \multicolumn{2}{c|}{\textbf{TechReport}} & \multicolumn{2}{c|}{\textit{Mean (std)}} & \multicolumn{2}{c}{\textit{Savings (vs.\ Full)}} \\
\midrule

\multicolumn{13}{c}{\textit{$K = 5$}} \\
\midrule
\hspace{3mm} Doc-Uniform & 50\% & \cellcolor{gray!20}71\% & 41\% & \cellcolor{gray!20}71\% & 21\% & \cellcolor{gray!20}25\% & 25\% & \cellcolor{gray!20}50\% & 34\% (11.7) & \cellcolor{gray!20}54\% (18.6) & 2.92$\times$ & \cellcolor{gray!20}1.84$\times$ \\
\hspace{3mm} Ball-carving & 28\% & \cellcolor{gray!20}44\% & 18\% & \cellcolor{gray!20}30\% & 16\% & \cellcolor{gray!20}20\% & 15\% & \cellcolor{gray!20}19\% & 19\% (5.3) & \cellcolor{gray!20}28\% (9.8) & 5.22$\times$ & \cellcolor{gray!20}3.54$\times$ \\
\hspace{3mm} \textbf{\method{} (Ours)} & \textbf{10\%} & \cellcolor{gray!20}\textbf{19\%} & \textbf{8\%} & \cellcolor{gray!20}\textbf{12\%} & \textbf{8\%} & \cellcolor{gray!20}\textbf{9\%} & \textbf{12\%} & \cellcolor{gray!20}\textbf{23\%} & \textbf{10\% (1.5)} & \cellcolor{gray!20}\textbf{15\% (5.7)} & \textbf{10.44$\times$} & \cellcolor{gray!20}\textbf{6.46$\times$} \\
\midrule
\multicolumn{13}{c}{\textit{$K = 10$}} \\
\midrule
\hspace{3mm} Doc-Uniform & 50\% & \cellcolor{gray!20}71\% & 31\% & \cellcolor{gray!20}50\% & 16\% & \cellcolor{gray!20}25\% & 25\% & \cellcolor{gray!20}50\% & 31\% (12.4) & \cellcolor{gray!20}49\% (16.0) & 3.28$\times$ & \cellcolor{gray!20}2.04$\times$ \\
\hspace{3mm} Ball-carving & 28\% & \cellcolor{gray!20}44\% & 18\% & \cellcolor{gray!20}30\% & 16\% & \cellcolor{gray!20}16\% & 15\% & \cellcolor{gray!20}19\% & 19\% (5.3) & \cellcolor{gray!20}27\% (10.9) & 5.22$\times$ & \cellcolor{gray!20}3.68$\times$ \\
\hspace{3mm} \textbf{\method{} (Ours)} & \textbf{11\%} & \cellcolor{gray!20}\textbf{21\%} & \textbf{9\%} & \cellcolor{gray!20}\textbf{13\%} & \textbf{9\%} & \cellcolor{gray!20}\textbf{9\%} & \textbf{14\%} & \cellcolor{gray!20}\textbf{14\%} & \textbf{10\% (2.0)} & \cellcolor{gray!20}\textbf{14\% (4.4)} & \textbf{9.53$\times$} & \cellcolor{gray!20}\textbf{7.07$\times$} \\
\midrule
\multicolumn{13}{c}{\textit{$K = 50$}} \\
\midrule
\hspace{3mm} Doc-Uniform & 41\% & \cellcolor{gray!20}71\% & 25\% & \cellcolor{gray!20}50\% & 16\% & \cellcolor{gray!20}25\% & 25\% & \cellcolor{gray!20}50\% & 27\% (8.8) & \cellcolor{gray!20}49\% (16.0) & 3.72$\times$ & \cellcolor{gray!20}2.04$\times$ \\
\hspace{3mm} Ball-carving & 28\% & \cellcolor{gray!20}44\% & 18\% & \cellcolor{gray!20}30\% & 16\% & \cellcolor{gray!20}16\% & 15\% & \cellcolor{gray!20}19\% & 19\% (5.3) & \cellcolor{gray!20}27\% (10.9) & 5.22$\times$ & \cellcolor{gray!20}3.68$\times$ \\
\hspace{3mm} \textbf{\method{} (Ours)} & \textbf{13\%} & \cellcolor{gray!20}\textbf{18\%} & \textbf{10\%} & \cellcolor{gray!20}\textbf{14\%} & \textbf{11\%} & \cellcolor{gray!20}\textbf{11\%} & \textbf{13\%} & \cellcolor{gray!20}\textbf{16\%} & \textbf{12\% (1.3)} & \cellcolor{gray!20}\textbf{15\% (2.7)} & \textbf{8.60$\times$} & \cellcolor{gray!20}\textbf{6.78$\times$} \\
\bottomrule
\end{tabular}%
}
\end{table}

\begin{table}[t]
\centering
\footnotesize
\setlength{\tabcolsep}{4pt}
\renewcommand{\arraystretch}{1.1}
\caption{\textbf{Per-dataset Efficiency Analysis: Recall@$K$ (BEIR).} We report the coverage budget required to achieve \textbf{90\%} (white) and \textbf{95\%} (gray) of Full-MaxSim\textquotesingle{}s \textbf{RECALL}@$K$ per corpus. Under \textbf{Average}, we report \textbf{mean coverage (std)} across datasets, and \textbf{Savings (vs.\ Full)} is the compute reduction factor relative to full reranking (i.e., $100\%/\textbf{Mean}$). Quora is shown as \textit{n/a} and excluded from the Average: its end-task metrics are zero across all methods under our harness (a qrels/doc-id mapping issue; Table~\ref{tab:dataset_stats}), so per-corpus task-metric budgets are undefined there. Quora retrieval fidelity is unaffected and reported via Overlap@$K$ (Table~\ref{tab:per_dataset_eff_overlap_beir_combinedK}).}
\label{tab:per_dataset_eff_recall_beir_combinedK}
\resizebox{\ifdim\width>\linewidth \linewidth\else \width\fi}{!}{%
\begin{tabular}{l|cc|cc|cc|cc|cc||cc|cc}
\toprule
\textbf{Task Domain} & \multicolumn{10}{c||}{\textbf{Text Retrieval Benchmarks (BEIR)}} & \multicolumn{4}{c}{\textbf{Average}} \\
\cmidrule(lr){1-1} \cmidrule(lr){2-11} \cmidrule(lr){12-15}
\textbf{Method} & \multicolumn{2}{c|}{\textbf{SciDocs}} & \multicolumn{2}{c|}{\textbf{Quora}} & \multicolumn{2}{c|}{\textbf{NQ}} & \multicolumn{2}{c|}{\textbf{HotpotQA}} & \multicolumn{2}{c|}{\textbf{ArguAna}} & \multicolumn{2}{c|}{\textit{Mean (std)}} & \multicolumn{2}{c}{\textit{Savings (vs.\ Full)}} \\
\midrule

\multicolumn{15}{c}{\textit{$K = 5$}} \\
\midrule
\multicolumn{15}{l}{\textit{\textbf{ColBERTv2}}} \\
\hspace{3mm} Doc-Uniform & 41\% & \cellcolor{gray!20}50\% & \multicolumn{2}{c|}{\textit{n/a}} & 22\% & \cellcolor{gray!20}22\% & 72\% & \cellcolor{gray!20}91\% & 50\% & \cellcolor{gray!20}50\% & 46\% (18.0) & \cellcolor{gray!20}53\% (24.6) & 2.16$\times$ & \cellcolor{gray!20}1.88$\times$ \\
\hspace{3mm} Ball-carving & 20\% & \cellcolor{gray!20}26\% & \multicolumn{2}{c|}{\textit{n/a}} & 17\% & \cellcolor{gray!20}17\% & 30\% & \cellcolor{gray!20}34\% & 36\% & \cellcolor{gray!20}44\% & 26\% (7.6) & \cellcolor{gray!20}30\% (10.0) & 3.88$\times$ & \cellcolor{gray!20}3.31$\times$ \\
\hspace{3mm} \textbf{\method{} (Ours)} & \textbf{14\%} & \cellcolor{gray!20}\textbf{14\%} & \multicolumn{2}{c|}{\textit{n/a}} & \textbf{13\%} & \cellcolor{gray!20}\textbf{13\%} & \textbf{13\%} & \cellcolor{gray!20}\textbf{13\%} & \textbf{13\%} & \cellcolor{gray!20}\textbf{13\%} & \textbf{13\% (0.4)} & \cellcolor{gray!20}\textbf{13\% (0.4)} & \textbf{7.55$\times$} & \cellcolor{gray!20}\textbf{7.55$\times$} \\
\midrule
\multicolumn{15}{l}{\textit{\textbf{Jina-ColBERT-V2}}} \\
\hspace{3mm} Doc-Uniform & 100\% & \cellcolor{gray!20}100\% & \multicolumn{2}{c|}{\textit{n/a}} & \multicolumn{2}{c|}{\textit{n/a}} & 50\% & \cellcolor{gray!20}50\% & 31\% & \cellcolor{gray!20}50\% & 60\% (29.1) & \cellcolor{gray!20}67\% (23.6) & 1.66$\times$ & \cellcolor{gray!20}1.50$\times$ \\
\hspace{3mm} Ball-carving & 18\% & \cellcolor{gray!20}23\% & \multicolumn{2}{c|}{\textit{n/a}} & \multicolumn{2}{c|}{\textit{n/a}} & 20\% & \cellcolor{gray!20}20\% & 19\% & \cellcolor{gray!20}24\% & 19\% (0.8) & \cellcolor{gray!20}22\% (1.7) & 5.26$\times$ & \cellcolor{gray!20}4.48$\times$ \\
\hspace{3mm} \textbf{\method{} (Ours)} & \textbf{15\%} & \cellcolor{gray!20}\textbf{24\%} & \multicolumn{2}{c|}{\textit{n/a}} & \multicolumn{2}{c|}{\textit{n/a}} & \textbf{7\%} & \cellcolor{gray!20}\textbf{7\%} & \textbf{7\%} & \cellcolor{gray!20}\textbf{7\%} & \textbf{10\% (3.8)} & \cellcolor{gray!20}\textbf{13\% (8.0)} & \textbf{10.34$\times$} & \cellcolor{gray!20}\textbf{7.89$\times$} \\
\midrule
\multicolumn{15}{c}{\textit{$K = 10$}} \\
\midrule
\multicolumn{15}{l}{\textit{\textbf{ColBERTv2}}} \\
\hspace{3mm} Doc-Uniform & 22\% & \cellcolor{gray!20}25\% & \multicolumn{2}{c|}{\textit{n/a}} & 16\% & \cellcolor{gray!20}22\% & 72\% & \cellcolor{gray!20}81\% & 25\% & \cellcolor{gray!20}72\% & 34\% (22.3) & \cellcolor{gray!20}50\% (26.7) & 2.96$\times$ & \cellcolor{gray!20}2.00$\times$ \\
\hspace{3mm} Ball-carving & 26\% & \cellcolor{gray!20}29\% & \multicolumn{2}{c|}{\textit{n/a}} & 12\% & \cellcolor{gray!20}12\% & 23\% & \cellcolor{gray!20}30\% & 36\% & \cellcolor{gray!20}36\% & 24\% (8.6) & \cellcolor{gray!20}27\% (8.9) & 4.12$\times$ & \cellcolor{gray!20}3.74$\times$ \\
\hspace{3mm} \textbf{\method{} (Ours)} & \textbf{15\%} & \cellcolor{gray!20}\textbf{15\%} & \multicolumn{2}{c|}{\textit{n/a}} & \textbf{13\%} & \cellcolor{gray!20}\textbf{13\%} & \textbf{13\%} & \cellcolor{gray!20}\textbf{13\%} & \textbf{13\%} & \cellcolor{gray!20}\textbf{13\%} & \textbf{14\% (0.9)} & \cellcolor{gray!20}\textbf{14\% (0.9)} & \textbf{7.41$\times$} & \cellcolor{gray!20}\textbf{7.41$\times$} \\
\midrule
\multicolumn{15}{l}{\textit{\textbf{Jina-ColBERT-V2}}} \\
\hspace{3mm} Doc-Uniform & 50\% & \cellcolor{gray!20}91\% & \multicolumn{2}{c|}{\textit{n/a}} & \multicolumn{2}{c|}{\textit{n/a}} & 31\% & \cellcolor{gray!20}50\% & 20\% & \cellcolor{gray!20}50\% & 34\% (12.4) & \cellcolor{gray!20}64\% (19.3) & 2.97$\times$ & \cellcolor{gray!20}1.57$\times$ \\
\hspace{3mm} Ball-carving & 15\% & \cellcolor{gray!20}18\% & \multicolumn{2}{c|}{\textit{n/a}} & \multicolumn{2}{c|}{\textit{n/a}} & 20\% & \cellcolor{gray!20}20\% & 19\% & \cellcolor{gray!20}19\% & 18\% (2.2) & \cellcolor{gray!20}19\% (0.8) & 5.56$\times$ & \cellcolor{gray!20}5.26$\times$ \\
\hspace{3mm} \textbf{\method{} (Ours)} & \textbf{22\%} & \cellcolor{gray!20}\textbf{60\%} & \multicolumn{2}{c|}{\textit{n/a}} & \multicolumn{2}{c|}{\textit{n/a}} & \textbf{34\%} & \cellcolor{gray!20}\textbf{34\%} & \textbf{7\%} & \cellcolor{gray!20}\textbf{7\%} & \textbf{21\% (11.0)} & \cellcolor{gray!20}\textbf{34\% (21.6)} & \textbf{4.76$\times$} & \cellcolor{gray!20}\textbf{2.97$\times$} \\
\midrule
\multicolumn{15}{c}{\textit{$K = 50$}} \\
\midrule
\multicolumn{15}{l}{\textit{\textbf{ColBERTv2}}} \\
\hspace{3mm} Doc-Uniform & 16\% & \cellcolor{gray!20}41\% & \multicolumn{2}{c|}{\textit{n/a}} & 12\% & \cellcolor{gray!20}16\% & 31\% & \cellcolor{gray!20}72\% & 25\% & \cellcolor{gray!20}25\% & 21\% (7.4) & \cellcolor{gray!20}38\% (21.3) & 4.76$\times$ & \cellcolor{gray!20}2.60$\times$ \\
\hspace{3mm} Ball-carving & 20\% & \cellcolor{gray!20}26\% & \multicolumn{2}{c|}{\textit{n/a}} & 12\% & \cellcolor{gray!20}12\% & 23\% & \cellcolor{gray!20}23\% & 26\% & \cellcolor{gray!20}26\% & 20\% (5.2) & \cellcolor{gray!20}22\% (5.8) & 4.94$\times$ & \cellcolor{gray!20}4.60$\times$ \\
\hspace{3mm} \textbf{\method{} (Ours)} & \textbf{14\%} & \cellcolor{gray!20}\textbf{17\%} & \multicolumn{2}{c|}{\textit{n/a}} & \textbf{13\%} & \cellcolor{gray!20}\textbf{13\%} & \textbf{13\%} & \cellcolor{gray!20}\textbf{14\%} & \textbf{14\%} & \cellcolor{gray!20}\textbf{14\%} & \textbf{14\% (0.5)} & \cellcolor{gray!20}\textbf{14\% (1.5)} & \textbf{7.41$\times$} & \cellcolor{gray!20}\textbf{6.90$\times$} \\
\midrule
\multicolumn{15}{l}{\textit{\textbf{Jina-ColBERT-V2}}} \\
\hspace{3mm} Doc-Uniform & 70\% & \cellcolor{gray!20}91\% & \multicolumn{2}{c|}{\textit{n/a}} & \multicolumn{2}{c|}{\textit{n/a}} & 25\% & \cellcolor{gray!20}31\% & 20\% & \cellcolor{gray!20}31\% & 38\% (22.5) & \cellcolor{gray!20}51\% (28.3) & 2.61$\times$ & \cellcolor{gray!20}1.96$\times$ \\
\hspace{3mm} Ball-carving & 23\% & \cellcolor{gray!20}23\% & \multicolumn{2}{c|}{\textit{n/a}} & \multicolumn{2}{c|}{\textit{n/a}} & 15\% & \cellcolor{gray!20}15\% & 19\% & \cellcolor{gray!20}19\% & 19\% (3.3) & \cellcolor{gray!20}19\% (3.3) & 5.26$\times$ & \cellcolor{gray!20}5.26$\times$ \\
\hspace{3mm} \textbf{\method{} (Ours)} & \textbf{16\%} & \cellcolor{gray!20}\textbf{27\%} & \multicolumn{2}{c|}{\textit{n/a}} & \multicolumn{2}{c|}{\textit{n/a}} & \textbf{10\%} & \cellcolor{gray!20}\textbf{12\%} & \textbf{9\%} & \cellcolor{gray!20}\textbf{10\%} & \textbf{12\% (3.1)} & \cellcolor{gray!20}\textbf{16\% (7.6)} & \textbf{8.57$\times$} & \cellcolor{gray!20}\textbf{6.12$\times$} \\
\bottomrule
\end{tabular}%
}
\end{table}

\begin{table}[t]
\centering
\footnotesize
\setlength{\tabcolsep}{4pt}
\renewcommand{\arraystretch}{1.1}
\caption{\textbf{Per-dataset Efficiency Analysis: Recall@$K$ (REAL-MM-RAG).} Coverage budget required to achieve \textbf{90\%} (white) and \textbf{95\%} (gray) of Full-MaxSim\textquotesingle{}s \textbf{Recall}@$K$ per corpus, on the Granite-Vision-Embedding (GVE) multimodal benchmarks. \textbf{Mean (std)} averages across the four corpora; \textbf{Savings} is $100\%/\textbf{Mean}$.}
\label{tab:per_dataset_eff_recall_mm_combinedK}
\resizebox{\ifdim\width>\linewidth \linewidth\else \width\fi}{!}{%
\begin{tabular}{l|cc|cc|cc|cc||cc|cc}
\toprule
\textbf{Task Domain} & \multicolumn{8}{c||}{\textbf{Multimodal (REAL-MM-RAG, GVE)}} & \multicolumn{4}{c}{\textbf{Average}} \\
\cmidrule(lr){1-1} \cmidrule(lr){2-9} \cmidrule(lr){10-13}
\textbf{Method} & \multicolumn{2}{c|}{\textbf{FinSlides}} & \multicolumn{2}{c|}{\textbf{FinReport}} & \multicolumn{2}{c|}{\textbf{TechSlides}} & \multicolumn{2}{c|}{\textbf{TechReport}} & \multicolumn{2}{c|}{\textit{Mean (std)}} & \multicolumn{2}{c}{\textit{Savings (vs.\ Full)}} \\
\midrule

\multicolumn{13}{c}{\textit{$K = 5$}} \\
\midrule
\hspace{3mm} Doc-Uniform & 41\% & \cellcolor{gray!20}50\% & 25\% & \cellcolor{gray!20}50\% & 16\% & \cellcolor{gray!20}21\% & 21\% & \cellcolor{gray!20}31\% & 26\% (9.3) & \cellcolor{gray!20}38\% (12.6) & 3.90$\times$ & \cellcolor{gray!20}2.64$\times$ \\
\hspace{3mm} Ball-carving & 21\% & \cellcolor{gray!20}44\% & 18\% & \cellcolor{gray!20}23\% & 16\% & \cellcolor{gray!20}16\% & 15\% & \cellcolor{gray!20}15\% & 17\% (2.4) & \cellcolor{gray!20}24\% (11.5) & 5.74$\times$ & \cellcolor{gray!20}4.11$\times$ \\
\hspace{3mm} \textbf{\method{} (Ours)} & \textbf{19\%} & \cellcolor{gray!20}\textbf{19\%} & \textbf{8\%} & \cellcolor{gray!20}\textbf{12\%} & \textbf{8\%} & \cellcolor{gray!20}\textbf{9\%} & \textbf{12\%} & \cellcolor{gray!20}\textbf{23\%} & \textbf{12\% (4.2)} & \cellcolor{gray!20}\textbf{15\% (5.7)} & \textbf{8.43$\times$} & \cellcolor{gray!20}\textbf{6.46$\times$} \\
\midrule
\multicolumn{13}{c}{\textit{$K = 10$}} \\
\midrule
\hspace{3mm} Doc-Uniform & 21\% & \cellcolor{gray!20}41\% & 25\% & \cellcolor{gray!20}50\% & 16\% & \cellcolor{gray!20}21\% & 21\% & \cellcolor{gray!20}31\% & 21\% (3.4) & \cellcolor{gray!20}36\% (10.9) & 4.84$\times$ & \cellcolor{gray!20}2.82$\times$ \\
\hspace{3mm} Ball-carving & 16\% & \cellcolor{gray!20}21\% & 18\% & \cellcolor{gray!20}18\% & 16\% & \cellcolor{gray!20}16\% & 15\% & \cellcolor{gray!20}15\% & 16\% (1.0) & \cellcolor{gray!20}17\% (2.4) & 6.18$\times$ & \cellcolor{gray!20}5.74$\times$ \\
\hspace{3mm} \textbf{\method{} (Ours)} & \textbf{21\%} & \cellcolor{gray!20}\textbf{21\%} & \textbf{9\%} & \cellcolor{gray!20}\textbf{13\%} & \textbf{9\%} & \cellcolor{gray!20}\textbf{9\%} & \textbf{14\%} & \cellcolor{gray!20}\textbf{26\%} & \textbf{13\% (5.0)} & \cellcolor{gray!20}\textbf{17\% (6.8)} & \textbf{7.62$\times$} & \cellcolor{gray!20}\textbf{5.79$\times$} \\
\midrule
\multicolumn{13}{c}{\textit{$K = 50$}} \\
\midrule
\hspace{3mm} Doc-Uniform & 11\% & \cellcolor{gray!20}16\% & 11\% & \cellcolor{gray!20}16\% & 11\% & \cellcolor{gray!20}16\% & 11\% & \cellcolor{gray!20}25\% & 11\% (0.1) & \cellcolor{gray!20}18\% (4.2) & 9.33$\times$ & \cellcolor{gray!20}5.49$\times$ \\
\hspace{3mm} Ball-carving & 12\% & \cellcolor{gray!20}16\% & 14\% & \cellcolor{gray!20}14\% & 16\% & \cellcolor{gray!20}16\% & 15\% & \cellcolor{gray!20}15\% & 14\% (1.3) & \cellcolor{gray!20}15\% (0.9) & 7.03$\times$ & \cellcolor{gray!20}6.59$\times$ \\
\hspace{3mm} \textbf{\method{} (Ours)} & \textbf{13\%} & \cellcolor{gray!20}\textbf{18\%} & \textbf{11\%} & \cellcolor{gray!20}\textbf{14\%} & \textbf{11\%} & \cellcolor{gray!20}\textbf{11\%} & \textbf{13\%} & \cellcolor{gray!20}\textbf{16\%} & \textbf{12\% (1.0)} & \cellcolor{gray!20}\textbf{15\% (2.7)} & \textbf{8.38$\times$} & \cellcolor{gray!20}\textbf{6.78$\times$} \\
\bottomrule
\end{tabular}%
}
\end{table}

\begin{table}[t]
\centering
\footnotesize
\setlength{\tabcolsep}{4pt}
\renewcommand{\arraystretch}{1.1}
\caption{\textbf{Per-dataset Efficiency Analysis: MRR@$K$ (BEIR).} We report the coverage budget required to achieve \textbf{90\%} (white) and \textbf{95\%} (gray) of Full-MaxSim\textquotesingle{}s \textbf{MRR}@$K$ per corpus. Under \textbf{Average}, we report \textbf{mean coverage (std)} across datasets, and \textbf{Savings (vs.\ Full)} is the compute reduction factor relative to full reranking (i.e., $100\%/\textbf{Mean}$). Quora is shown as \textit{n/a} and excluded from the Average: its end-task metrics are zero across all methods under our harness (a qrels/doc-id mapping issue; Table~\ref{tab:dataset_stats}), so per-corpus task-metric budgets are undefined there. Quora retrieval fidelity is unaffected and reported via Overlap@$K$ (Table~\ref{tab:per_dataset_eff_overlap_beir_combinedK}).}
\label{tab:per_dataset_eff_mrr_beir_combinedK}
\resizebox{\ifdim\width>\linewidth \linewidth\else \width\fi}{!}{%
\begin{tabular}{l|cc|cc|cc|cc|cc||cc|cc}
\toprule
\textbf{Task Domain} & \multicolumn{10}{c||}{\textbf{Text Retrieval Benchmarks (BEIR)}} & \multicolumn{4}{c}{\textbf{Average}} \\
\cmidrule(lr){1-1} \cmidrule(lr){2-11} \cmidrule(lr){12-15}
\textbf{Method} & \multicolumn{2}{c|}{\textbf{SciDocs}} & \multicolumn{2}{c|}{\textbf{Quora}} & \multicolumn{2}{c|}{\textbf{NQ}} & \multicolumn{2}{c|}{\textbf{HotpotQA}} & \multicolumn{2}{c|}{\textbf{ArguAna}} & \multicolumn{2}{c|}{\textit{Mean (std)}} & \multicolumn{2}{c}{\textit{Savings (vs.\ Full)}} \\
\midrule

\multicolumn{15}{c}{\textit{$K = 5$}} \\
\midrule
\multicolumn{15}{l}{\textit{\textbf{ColBERTv2}}} \\
\hspace{3mm} Doc-Uniform & 22\% & \cellcolor{gray!20}72\% & \multicolumn{2}{c|}{\textit{n/a}} & 22\% & \cellcolor{gray!20}25\% & 50\% & \cellcolor{gray!20}50\% & 41\% & \cellcolor{gray!20}50\% & 34\% (12.2) & \cellcolor{gray!20}49\% (16.6) & 2.96$\times$ & \cellcolor{gray!20}2.03$\times$ \\
\hspace{3mm} Ball-carving & 20\% & \cellcolor{gray!20}20\% & \multicolumn{2}{c|}{\textit{n/a}} & 17\% & \cellcolor{gray!20}17\% & 23\% & \cellcolor{gray!20}30\% & 26\% & \cellcolor{gray!20}36\% & 22\% (3.4) & \cellcolor{gray!20}26\% (7.6) & 4.65$\times$ & \cellcolor{gray!20}3.88$\times$ \\
\hspace{3mm} \textbf{\method{} (Ours)} & \textbf{13\%} & \cellcolor{gray!20}\textbf{13\%} & \multicolumn{2}{c|}{\textit{n/a}} & \textbf{13\%} & \cellcolor{gray!20}\textbf{13\%} & \textbf{13\%} & \cellcolor{gray!20}\textbf{13\%} & \textbf{13\%} & \cellcolor{gray!20}\textbf{13\%} & \textbf{13\% (0.0)} & \cellcolor{gray!20}\textbf{13\% (0.0)} & \textbf{7.69$\times$} & \cellcolor{gray!20}\textbf{7.69$\times$} \\
\midrule
\multicolumn{15}{l}{\textit{\textbf{Jina-ColBERT-V2}}} \\
\hspace{3mm} Doc-Uniform & 91\% & \cellcolor{gray!20}100\% & \multicolumn{2}{c|}{\textit{n/a}} & \multicolumn{2}{c|}{\textit{n/a}} & 50\% & \cellcolor{gray!20}70\% & 50\% & \cellcolor{gray!20}50\% & 64\% (19.3) & \cellcolor{gray!20}73\% (20.5) & 1.57$\times$ & \cellcolor{gray!20}1.36$\times$ \\
\hspace{3mm} Ball-carving & 15\% & \cellcolor{gray!20}18\% & \multicolumn{2}{c|}{\textit{n/a}} & \multicolumn{2}{c|}{\textit{n/a}} & 20\% & \cellcolor{gray!20}26\% & 19\% & \cellcolor{gray!20}24\% & 18\% (2.2) & \cellcolor{gray!20}23\% (3.4) & 5.56$\times$ & \cellcolor{gray!20}4.41$\times$ \\
\hspace{3mm} \textbf{\method{} (Ours)} & \textbf{10\%} & \cellcolor{gray!20}\textbf{10\%} & \multicolumn{2}{c|}{\textit{n/a}} & \multicolumn{2}{c|}{\textit{n/a}} & \textbf{7\%} & \cellcolor{gray!20}\textbf{7\%} & \textbf{7\%} & \cellcolor{gray!20}\textbf{7\%} & \textbf{8\% (1.4)} & \cellcolor{gray!20}\textbf{8\% (1.4)} & \textbf{12.50$\times$} & \cellcolor{gray!20}\textbf{12.50$\times$} \\
\midrule
\multicolumn{15}{c}{\textit{$K = 10$}} \\
\midrule
\multicolumn{15}{l}{\textit{\textbf{ColBERTv2}}} \\
\hspace{3mm} Doc-Uniform & 22\% & \cellcolor{gray!20}72\% & \multicolumn{2}{c|}{\textit{n/a}} & 22\% & \cellcolor{gray!20}50\% & 50\% & \cellcolor{gray!20}50\% & 41\% & \cellcolor{gray!20}50\% & 34\% (12.2) & \cellcolor{gray!20}56\% (9.5) & 2.96$\times$ & \cellcolor{gray!20}1.80$\times$ \\
\hspace{3mm} Ball-carving & 20\% & \cellcolor{gray!20}20\% & \multicolumn{2}{c|}{\textit{n/a}} & 12\% & \cellcolor{gray!20}17\% & 23\% & \cellcolor{gray!20}30\% & 26\% & \cellcolor{gray!20}36\% & 20\% (5.2) & \cellcolor{gray!20}26\% (7.6) & 4.94$\times$ & \cellcolor{gray!20}3.88$\times$ \\
\hspace{3mm} \textbf{\method{} (Ours)} & \textbf{13\%} & \cellcolor{gray!20}\textbf{13\%} & \multicolumn{2}{c|}{\textit{n/a}} & \textbf{13\%} & \cellcolor{gray!20}\textbf{13\%} & \textbf{13\%} & \cellcolor{gray!20}\textbf{13\%} & \textbf{13\%} & \cellcolor{gray!20}\textbf{13\%} & \textbf{13\% (0.0)} & \cellcolor{gray!20}\textbf{13\% (0.0)} & \textbf{7.69$\times$} & \cellcolor{gray!20}\textbf{7.69$\times$} \\
\midrule
\multicolumn{15}{l}{\textit{\textbf{Jina-ColBERT-V2}}} \\
\hspace{3mm} Doc-Uniform & 81\% & \cellcolor{gray!20}100\% & \multicolumn{2}{c|}{\textit{n/a}} & \multicolumn{2}{c|}{\textit{n/a}} & 50\% & \cellcolor{gray!20}70\% & 50\% & \cellcolor{gray!20}70\% & 60\% (14.6) & \cellcolor{gray!20}80\% (14.1) & 1.66$\times$ & \cellcolor{gray!20}1.25$\times$ \\
\hspace{3mm} Ball-carving & 15\% & \cellcolor{gray!20}18\% & \multicolumn{2}{c|}{\textit{n/a}} & \multicolumn{2}{c|}{\textit{n/a}} & 20\% & \cellcolor{gray!20}20\% & 19\% & \cellcolor{gray!20}24\% & 18\% (2.2) & \cellcolor{gray!20}21\% (2.5) & 5.56$\times$ & \cellcolor{gray!20}4.84$\times$ \\
\hspace{3mm} \textbf{\method{} (Ours)} & \textbf{22\%} & \cellcolor{gray!20}\textbf{22\%} & \multicolumn{2}{c|}{\textit{n/a}} & \multicolumn{2}{c|}{\textit{n/a}} & \textbf{7\%} & \cellcolor{gray!20}\textbf{7\%} & \textbf{7\%} & \cellcolor{gray!20}\textbf{7\%} & \textbf{12\% (7.1)} & \cellcolor{gray!20}\textbf{12\% (7.1)} & \textbf{8.33$\times$} & \cellcolor{gray!20}\textbf{8.33$\times$} \\
\midrule
\multicolumn{15}{c}{\textit{$K = 50$}} \\
\midrule
\multicolumn{15}{l}{\textit{\textbf{ColBERTv2}}} \\
\hspace{3mm} Doc-Uniform & 22\% & \cellcolor{gray!20}81\% & \multicolumn{2}{c|}{\textit{n/a}} & 22\% & \cellcolor{gray!20}25\% & 50\% & \cellcolor{gray!20}50\% & 41\% & \cellcolor{gray!20}50\% & 34\% (12.2) & \cellcolor{gray!20}52\% (19.9) & 2.96$\times$ & \cellcolor{gray!20}1.94$\times$ \\
\hspace{3mm} Ball-carving & 20\% & \cellcolor{gray!20}20\% & \multicolumn{2}{c|}{\textit{n/a}} & 12\% & \cellcolor{gray!20}17\% & 23\% & \cellcolor{gray!20}30\% & 26\% & \cellcolor{gray!20}26\% & 20\% (5.2) & \cellcolor{gray!20}23\% (5.1) & 4.94$\times$ & \cellcolor{gray!20}4.30$\times$ \\
\hspace{3mm} \textbf{\method{} (Ours)} & \textbf{13\%} & \cellcolor{gray!20}\textbf{13\%} & \multicolumn{2}{c|}{\textit{n/a}} & \textbf{13\%} & \cellcolor{gray!20}\textbf{13\%} & \textbf{13\%} & \cellcolor{gray!20}\textbf{13\%} & \textbf{13\%} & \cellcolor{gray!20}\textbf{13\%} & \textbf{13\% (0.0)} & \cellcolor{gray!20}\textbf{13\% (0.0)} & \textbf{7.69$\times$} & \cellcolor{gray!20}\textbf{7.69$\times$} \\
\midrule
\multicolumn{15}{l}{\textit{\textbf{Jina-ColBERT-V2}}} \\
\hspace{3mm} Doc-Uniform & 81\% & \cellcolor{gray!20}100\% & \multicolumn{2}{c|}{\textit{n/a}} & \multicolumn{2}{c|}{\textit{n/a}} & 50\% & \cellcolor{gray!20}70\% & 50\% & \cellcolor{gray!20}70\% & 60\% (14.6) & \cellcolor{gray!20}80\% (14.1) & 1.66$\times$ & \cellcolor{gray!20}1.25$\times$ \\
\hspace{3mm} Ball-carving & 15\% & \cellcolor{gray!20}18\% & \multicolumn{2}{c|}{\textit{n/a}} & \multicolumn{2}{c|}{\textit{n/a}} & 20\% & \cellcolor{gray!20}20\% & 24\% & \cellcolor{gray!20}24\% & 20\% (3.7) & \cellcolor{gray!20}21\% (2.5) & 5.08$\times$ & \cellcolor{gray!20}4.84$\times$ \\
\hspace{3mm} \textbf{\method{} (Ours)} & \textbf{10\%} & \cellcolor{gray!20}\textbf{13\%} & \multicolumn{2}{c|}{\textit{n/a}} & \multicolumn{2}{c|}{\textit{n/a}} & \textbf{10\%} & \cellcolor{gray!20}\textbf{10\%} & \textbf{8\%} & \cellcolor{gray!20}\textbf{9\%} & \textbf{9\% (0.9)} & \cellcolor{gray!20}\textbf{11\% (1.7)} & \textbf{10.71$\times$} & \cellcolor{gray!20}\textbf{9.38$\times$} \\
\bottomrule
\end{tabular}%
}
\end{table}

\begin{table}[t]
\centering
\footnotesize
\setlength{\tabcolsep}{4pt}
\renewcommand{\arraystretch}{1.1}
\caption{\textbf{Per-dataset Efficiency Analysis: MRR@$K$ (REAL-MM-RAG).} Coverage budget required to achieve \textbf{90\%} (white) and \textbf{95\%} (gray) of Full-MaxSim\textquotesingle{}s \textbf{MRR}@$K$ per corpus, on the Granite-Vision-Embedding (GVE) multimodal benchmarks. \textbf{Mean (std)} averages across the four corpora; \textbf{Savings} is $100\%/\textbf{Mean}$.}
\label{tab:per_dataset_eff_mrr_mm_combinedK}
\resizebox{\ifdim\width>\linewidth \linewidth\else \width\fi}{!}{%
\begin{tabular}{l|cc|cc|cc|cc||cc|cc}
\toprule
\textbf{Task Domain} & \multicolumn{8}{c||}{\textbf{Multimodal (REAL-MM-RAG, GVE)}} & \multicolumn{4}{c}{\textbf{Average}} \\
\cmidrule(lr){1-1} \cmidrule(lr){2-9} \cmidrule(lr){10-13}
\textbf{Method} & \multicolumn{2}{c|}{\textbf{FinSlides}} & \multicolumn{2}{c|}{\textbf{FinReport}} & \multicolumn{2}{c|}{\textbf{TechSlides}} & \multicolumn{2}{c|}{\textbf{TechReport}} & \multicolumn{2}{c|}{\textit{Mean (std)}} & \multicolumn{2}{c}{\textit{Savings (vs.\ Full)}} \\
\midrule

\multicolumn{13}{c}{\textit{$K = 5$}} \\
\midrule
\hspace{3mm} Doc-Uniform & 71\% & \cellcolor{gray!20}71\% & 41\% & \cellcolor{gray!20}71\% & 21\% & \cellcolor{gray!20}41\% & 31\% & \cellcolor{gray!20}50\% & 41\% (18.7) & \cellcolor{gray!20}58\% (13.0) & 2.46$\times$ & \cellcolor{gray!20}1.72$\times$ \\
\hspace{3mm} Ball-carving & 28\% & \cellcolor{gray!20}44\% & 18\% & \cellcolor{gray!20}30\% & 16\% & \cellcolor{gray!20}20\% & 15\% & \cellcolor{gray!20}19\% & 19\% (5.3) & \cellcolor{gray!20}28\% (9.8) & 5.22$\times$ & \cellcolor{gray!20}3.54$\times$ \\
\hspace{3mm} \textbf{\method{} (Ours)} & \textbf{10\%} & \cellcolor{gray!20}\textbf{19\%} & \textbf{8\%} & \cellcolor{gray!20}\textbf{12\%} & \textbf{8\%} & \cellcolor{gray!20}\textbf{9\%} & \textbf{12\%} & \cellcolor{gray!20}\textbf{23\%} & \textbf{10\% (1.5)} & \cellcolor{gray!20}\textbf{15\% (5.7)} & \textbf{10.44$\times$} & \cellcolor{gray!20}\textbf{6.46$\times$} \\
\midrule
\multicolumn{13}{c}{\textit{$K = 10$}} \\
\midrule
\hspace{3mm} Doc-Uniform & 71\% & \cellcolor{gray!20}71\% & 41\% & \cellcolor{gray!20}71\% & 21\% & \cellcolor{gray!20}25\% & 31\% & \cellcolor{gray!20}50\% & 41\% (18.7) & \cellcolor{gray!20}54\% (18.6) & 2.46$\times$ & \cellcolor{gray!20}1.84$\times$ \\
\hspace{3mm} Ball-carving & 28\% & \cellcolor{gray!20}44\% & 18\% & \cellcolor{gray!20}30\% & 16\% & \cellcolor{gray!20}20\% & 15\% & \cellcolor{gray!20}19\% & 19\% (5.3) & \cellcolor{gray!20}28\% (9.8) & 5.22$\times$ & \cellcolor{gray!20}3.54$\times$ \\
\hspace{3mm} \textbf{\method{} (Ours)} & \textbf{11\%} & \cellcolor{gray!20}\textbf{21\%} & \textbf{9\%} & \cellcolor{gray!20}\textbf{13\%} & \textbf{9\%} & \cellcolor{gray!20}\textbf{9\%} & \textbf{14\%} & \cellcolor{gray!20}\textbf{14\%} & \textbf{10\% (2.0)} & \cellcolor{gray!20}\textbf{14\% (4.5)} & \textbf{9.53$\times$} & \cellcolor{gray!20}\textbf{7.09$\times$} \\
\midrule
\multicolumn{13}{c}{\textit{$K = 50$}} \\
\midrule
\hspace{3mm} Doc-Uniform & 71\% & \cellcolor{gray!20}71\% & 41\% & \cellcolor{gray!20}71\% & 21\% & \cellcolor{gray!20}25\% & 31\% & \cellcolor{gray!20}50\% & 41\% (18.7) & \cellcolor{gray!20}54\% (18.6) & 2.46$\times$ & \cellcolor{gray!20}1.84$\times$ \\
\hspace{3mm} Ball-carving & 28\% & \cellcolor{gray!20}44\% & 18\% & \cellcolor{gray!20}30\% & 16\% & \cellcolor{gray!20}20\% & 15\% & \cellcolor{gray!20}19\% & 19\% (5.3) & \cellcolor{gray!20}28\% (9.8) & 5.22$\times$ & \cellcolor{gray!20}3.54$\times$ \\
\hspace{3mm} \textbf{\method{} (Ours)} & \textbf{13\%} & \cellcolor{gray!20}\textbf{18\%} & \textbf{9\%} & \cellcolor{gray!20}\textbf{11\%} & \textbf{11\%} & \cellcolor{gray!20}\textbf{11\%} & \textbf{11\%} & \cellcolor{gray!20}\textbf{16\%} & \textbf{11\% (1.3)} & \cellcolor{gray!20}\textbf{14\% (3.2)} & \textbf{9.08$\times$} & \cellcolor{gray!20}\textbf{7.11$\times$} \\
\bottomrule
\end{tabular}%
}
\end{table}

\subsection{Headline Speedup at Other $\alpha_{\mathrm{ef}}$ and $K$}
\label{sec:app_headline_variants}

This subsection extends the headline figure from the main paper (Figure~\ref{fig:headline_speedup}, which fixes the deployed knob $\alpha_{\mathrm{ef}}{=}0.2$, $K{=}5$) by sweeping the calibration knob $\alpha_{\mathrm{ef}}$ and the target rank $K$ while holding everything else identical (\textsc{Cpu-S} = AMD EPYC 7763, $1$/$8$/$16$ threads, BEIR + REAL-MM-RAG, ColBERTv2 / GVE). Each figure uses the same bar/diamond layout as the body figure: bars are CB-vs-Full speedup at $1$/$8$/$16$ threads per corpus; right-axis purple diamonds are Overlap@$K$ vs.\ Full-MaxSim's exhaustive top-$K$.

\begin{figure}[!htbp]
\centering
\includegraphics[width=\linewidth]{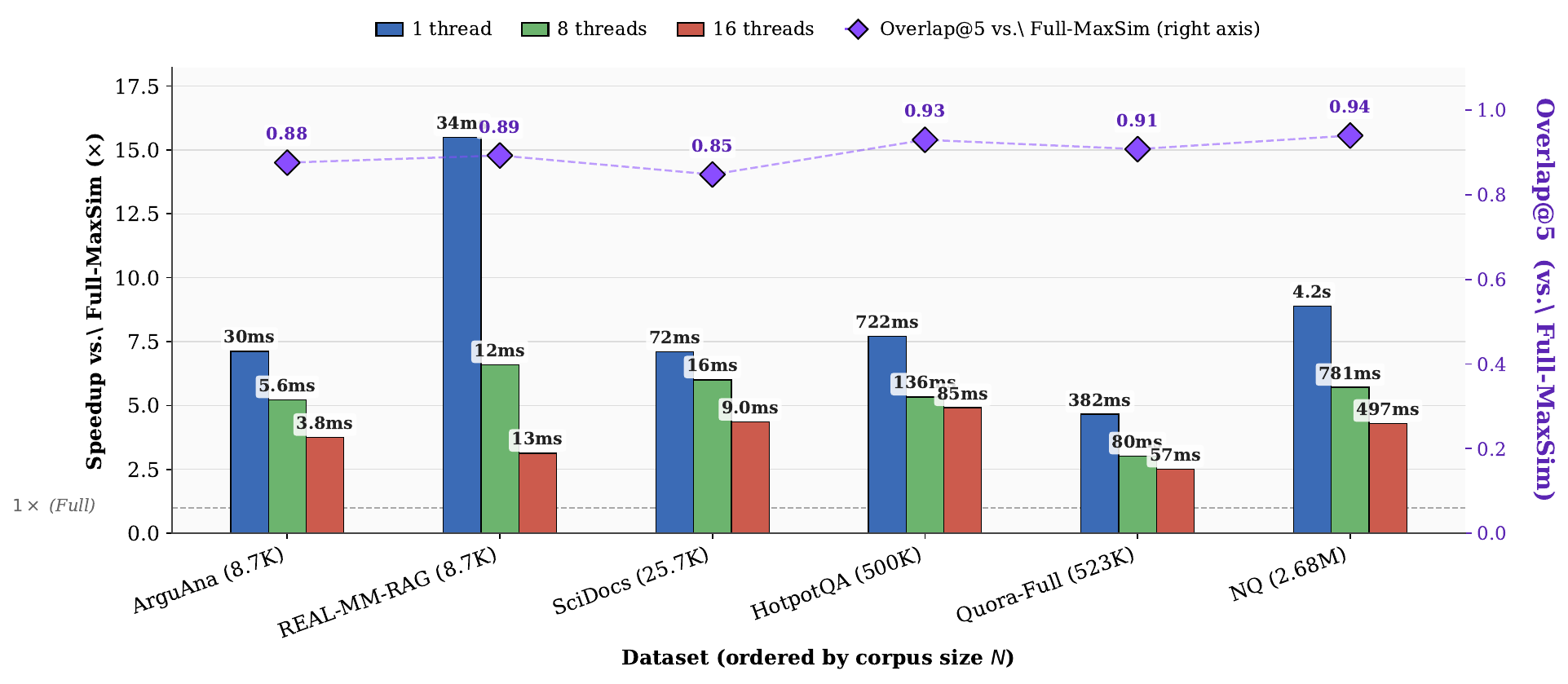}
\caption{Wall-clock speedup at $\alpha_{\mathrm{ef}}{=}0.1$ (aggressive corner): higher speedup than the deployed knob but Ov@$5$ falls as low as $0.76$ on the toughest corpus (full per-corpus ranges in Table~\ref{tab:alpha_guidance}).}
\label{fig:headline_K5_a01}
\end{figure}

\begin{figure}[!htbp]
\centering
\includegraphics[width=\linewidth]{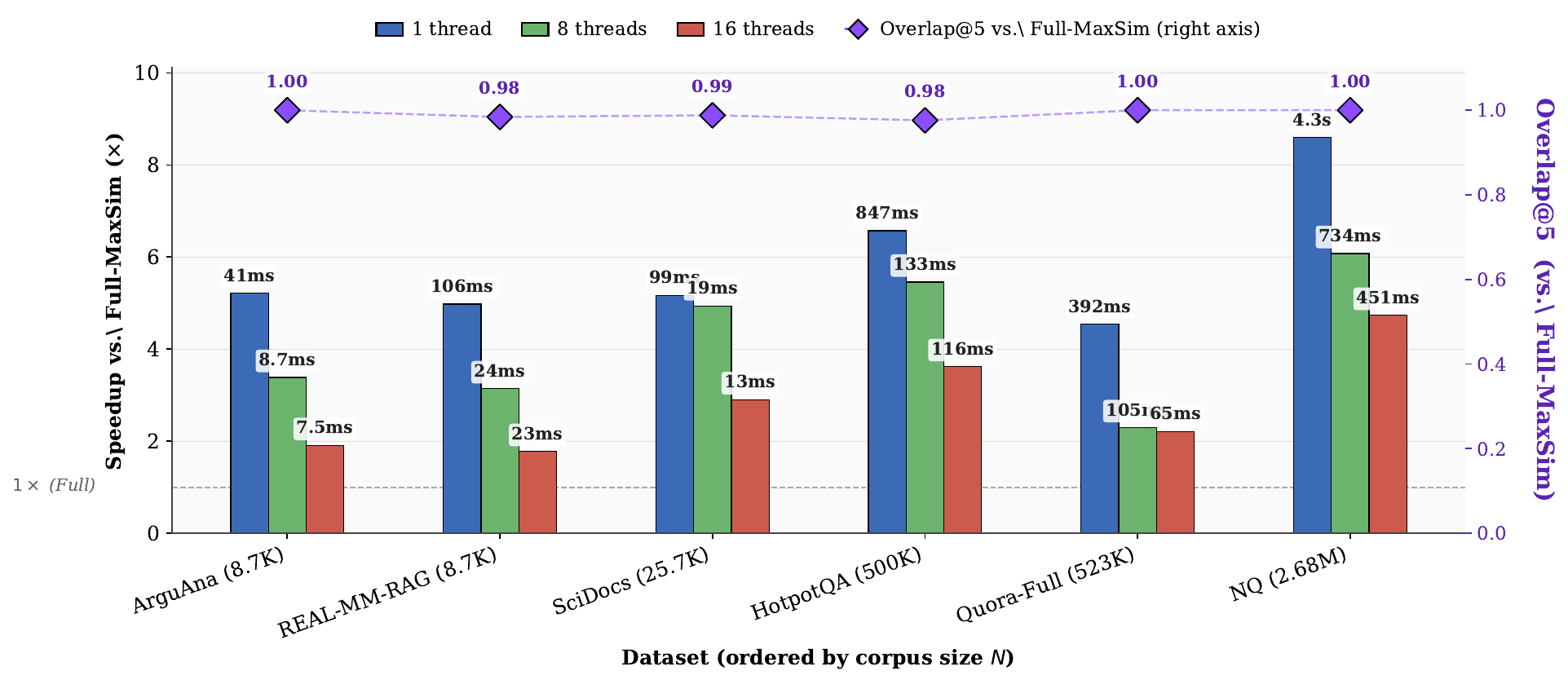}
\caption{Wall-clock speedup at $\alpha_{\mathrm{ef}}{=}0.3$ (conservative): a slightly tighter knob than deployed; Ov@$5\geq 0.98$ on every corpus at modestly reduced speedup.}
\label{fig:headline_K5_a03}
\end{figure}

\begin{figure}[!htbp]
\centering
\includegraphics[width=\linewidth]{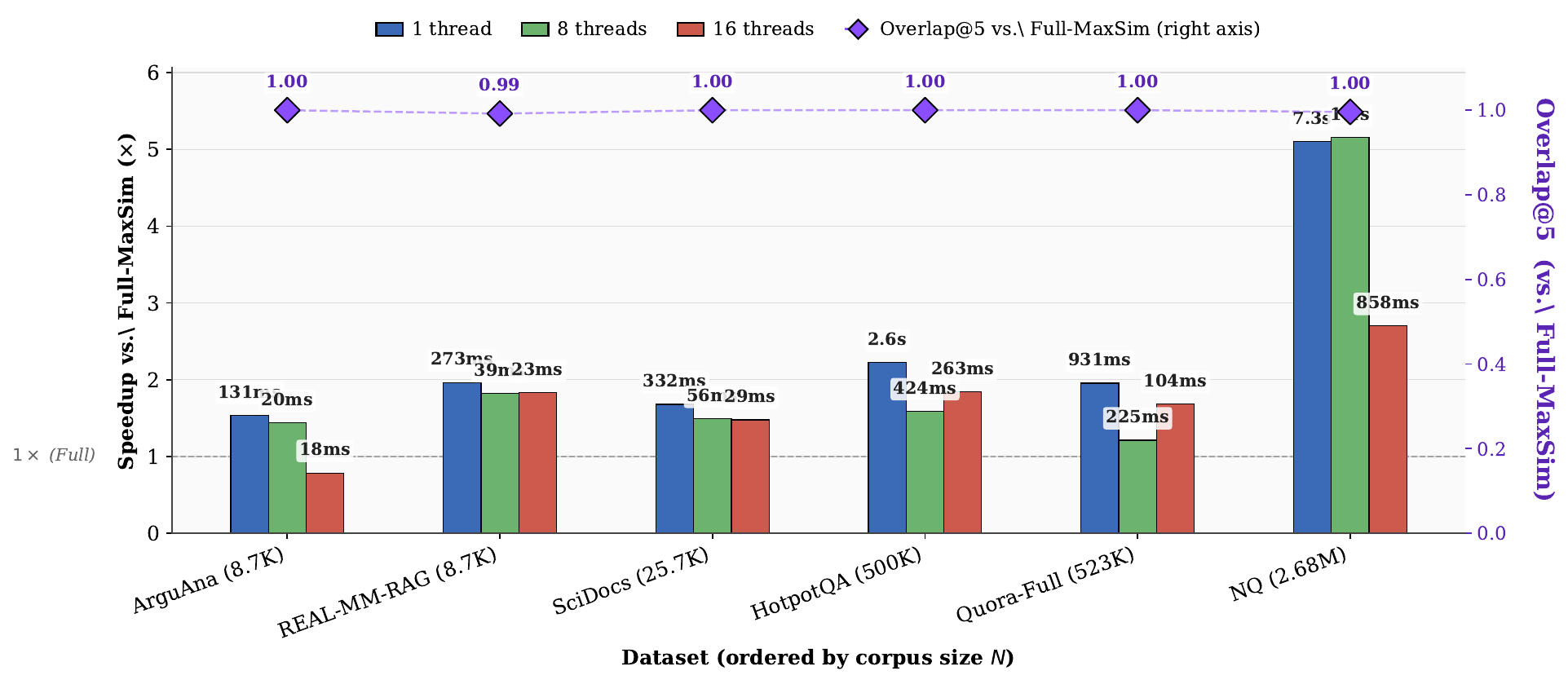}
\caption{Wall-clock speedup at $\alpha_{\mathrm{ef}}{=}1$ ($\delta$-PAC corner of Theorem~\ref{thm:pac}): Ov@$5{=}1.0$ on every text corpus, at the cost of $\sim\!28$--$59\%$ coverage. Even at the PAC-valid setting \method{} delivers nontrivial speedups over Full-MaxSim.}
\label{fig:headline_K5_a1}
\end{figure}

\paragraph{$\alpha_{\mathrm{ef}}$ sweep at $K{=}5$.}
Smaller $\alpha_{\mathrm{ef}}$ tightens the decision radius and eliminates more aggressively, trading lower Overlap@$5$ for higher speedup. The deployed default ($\alpha_{\mathrm{ef}}{=}0.2$, body figure) is bracketed by Figure~\ref{fig:headline_K5_a01} (aggressive, $\alpha_{\mathrm{ef}}{=}0.1$), Figure~\ref{fig:headline_K5_a03} (conservative, $\alpha_{\mathrm{ef}}{=}0.3$), and Figure~\ref{fig:headline_K5_a1} ($\delta$-PAC corner $\alpha_{\mathrm{ef}}{=}1$, Theorem~\ref{thm:pac}). Across the three settings the qualitative ranking of corpora is preserved; the deployed setting strikes the practical Pareto trade-off between speedup and Overlap@$5$.


\begin{figure}[!htbp]
\centering
\includegraphics[width=\linewidth]{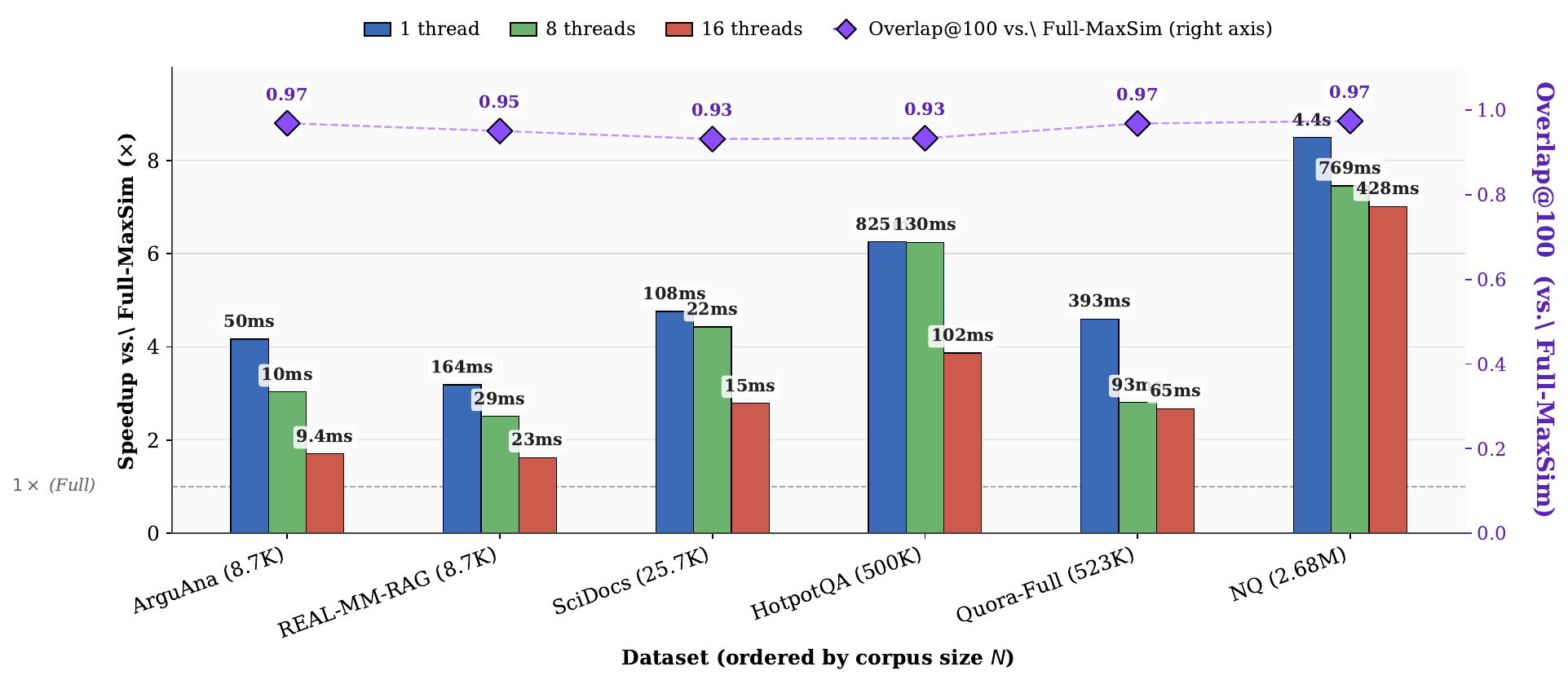}
\caption{Wall-clock speedup at the deployed $\alpha_{\mathrm{ef}}{=}0.2$, $K{=}100$ (recall-class setting): \method{} retains substantial speedup at the larger top-$K$ target with Ov@$100\geq 0.93$ on every corpus.}
\label{fig:headline_K100}
\end{figure}

\paragraph{$K$ sweep at $\alpha_{\mathrm{ef}}{=}0.2$.}
Holding the deployed knob fixed and growing $K$ to a recall-class setting ($K{=}100$). Figure~\ref{fig:headline_K100} confirms the sub-linear-in-$K$ behaviour of \method{} reported quantitatively in §5.3 (Table~\ref{tab:k_sensitivity}): the speedup degrades modestly relative to the body $K{=}5$ headline because the elimination phase has to discriminate among more candidates, but the Overlap@$100$ stays $\geq 0.93$ on every corpus.

\subsection{Extended Retrieval Effectiveness (Top-1 Analysis)}
\label{sec:app_effectiveness_k1}

The Top-1 ranking regime (Recall@1, nDCG@1, MRR@1) is a stricter test of fidelity: a single mis-identified document costs the full point. Table~\ref{tab:retrieval_effectiveness} reports retrieval effectiveness at varying coverage levels averaged across BEIR and REAL-MM-RAG. The Top-1 trends mirror those at $K=5$ (Table~\ref{tab:main_benchmark_universal_eff_combined}): \method{} maintains near-lossless performance compared to the Full ColBERT baseline, even when pruning significantly more aggressively than non-adaptive methods. At lower coverage budgets, the gap between \method{} and the non-adaptive baselines (Doc-Uniform and Ball-carving) widens, highlighting the necessity of variance-aware sampling for correctly identifying the single best document with high confidence.

\begin{table}[t]
\centering
\footnotesize
\setlength{\tabcolsep}{4pt}
\renewcommand{\arraystretch}{1.1}
\caption{Retrieval effectiveness at different coverage levels averaged across all nine corpora: \textit{REAL-MM-RAG} (Fin.\ Reports, Fin.\ Slides, Tech.\ Reports, Tech.\ Slides; Granite Vision Embedding) and \textit{BEIR} (ArguAna, Quora, SciDocs, NQ-$2.68$M, HotpotQA-$500$\,K; ColBERTv2). Each row is the corpus-mean of the metric (qrel-filtered queries on Quora-Full and HotpotQA-$500$\,K to keep at-least-one-relevant-doc-in-subset; see Appendix~\ref{app:datasets_models}). Full reranking at $100\%$ coverage serves as the reference.}
\label{tab:retrieval_effectiveness}
\resizebox{\ifdim\width>\linewidth \linewidth\else \width\fi}{!}{%
\begin{tabular}{lccccc}
\toprule
Method & Coverage & Recall@1 & nDCG@1 & MRR@1 \\
\midrule
Full ColBERT & 100\% & 0.41 & 0.51 & 0.51 \\
\midrule
Col-Bandit & 20\% & 0.40 & 0.50 & 0.50 \\
Col-Bandit & 40\% & 0.41 & 0.50 & 0.50 \\
\midrule
Ball-carving & 20\% & 0.33 & 0.42 & 0.42 \\
Doc-Uniform & 20\% & 0.23 & 0.28 & 0.28 \\
Ball-carving & 40\% & 0.37 & 0.47 & 0.47 \\
Doc-Uniform & 40\% & 0.31 & 0.38 & 0.38 \\
\midrule
\multicolumn{5}{l}{\textit{Relative Retention at 20\% Coverage (vs.\ Full ColBERT)}} \\
\midrule
Col-Bandit & -- & 98.9\% & 98.7\% & 98.7\% \\
Ball-carving & -- & 81.0\% & 82.1\% & 82.1\% \\
Doc-Uniform & -- & 55.9\% & 55.6\% & 55.6\% \\
\midrule
\multicolumn{5}{l}{\textit{Relative Retention at 40\% Coverage (vs.\ Full ColBERT)}} \\
\midrule
Col-Bandit & -- & 99.1\% & 98.9\% & 98.9\% \\
Ball-carving & -- & 90.8\% & 91.9\% & 91.9\% \\
Doc-Uniform & -- & 74.9\% & 74.6\% & 74.6\% \\
\bottomrule
\end{tabular}
}
\end{table}

\subsection{Per-Corpus Retrieval Effectiveness on REAL-MM-RAG}
\label{sec:app_effectiveness_k5_realmm}

Table~\ref{tab:retrieval_real_mm_all} breaks retrieval effectiveness down per corpus on the four REAL-MM-RAG sub-corpora (FinReport, FinSlides, TechReport, TechSlides), reporting Recall@$5$, nDCG@$5$, and MRR@$5$ at two coverage operating points ($\sim\!20\%$ and $\sim\!40\%$). At the $\sim\!20\%$ point \method{} retains $\geq 97.8\%$ of Full ColBERT's Recall@$5$/nDCG@$5$/MRR@$5$ on every corpus, versus $78\text{--}95\%$ for Ball-carving and Doc-Uniform; the gap widens as the coverage budget tightens, underscoring the value of variance-aware sampling for high-fidelity ranking.

\begin{table}[t]
\centering
\footnotesize
\setlength{\tabcolsep}{5pt}
\renewcommand{\arraystretch}{1.05}
\caption{\textbf{Per-corpus retrieval effectiveness on the four REAL-MM-RAG sub-corpora} (Granite Vision Embedding). Full ColBERT at $100\%$ coverage is the reference; for each method we report Recall@$5$ / nDCG@$5$ / MRR@$5$ at two operating points ($\sim\!20\%$ and $\sim\!40\%$ coverage). At $\sim\!20\%$ coverage \method{} retains $\geq 97.8\%$ of Full ColBERT on every metric and corpus, versus $78\text{--}95\%$ for the baselines.}
\label{tab:retrieval_real_mm_all}
\resizebox{\ifdim\width>\linewidth \linewidth\else \width\fi}{!}{%
\begin{tabular}{l c ccc}
\toprule
Method & Coverage & Recall@5 & nDCG@5 & MRR@5 \\
\midrule
\multicolumn{5}{l}{\textit{\textbf{Financial Reports}}} \\
Full ColBERT & 100\% & 0.91 & 0.76 & 0.70 \\
\textbf{\method{}} & 23\% & 0.91 & 0.76 & 0.70 \\
Ball-carving & 18\% & 0.82 & 0.69 & 0.64 \\
Doc-Uniform & 21\% & 0.80 & 0.64 & 0.59 \\
\textbf{\method{}} & 39\% & 0.91 & 0.76 & 0.70 \\
Ball-carving & 44\% & 0.92 & 0.77 & 0.72 \\
Doc-Uniform & 41\% & 0.82 & 0.68 & 0.64 \\
\midrule
\multicolumn{5}{l}{\textit{\textbf{Financial Slides}}} \\
Full ColBERT & 100\% & 0.93 & 0.77 & 0.72 \\
\textbf{\method{}} & 19\% & 0.91 & 0.76 & 0.72 \\
Ball-carving & 21\% & 0.88 & 0.68 & 0.61 \\
Doc-Uniform & 21\% & 0.78 & 0.61 & 0.55 \\
\textbf{\method{}} & 44\% & 0.93 & 0.77 & 0.72 \\
Ball-carving & 44\% & 0.90 & 0.74 & 0.69 \\
Doc-Uniform & 41\% & 0.84 & 0.67 & 0.61 \\
\midrule
\multicolumn{5}{l}{\textit{\textbf{Technical Reports}}} \\
Full ColBERT & 100\% & 0.94 & 0.86 & 0.83 \\
\textbf{\method{}} & 23\% & 0.92 & 0.85 & 0.82 \\
Ball-carving & 19\% & 0.93 & 0.84 & 0.81 \\
Doc-Uniform & 21\% & 0.85 & 0.72 & 0.68 \\
\textbf{\method{}} & 36\% & 0.93 & 0.86 & 0.83 \\
Ball-carving & 34\% & 0.96 & 0.86 & 0.82 \\
Doc-Uniform & 41\% & 0.92 & 0.81 & 0.77 \\
\midrule
\multicolumn{5}{l}{\textit{\textbf{Technical Slides}}} \\
Full ColBERT & 100\% & 1.00 & 0.95 & 0.93 \\
\textbf{\method{}} & 22\% & 1.00 & 0.95 & 0.93 \\
Ball-carving & 20\% & 0.98 & 0.92 & 0.89 \\
Doc-Uniform & 21\% & 0.96 & 0.88 & 0.85 \\
\textbf{\method{}} & 36\% & 1.00 & 0.95 & 0.93 \\
Ball-carving & 47\% & 1.00 & 0.94 & 0.92 \\
Doc-Uniform & 41\% & 0.99 & 0.93 & 0.91 \\
\bottomrule
\end{tabular}
}
\end{table}

\subsection{Runtime Overhead Breakdown}
\label{app:overhead_breakdown}

Table~\ref{tab:overhead_breakdown} decomposes per-query \method{} latency into three stages: the MaxSim micro-kernel calls during elimination rounds (kernel internals in Appendix~\ref{app:impl_details}); the elimination decision points themselves (LCB/UCB updates and survivor-set bookkeeping); and the final $K{+}M$ exhaustive-aligned rescore. Across the eight corpora profiled here (HotpotQA omitted from this profiling run) at the deployed knob, the kernel dominates ($\geq 96\%$ of total runtime); elimination plus rescore together stay under $4\%$. This is consistent with the speedup model used throughout the main text: the wallclock benefit comes from calling the kernel on $\mathrm{Cov}\%$ of the $(d, t)$ cells rather than $100\%$, not from cheaper bookkeeping.

\begin{table}[t]
\centering
\renewcommand{\arraystretch}{1.1}
\caption{\textbf{\method{} per-stage runtime breakdown on \textsc{Cpu-S} at $K{=}5$, $1$ thread.} Deployed knob ($\alpha_{\mathrm{ef}}{=}0.2$, $M{=}5$). The kernel column is the MaxSim micro-kernel (kernel internals in Appendix~\ref{app:impl_details}); Elim and Rescore together stay under $4\%$ of total runtime on every corpus. This is an independent single-thread profiling run, so absolute ms differ slightly from the wall-clock means of Table~\ref{tab:wallclock_cpu}; the takeaway is the stage \emph{proportions}, not the absolute latencies.}
\label{tab:overhead_breakdown}
\footnotesize
\setlength{\tabcolsep}{4pt}
\resizebox{\ifdim\width>\linewidth \linewidth\else \width\fi}{!}{%
\begin{tabular}{lrrrrrrr}
\toprule
Dataset & Cov & Kernel (ms) & Elim (ms) & Rescore (ms) & Total (ms) & Kern \% & Bookkeep \% \\
\midrule
ArguAna & 14.7\% & 36.78 & 0.318 & 0.246 & 37.35 & 98.5\% & 1.51\% \\
TechSlides & 21.0\% & 45.87 & 0.124 & 1.661 & 47.66 & 96.3\% & 3.75\% \\
TechReport & 21.9\% & 46.50 & 0.126 & 1.753 & 48.37 & 96.1\% & 3.88\% \\
FinSlides & 17.7\% & 54.83 & 0.156 & 1.929 & 56.92 & 96.3\% & 3.66\% \\
FinReport & 23.0\% & 70.89 & 0.182 & 1.936 & 73.00 & 97.1\% & 2.90\% \\
SciDocs & 14.3\% & 81.38 & 0.616 & 0.188 & 82.18 & 99.0\% & 0.98\% \\
Quora & 12.6\% & 360.62 & 9.143 & 0.020 & 369.78 & 97.5\% & 2.48\% \\
NQ-2.68M & 12.6\% & 4735.23 & 48.310 & 0.097 & 4783.64 & 99.0\% & 1.01\% \\
\bottomrule
\end{tabular}
}
\end{table}

\paragraph{Deviations from the $100/\mathrm{Cov}\%$ heuristic.}
The realised speedup deviates from the first-order heuristic in both directions: thread scaling, memory traffic, and the elimination/rescore phases combine to make small corpora track the heuristic only loosely. For example, Quora at $13\%$ Cov yields only $4.7\times$ speedup at $1$\,t (rescore overhead dominates relative to a small kernel call), while \textsc{MM} at $21\%$ Cov yields $8.1\times$ at $1$\,t (the cell-skip pattern aligns favourably with the SIMD register tile). Tightening $\alpha_{\mathrm{ef}}$ from the deployed $0.2$ up to the $\alpha_{\mathrm{ef}}{=}1$ corner trades additional coverage for higher fidelity, as shown by the full Pareto frontier in Figure~\ref{fig:results_grid_}.

\subsection{Multi-Thread Wall-Clock Scaling}
\label{app:wallclock_threads}

The main-paper wall-clock tables (Tables~\ref{tab:wallclock_cpu} and~\ref{tab:m1_neon}) report single-thread latency for compactness. Tables~\ref{tab:wallclock_cpu_threads} and~\ref{tab:m1_neon_threads} give the full $1$t\,/\,$8$t breakdown on \textsc{Cpu-S} (AMD EPYC 7763, AVX2) and \textsc{Cpu-M1} (Apple M1 Max, NEON) respectively. Thread scaling is sub-linear because the $K{+}M$ rescore and elimination bookkeeping do not parallelise as cleanly as the dense kernel sweep, but \method{} retains a substantial multi-thread speedup on every corpus.

\begin{table}[t]
\centering
\caption{\textbf{Wall-clock CPU benchmarks on \textsc{Cpu-S}, multi-thread breakdown} (AMD EPYC 7763, AVX2; $\alpha_{\mathrm{ef}}{=}0.2$, $M{=}5$, $\delta{=}0.01$). Per-query latency in ms ($1$t\,/\,$8$t). Single-thread summary in the main paper (Table~\ref{tab:wallclock_cpu}).}
\label{tab:wallclock_cpu_threads}
\footnotesize
\setlength{\tabcolsep}{4pt}
\renewcommand{\arraystretch}{1.1}
\resizebox{\ifdim\width>\linewidth \linewidth\else \width\fi}{!}{%
\begin{tabular}{lr|ccccc|ccccc}
\toprule
 & & \multicolumn{5}{c|}{\textbf{$K{=}5$}} & \multicolumn{5}{c}{\textbf{$K{=}100$}} \\
\cmidrule(lr){3-7}\cmidrule(lr){8-12}
\textbf{Dataset} & \textbf{$N$} & \textbf{Ov@$5$} & \textbf{Cov} & \textbf{Full (ms)} & \textbf{CB (ms)} & \textbf{Sp.} & \textbf{Ov@$100$} & \textbf{Cov} & \textbf{Full (ms)} & \textbf{CB (ms)} & \textbf{Sp.} \\
 & & & & $1$t / $8$t & $1$t / $8$t & $1$t / $8$t & & & $1$t / $8$t & $1$t / $8$t & $1$t / $8$t \\
\midrule
ArguAna   & $8.7$\,K   & 0.98 & 14\% & 211 / 29           & \textbf{33 / 7}        & 6.4$\times$ / 4.2$\times$ & 0.97 & 20\% & 210 / 32           & \textbf{50 / 10}       & 4.2$\times$ / 3.0$\times$ \\
SciDocs   & $25.7$\,K  & 0.96 & 14\% & 513 / 96           & \textbf{81 / 18}       & 6.3$\times$ / 5.3$\times$ & 0.93 & 18\% & 515 / 96           & \textbf{108 / 22}      & 4.8$\times$ / 4.4$\times$ \\
HotpotQA  & $500$\,K   & 0.93 & 13\% & 5{,}568 / 728      & \textbf{757 / 128}     & 7.4$\times$ / 5.7$\times$ & 0.93 & 14\% & 5{,}153 / 809      & \textbf{825 / 130}     & 6.2$\times$ / 6.2$\times$ \\
Quora     & $522.9$\,K & 0.99 & 13\% & 1{,}778 / 242      & \textbf{381 / 99}      & 4.7$\times$ / 2.4$\times$ & 0.97 & 13\% & 1{,}806 / 260      & \textbf{393 / 93}      & 4.6$\times$ / 2.8$\times$ \\
NQ  & $2.68$\,M  & 0.98 & 13\% & 37{,}283 / 4{,}459 & \textbf{4{,}222 / 714} & 8.8$\times$ / 6.2$\times$ & 0.97 & 13\% & 37{,}224 / 5{,}731 & \textbf{4{,}384 / 769} & 8.5$\times$ / 7.5$\times$ \\
MM & $8.6$\,K   & 0.97 & 21\% & 527 / 76           & \textbf{65 / 20}       & 8.1$\times$ / 3.8$\times$ & 0.95 & 47\% & 523 / 73           & \textbf{164 / 29}      & 3.2$\times$ / 2.5$\times$ \\
\midrule
\rowcolor{gray!15}\textbf{Mean} & --- & --- & --- & --- & --- & \textbf{7.0$\times$ / 4.6$\times$} & --- & --- & --- & --- & \textbf{5.3$\times$ / 4.4$\times$} \\
\bottomrule
\end{tabular}%
}
\end{table}

\begin{table}[t]
\centering
\caption{\textbf{Cross-platform benchmark on \textsc{Cpu-M1}, multi-thread breakdown} (Apple M1 Max, NEON; deployed knob). Per-query latency in ms ($1$t\,/\,$8$t); vs.\ m-cpu~\citep{maxsimcpu2025mxbai} is $1$t only. Single-thread summary in the main paper (Table~\ref{tab:m1_neon}).}
\label{tab:m1_neon_threads}
\footnotesize
\setlength{\tabcolsep}{4pt}
\renewcommand{\arraystretch}{1.1}
\resizebox{\ifdim\width>\linewidth \linewidth\else \width\fi}{!}{%
\begin{tabular}{lr|cccccc|cccccc}
\toprule
 & & \multicolumn{6}{c|}{\textbf{$K{=}5$}} & \multicolumn{6}{c}{\textbf{$K{=}100$}} \\
\cmidrule(lr){3-8}\cmidrule(lr){9-14}
\textbf{Dataset} & \textbf{$N$} & \textbf{Ov@$5$} & \textbf{Cov} & \textbf{Full (ms)} & \textbf{CB (ms)} & \textbf{Sp.\ vs Full} & \textbf{vs m-cpu} & \textbf{Ov@$100$} & \textbf{Cov} & \textbf{Full (ms)} & \textbf{CB (ms)} & \textbf{Sp.\ vs Full} & \textbf{vs m-cpu} \\
 & & & & $1$t / $8$t & $1$t / $8$t & $1$t / $8$t & $1$t & & & $1$t / $8$t & $1$t / $8$t & $1$t / $8$t & $1$t \\
\midrule
ArguAna  & $8.7$\,K   & 0.95 & 14\% & 284 / 44          & \textbf{23 / 6}    & 12.3$\times$ / 7.3$\times$ & 4.6$\times$ & 0.96 & 20\% & 289 / 46          & \textbf{36 / 8}    & 8.0$\times$ / 5.8$\times$  & 2.8$\times$ \\
SciDocs  & $25.7$\,K  & 0.92 & 14\% & 821 / 120         & \textbf{68 / 14}   & 12.1$\times$ / 8.6$\times$ & 6.0$\times$ & 0.93 & 18\% & 832 / 129         & \textbf{93 / 19}   & 8.9$\times$ / 6.8$\times$  & 4.4$\times$ \\
MM       & $8.6$\,K   & 0.99 & 21\% & 1{,}636 / 241     & \textbf{103 / 18}  & 15.9$\times$ / 13.4$\times$ & 4.0$\times$ & 0.97 & 47\% & 1{,}657 / 238     & \textbf{170 / 27}  & 9.7$\times$ / 8.8$\times$  & 2.4$\times$ \\
HotpotQA & $500$\,K   & 0.90 & 13\% & 7{,}394 / 1{,}673 & \textbf{660 / 140} & 11.2$\times$ / 11.9$\times$ & 4.7$\times$ & 0.94 & 14\% & 7{,}215 / 1{,}449 & \textbf{766 / 218} & 9.4$\times$ / 6.6$\times$  & 4.3$\times$ \\
\midrule
\rowcolor{gray!15}\textbf{Mean} & --- & --- & --- & --- & --- & \textbf{12.9$\times$ / 10.3$\times$} & \textbf{4.8$\times$} & --- & --- & --- & --- & \textbf{9.0$\times$ / 7.0$\times$} & \textbf{3.5$\times$} \\
\bottomrule
\end{tabular}%
}
\end{table}

\subsection{Operating-Point, $K$-Sensitivity, and Commodity-CPU Tables}
\label{sec:app_extra_tables}

These three tables support claims made in \S\ref{sec:main_results} and are placed here for space. Table~\ref{tab:k_sensitivity} reports the $K$-sensitivity sweep ($K\!\in\!\{5,100,500,1000\}$ on \textsc{Cpu-S}, single thread) discussed under ``Sub-linear scaling in $K$''. Table~\ref{tab:alpha_guidance} gives the operating-point guidance for the calibration knob $\alpha_{\mathrm{ef}}$ referenced in ``Choosing the calibration knob $\alpha_{\mathrm{ef}}$''. Table~\ref{tab:cpu_vs_a100} is the commodity-CPU-vs-GPU architecture-level pairing ($16$-thread \method{} on \textsc{Cpu-S} vs.\ A100 PyTorch Full-MaxSim) discussed under ``Commodity CPU vs.\ GPU''.

\begin{table}[t]
\centering
\centering
\footnotesize
\setlength{\tabcolsep}{4pt}
\renewcommand{\arraystretch}{1.1}
\caption{\textbf{$K$-sensitivity} on \textsc{Cpu-S} (1 thread, ms) at the deployed knob ($\alpha_{\mathrm{ef}}{=}0.2$, $M{=}5$). \method{} latency grows sub-linearly in $K$.}
\label{tab:k_sensitivity}
\resizebox{\ifdim\width>\linewidth \linewidth\else \width\fi}{!}{%
\begin{tabular}{r|rr|rr|rr}
\toprule
 & \multicolumn{2}{c|}{\textbf{ArguAna}} & \multicolumn{2}{c|}{\textbf{SciDocs}} & \multicolumn{2}{c}{\textbf{HotpotQA}} \\
\cmidrule(lr){2-3}\cmidrule(lr){4-5}\cmidrule(lr){6-7}
$K$ & Full & CB & Full & CB & Full & CB \\
\midrule
$5$    & 214 & \textbf{34}  & 513 & \textbf{80}  & 4{,}885 & \textbf{742} \\
$100$  & 223 & \textbf{47}  & 551 & \textbf{105} & 5{,}180 & \textbf{811} \\
$500$  & 286 & \textbf{121} & 541 & \textbf{156} & 5{,}171 & \textbf{911} \\
$1000$ & 243 & \textbf{154} & 585 & \textbf{217} & 5{,}447 & \textbf{991} \\
\bottomrule
\end{tabular}
}

\end{table}

\begin{table}[t]
\centering
\centering
\renewcommand{\arraystretch}{1.1}
\footnotesize
\setlength{\tabcolsep}{4pt}
\caption{\textbf{Operating-point guidance for $\alpha_{\mathrm{ef}}$} (the single Pareto knob). Coverage and Overlap@$5$ \emph{ranges} across four BEIR text corpora at $K{=}5$, $M{=}5$, $1$ thread. Smaller $\alpha_{\mathrm{ef}}$ cuts more aggressively (lower coverage, lower fidelity); $\alpha_{\mathrm{ef}}{=}1$ is the $\delta$-PAC corner of Theorem~\ref{thm:pac}.}
\label{tab:alpha_guidance}
\resizebox{\ifdim\width>\linewidth \linewidth\else \width\fi}{!}{%
\begin{tabular}{lccc}
\toprule
$\alpha_{\mathrm{ef}}$ & \textbf{Regime} & \textbf{Cov} & \textbf{Ov@$5$} \\
\midrule
$0.10$ & Aggressive   & $12.5$--$12.9\%$ & $0.76$--$0.91$ \\
$0.20$ & \textbf{Deployed} & $\mathbf{12.6}$--$\mathbf{14.4\%}$ & $\mathbf{0.92}$--$\mathbf{0.99}$ \\
$0.30$ & Conservative & $12.9$--$17.4\%$ & $0.98$--$1.00$ \\
$1.00$ & Certified    & $27.7$--$58.9\%$ & $1.00$ \\
\bottomrule
\end{tabular}%
}

\end{table}

\begin{table}[t]
\centering
\centering
\footnotesize
\setlength{\tabcolsep}{4pt}
\renewcommand{\arraystretch}{1.1}
\caption{\textbf{Commodity CPU \method{} vs.\ A100 80GB GPU Full-MaxSim} at $K{=}5$, $\alpha_{\mathrm{ef}}{=}0.2$. Per-query latency (ms): \textsc{Gpu} runs PyTorch dense Full-MaxSim, \textsc{Cpu-S} runs $16$-thread \method{}. \textsc{Cpu-S} matches GPU within $1.06$--$1.41\times$ on $N\geq 500$\,K text corpora.}
\label{tab:cpu_vs_a100}
\resizebox{\ifdim\width>\linewidth \linewidth\else \width\fi}{!}{%
\begin{tabular}{lrrrc}
\toprule
\textbf{Dataset} & $N$ & \textsc{Gpu} Full & \textsc{Cpu-S} CB & \textbf{Ov@5} \\
 & & (ms) & 16t (ms) & \\
\midrule
ArguAna   & $8.7$\,K   & 1.9   & 7.0   & 0.98 \\
SciDocs   & $25.7$\,K  & 5.4   & 11.5  & 0.96 \\
HotpotQA  & $500$\,K   & 65.9  & 92.7  & 0.92 \\
Quora     & $522.9$\,K & 62.6  & 66.6  & 0.99 \\
NQ-2.68M  & $2.68$\,M  & \textbf{OOM} & 472   & 0.98 \\
\bottomrule
\end{tabular}%
}

\end{table}

\subsection{Full Three-Panel Pareto Frontier}
\label{app:pareto_3panel}

\begin{figure}[t]
\centering
\includegraphics[width=\linewidth]{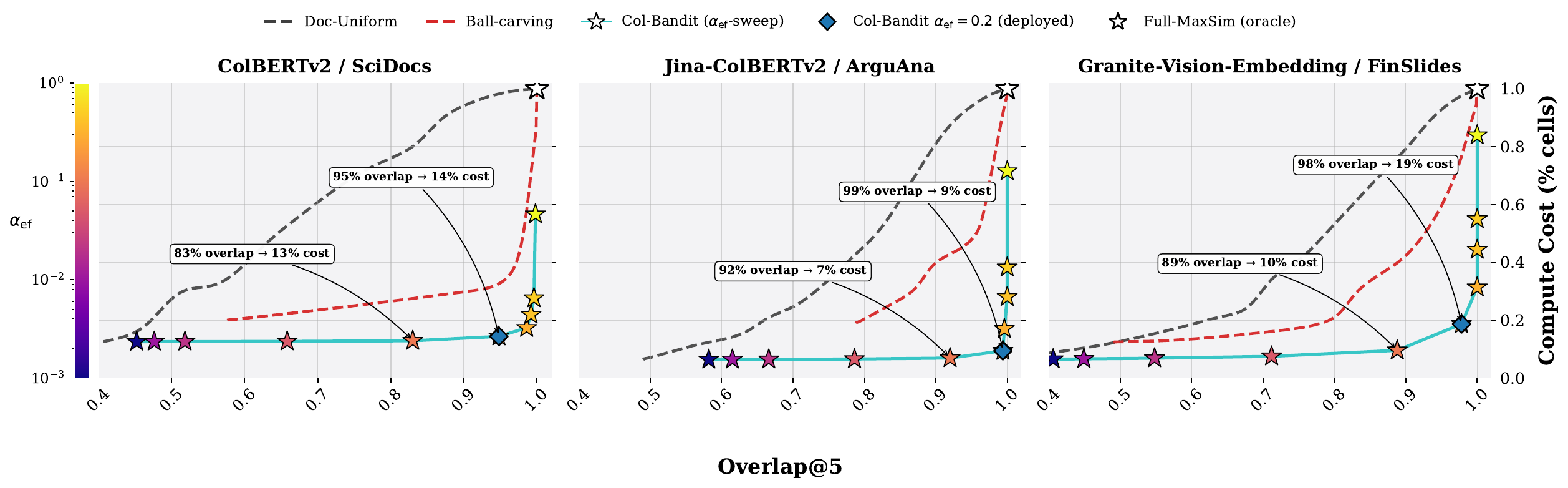}
\caption{\textbf{Quality--coverage Pareto frontier ($K{=}5$).} \method{} (sweeping $\alpha_{\mathrm{ef}}$) vs.\ Doc-Uniform and Ball-carving (MUVERA App.~C.3) on ColBERTv2/SciDocs, Jina-ColBERTv2/ArguAna, and GVE/FinSlides. \method{} dominates both baselines, reaching high fidelity at $\leq 20\%$ of the cell budget; deployed knob $\alpha_{\mathrm{ef}}{=}0.2$ marked, Full-MaxSim oracle at top-right.}
\label{fig:results_grid_}
\end{figure}

The main paper (Figure~\ref{fig:pareto_finslides}) shows the GVE/FinSlides panel for space. Figure~\ref{fig:results_grid_} gives the full three-panel comparison across one small-text (ColBERTv2/SciDocs), one large-text (Jina-ColBERTv2/ArguAna), and one multimodal (GVE/FinSlides) corpus.

\subsection{Matryoshka Dimension Reduction}
\label{app:matryoshka}

Jina-ColBERTv2's $d{=}128 \to d{=}64$ Matryoshka projection halves the per-token storage with no impact on \method{}'s elimination dynamics. Table~\ref{tab:matryoshka} sweeps the four BEIR text corpora at the deployed knob ($\alpha_{\mathrm{ef}}{=}0.2$, $K{=}5$). Coverage is essentially invariant to dimension (shifts $\leq 1.7$ percentage points on three of four corpora; on Quora-Full coverage actually \emph{drops} $5.9$ pp at $d{=}64$ with no quality loss). Ov@$5$ retention is $\geq 0.95$ on every cell; nDCG@$5$ retention is $100\%$ on six of eight cells. The dimension and cell-skipping axes are orthogonal: halving $d$ does not change which documents \method{} eliminates.

\begin{table}[t]
  \centering
  \caption{%
    \method{} with $\alpha_{\mathrm{ef}}{=}0.2$ applied to Jina-ColBERT-v2 at
    two Matryoshka projection sizes ($d{=}128$ and $d{=}64$).
    Coverage, overlap, and nDCG are nearly identical across dimensions,
    confirming that dimension reduction and CB-NK's cell-skipping are
    orthogonal compression axes that compose without loss.%
  }
  \label{tab:matryoshka}
  \footnotesize
  \setlength{\tabcolsep}{4pt}
  \renewcommand{\arraystretch}{1.1}
  \resizebox{\ifdim\width>\linewidth \linewidth\else \width\fi}{!}{%
\begin{tabular}{llrrrrrr}
    \toprule
    \textbf{Dataset} & \textbf{Dim} &
    \textbf{Cov\%} & \textbf{Ov@5} & \textbf{Ov@100} &
    \textbf{nDCG@5} & \textbf{nDCG@100} &
    \textbf{nDCG@5\,ret.} \\
    \midrule
    \multirow{2}{*}{ArguAna}
      & 128 & 24.8 & 0.996 & 0.987 & 0.2858 & 0.3986 & 100.0\% \\
      &  64 & 25.2 & 0.997 & 0.986 & 0.2922 & 0.3942 & 100.0\% \\
    \midrule
    \multirow{2}{*}{SciDocs}
      & 128 & 42.0 & 0.953 & 0.921 & 0.1491 & 0.2550 &  95.2\% \\
      &  64 & 41.7 & 0.947 & 0.923 & 0.1437 & 0.2448 &  93.5\% \\
    \midrule
    \multirow{2}{*}{HotpotQA-500K}
      & 128 & 53.0 & 0.963 & 0.950 & 0.2464 & 0.2557 & 100.0\% \\
      &  64 & 54.6 & 0.975 & 0.962 & 0.2396 & 0.2461 &  99.2\% \\
    \midrule
    \multirow{2}{*}{Quora-Full}
      & 128 & 35.0 & 0.983 & 0.953 & 0.5601 & 0.5684 & 100.0\% \\
      &  64 & 29.1 & 0.987 & 0.954 & 0.5526 & 0.5602 & 100.0\% \\
    \bottomrule
  \end{tabular}
}
  \par\smallskip
  \noindent\footnotesize
  \textit{nDCG@5\,ret.} = CB-NK nDCG@5 / full-\textsc{MaxSim} nDCG@5
  at the same dimension.
  Full-\textsc{MaxSim} references (d=128): ArguAna 0.2858,
  SciDocs 0.1567, HotpotQA-500K 0.2464, Quora-Full 0.5601.
  Full-\textsc{MaxSim} references (d=64): ArguAna 0.2922,
  SciDocs 0.1536, HotpotQA-500K 0.2416, Quora-Full 0.5526.
  nDCG@$5$ can equal the full reference even at Overlap@$5<1$ when the displaced
  top-$5$ documents are non-relevant (common on single-relevant corpora such as
  ArguAna), so $100\%$ retention does not imply an identical top-$5$ set.
\end{table}

\subsection{Ward Token-Axis Pooling}
\label{app:ward}

Ward hierarchical clustering~\citep{clavie2024reducing} pools each document's $L_d$ token embeddings into $L_d / k$ cluster centroids ($k\in\{2,4,8\}$), with the clustering applied per-document on the actual non-padded token count. This is orthogonal to \method{}: pooling reduces the number of cells per row at encode time; \method{} then eliminates documents based on the (smaller) per-row score at query time. We benchmark the stack on $\textsc{Cpu-S}$ for SciDocs (text) and TechSlides (multimodal) at the deployed knob.

Table~\ref{tab:ward_compose} reports both \emph{Fidelity Ov@$5$} (against the exhaustive top-$5$ on the pooled corpus) and \emph{End-to-End Ov@$5$} (against the exhaustive top-$5$ on the original unpooled corpus), with one row per pool factor showing Ward-Full alone (100\% coverage) and one row showing Ward-Full $+$ \method{}. The two-row layout makes the loss decomposition explicit: Ward-Full's E2E gap captures \emph{pooling-induced} information loss; the $\Delta$ from Ward-Full to Ward-Full $+$ \method{} captures \method{}'s \emph{elimination} cost. The numbers show pooling dominates: across the six cells, pooling contributes $9$--$41$\,pp of E2E loss (median $20$\,pp), while \method{}'s elimination contributes $0.8$--$4.0$\,pp (median $2.0$\,pp). \method{}'s fidelity against the pooled-corpus exhaustive top-$5$ holds $\geq 0.948$ on every cell, with coverage in the deployed $14$--$21\%$ range and $5.3$--$7.9\times$ wall-clock speedup vs Ward-Full on the pooled corpus.

\begin{table}[t]
\centering
\footnotesize
\setlength{\tabcolsep}{4pt}
\renewcommand{\arraystretch}{1.1}
\caption{\textbf{Ward token-axis pooling composes with \method{}.} Per-query latency on \textsc{Cpu-S} at the deployed knob ($\alpha_{\mathrm{ef}}{=}0.2$, $K{=}5$). \emph{Fid Ov@5} is overlap with Full-MaxSim on the \textit{pooled} corpus (isolates \method{}'s elimination loss); \emph{E2E Ov@5} is overlap with Full-MaxSim on the \textit{unpooled} corpus (combined effect). The end-to-end loss decomposes additively: pooling-induced loss (gap between unpooled Full and the Ward-Full row) and \method{}'s elimination cost (gap between the Ward-Full row and the +\method{} row). Pooling is per-document on actual (non-padded) length; pooling dominates by $\sim$$10\times$.}
\label{tab:ward_compose}
\resizebox{\ifdim\width>\linewidth \linewidth\else \width\fi}{!}{%
\begin{tabular}{l l c c c r r}
\toprule
\textbf{Corpus} & \textbf{Method} & \textbf{Cov\,(\%)} & \textbf{Fid\,Ov@5} & \textbf{E2E\,Ov@5} & \textbf{Lat 1t} & \textbf{Lat 8t} \\
\midrule
\multirow{6}{*}{SciDocs (text)}
 & Ward $2\times$ Full              & 100  & ---   & 0.912 & 376 & 46 \\
 & \quad + \method{} ($2\times$)    & 14.4 & 0.952 & 0.872 & \textbf{58.8} & \textbf{9.4} \\
 & Ward $4\times$ Full              & 100  & ---   & 0.788 & 211 & 23 \\
 & \quad + \method{} ($4\times$)    & 14.2 & 0.948 & 0.760 & \textbf{39.5} & \textbf{6.6} \\
 & Ward $8\times$ Full              & 100  & ---   & 0.592 & 154 & 14 \\
 & \quad + \method{} ($8\times$)    & 14.0 & 0.960 & 0.572 & \textbf{25.1} & \textbf{4.9} \\
\midrule
\multirow{6}{*}{TechSlides (MM)}
 & Ward $2\times$ Full              & 100  & ---   & 0.884 & 262 & 48 \\
 & \quad + \method{} ($2\times$)    & 20.9 & 0.968 & 0.864 & \textbf{33.3} & \textbf{10.7} \\
 & Ward $4\times$ Full              & 100  & ---   & 0.816 & 137 & 27 \\
 & \quad + \method{} ($4\times$)    & 21.1 & 0.980 & 0.800 & \textbf{19.7} & \textbf{6.3} \\
 & Ward $8\times$ Full              & 100  & ---   & 0.672 & 72  & 14 \\
 & \quad + \method{} ($8\times$)    & 20.9 & 0.980 & 0.664 & \textbf{11.9} & \textbf{3.5} \\
\bottomrule
\end{tabular}
}
\end{table}

\FloatBarrier
\section{Theoretical Validity in Uniform-Sampling Mode (Special Case)}
\label{app:uniform_pac}

\paragraph{Objective.} We seek an algorithm that adaptively reveals entries of $H$ and returns $\widehat{\calT}_K$ satisfying $\Prob(\widehat{\calT}_K=\calT^\star_K)\ge 1-\delta$ for user-defined $\delta\in(0,1)$, while minimising the expected coverage $\E[\gamma(\Omega)]$ at termination.

\begin{theorem}[$\delta$-PAC correctness at $\alpha_{\mathrm{ef}}{=}1$, conditional on Eq.~\ref{eq:effective_radius}]
\label{thm:pac}
For any $\delta\in(0,1)$ and $\alpha_{\mathrm{ef}}{=}1$, under uniform-without-replacement reveals within each row, and \emph{provided} the radius of Eq.~\ref{eq:effective_radius} is a valid per-row $(1{-}\delta_{i,n})$ confidence radius (Remark~\ref{rem:pac_scope} discusses the $c{=}1$ and omitted-$O(1/n)$ simplifications under which we deploy it), Algorithm~\ref{alg:progressive} returns a set $\widehat{\calT}_K$ satisfying
\begin{equation}
\Prob\!\left(\widehat{\calT}_K \;=\; \calT^\star_K\right) \;\ge\; 1 - \delta.
\label{eq:pac_guarantee}
\end{equation}
\end{theorem}

\begin{remark}[Scope of the guarantee]
\label{rem:pac_scope}
Theorem~\ref{thm:pac} holds for the decision radius of Eq.~\ref{eq:effective_radius} \emph{as instantiated} at $\alpha_{\mathrm{ef}}{=}1$: the Bernstein--Serfling pre-factor is set to $c{=}1$ and the standard $O(1/n)$ lower-order term is omitted. These simplifications are folded into the radius rather than proved conservative. Where the omitted bias is largest (small $n_i$) the deterministic hard bounds $[LB^{\mathrm{hard}}_i, UB^{\mathrm{hard}}_i]$ tend empirically to be the binding constraint (Appendix~\ref{app:variance_details}; an observation, not a conservativeness proof), and the $\alpha_{\mathrm{ef}}{=}1$ corner empirically attains Overlap@$5{=}1.00$ on every text corpus (Appendix~\ref{sec:app_headline_variants}), consistent with $\delta{=}0.01$. The theorem also concerns the \emph{exact-precision} reveal model: the deployed CB-NK kernel makes elimination decisions on int8-quantised estimates (Appendix~\ref{app:impl_details}), a further empirical relaxation for which we make no PAC claim. We therefore describe $\alpha_{\mathrm{ef}}{=}1$ as $\delta$-PAC \emph{under the radius of Eq.~\ref{eq:effective_radius} in the exact-precision model}, and treat the deployed $\alpha_{\mathrm{ef}}{=}0.2$ (with int8 elimination) as a calibrated relaxation (Limitations).
\end{remark}

We state a special case in which the simplified empirical Bernstein--Serfling-style radius used in Eq.~\ref{eq:effective_radius} is $\delta$-valid when $\alpha_{\mathrm{ef}}=1$.
Concretely, a single permutation $\pi$ of $[T]$ is drawn uniformly and \emph{shared} across all surviving documents; revealing the first $n_i$ entries of $\pi$ for document $i$ makes $\mathcal{O}_i$ a uniform-without-replacement size-$n_i$ subset of $[T]$ \emph{marginally} for each row. The per-row inequality below depends only on this marginal, so the cross-row coupling induced by the shared $\pi$ is immaterial (it is handled by the union-bound argument that follows).

Fix a document $i$ and let $\mathcal{O}_i$ be the set of revealed token indices with $n_i=|\mathcal{O}_i|$.
Define the row mean and sum
\[
\mu_i \triangleq \frac{1}{T}\sum_{t=1}^T H_{i,t},
\qquad
S_i \triangleq \sum_{t=1}^T H_{i,t}=T\mu_i,
\]
and the empirical mean/standard deviation over the revealed entries
\[
\widehat{\mu}_i=\frac{1}{n_i}\sum_{t\in\mathcal{O}_i}H_{i,t},
\qquad
\widehat{\sigma}_i^2=\frac{1}{n_i-1}\sum_{t\in\mathcal{O}_i}(H_{i,t}-\widehat{\mu}_i)^2.
\]
Under uniform-without-replacement sampling within the row and bounded support $H_{i,t}\in[a,b]$, an empirical Bernstein--Serfling inequality \citep{bardenet2015concentration} implies that, for any fixed $(i,n)$,
\[
\Pr\!\left(\left|S_i-T\widehat{\mu}_i\right| \;\le\; 
T\widehat{\sigma}_i\sqrt{\frac{2\log(c/\delta_{i,n})}{n}}\sqrt{\rho_n}
\right)\;\ge\;1-\delta_{i,n}.
\]
To obtain a time-uniform statement over all documents and all sample sizes, set $\delta_{i,n}=\delta/(NT)$ and union bound over $i\in[N]$ and $n\in[T]$.
The single shared permutation $\pi$ couples the reveals across rows, but this does not affect the argument: the per-row inequality requires only that each row's revealed set be \emph{marginally} a uniform-without-replacement sample of its $T$ tokens, which holds under a uniformly drawn $\pi$, and Boole's inequality holds for arbitrarily dependent events, so the union over rows remains valid regardless of the cross-row coupling.
Therefore, with probability at least $1-\delta$, simultaneously for all $i$ and all $n$,
\[
S_i \in \Big[T\widehat{\mu}_i \pm r^{\mathrm{th}}_i(n)\Big],
\quad
r^{\mathrm{th}}_i(n)\triangleq
T\widehat{\sigma}_i\sqrt{\frac{2\log(cNT/\delta)}{n}}\sqrt{\rho_n}.
\]
In this uniform-within-row mode, choosing $\alpha_{\mathrm{ef}}=1$ in Eq.~\ref{eq:effective_radius} recovers the above theoretical form (up to the constant $c$), justifying its use as a $\delta$-valid decision radius.

\paragraph{Proof sketch of Theorem~\ref{thm:pac}.}
We give the high-level argument for the PAC-valid setting ($\alpha_{\mathrm{ef}}{=}1$, uniform-without-replacement reveals each round).

\textbf{(1) Per-cell concentration.} Each MaxSim cell $H_{i,t}\in[a,b]$ is bounded, and within a row the unrevealed entries form a finite population of size $|\calU_i|$ sampled without replacement.
The empirical Bernstein--Serfling inequality of \citet{bardenet2015concentration} (Theorem~4.3) therefore yields, for any fixed document $i$ and reveal count $n_i$, a high-probability two-sided radius for the partial-sum estimator $T\widehat{\mu}_i$ around the row-sum $S_i$.

\textbf{(2) Per-round event.} At round $r$, define $A_r$ as the event that, simultaneously for every surviving document $i\in\calA_r$, the true score satisfies $\LCB_i \le S_i \le \UCB_i$.
With $\alpha_{\mathrm{ef}}=1$ the radius in Eq.~\ref{eq:effective_radius} matches the Bardenet--Maillard form, so $\Prob[A_r] \ge 1-\delta_r$ when the per-round budget $\delta_r$ is plugged into the confidence parameter.

\textbf{(3) Time-uniform union bound.} Allocating a per-cell budget $\delta/(NT)$ and union-bounding over all $N$ documents and $T$ reveal counts makes every $[\LCB_i,\UCB_i]$ valid \emph{simultaneously} for all $(i,n)$. Because this holds uniformly over reveal counts, it also holds at the data-dependent stopping time, so the elimination loop needs no separate optional-stopping argument and $\Prob\!\left[\bigcap_{r} A_r\right]\ge 1-\delta$.

\textbf{(4) Elimination correctness.} On $\bigcap_r A_r$, every surviving document obeys $\LCB_i\le S_i\le \UCB_i$.
If at round $r$ document $d$ has $\UCB_d<\tau_r$, where $\tau_r$ is the $K$-th largest $\LCB_i$ over $\calA_r$, then at least $K$ documents $j$ satisfy $\LCB_j\ge\tau_r>\UCB_d$; on $\bigcap_r A_r$ each obeys $S_j\ge\LCB_j>\UCB_d\ge S_d$, so $d$ is outranked by $\ge K$ documents, hence $d\notin\calT^\star_K$ and removing it from $\calA_{r+1}$ is safe.
Following the Successive Elimination argument of \citet{kalyanakrishnan2012pac,audibert2010best}, repeated application of this rule never discards a true top-$K$ document on the high-probability event.

\textbf{(5) Termination and rescore.} The loop exits with $|\calA_r|\le K{+}M$ or $(r{-}1)B \ge T$.
In either case, the $K$-margin rescore (Algorithm~\ref{alg:progressive_full}) computes the exact score $S_i$ for every survivor using all $T$ tokens, and the final $\arg\mathrm{topK}$ over $\calA_r$ is taken on these exact values.
On $\bigcap_r A_r$ the survivors include every member of $\calT^\star_K$, so the returned set equals $\calT^\star_K$.

\textbf{(6) Combining.} Equation~\eqref{eq:pac_guarantee} follows from $\Prob\!\left[\bigcap_r A_r\right]\ge 1-\delta$ and the deterministic implication ``$\bigcap_r A_r \Rightarrow \widehat{\calT}_K=\calT^\star_K$''.

\paragraph{Empirical sanity check.}
At $\alpha_{\mathrm{ef}}{=}1$ on every text corpus, Algorithm~\ref{alg:progressive} recovers $\widehat{\calT}_K = \calT^\star_K$ with measured Overlap@$5 \ge 99\%$ (Table~\ref{tab:alpha_guidance}, $\alpha_{\mathrm{ef}}{=}1$ row), exceeding the $1-\delta = 0.99$ threshold for $\delta = 0.01$. Increasing coverage drives Overlap@5 toward $1.0$ on every corpus, consistent with the radius collapsing as $n_i\to T$.

\FloatBarrier
\section{Implementation Details and Parameter Selection}
\label{app:implementation}

\subsection{Detailed Algorithm Listing}
\label{app:algo_detailed}

Algorithm~\ref{alg:progressive} in the main text is condensed for space. Algorithm~\ref{alg:progressive_full} gives the fully-annotated version, making the per-round bookkeeping ($\Omega$, $\widehat{S}_i$, the explicit batch set $\calB_r$) and the two-phase structure (elimination loop followed by exact rescore) explicit.

{\algrenewcommand\algorithmiccomment[1]{\hfill{\scriptsize$\triangleright$\ #1}}%
\begin{algorithm}[t]
\caption{\method{}: batched progressive elimination for top-$K$ identification on the late-interaction matrix (detailed).}
\label{alg:progressive_full}
\small
\setlength{\itemsep}{0pt}
\begin{algorithmic}[1]
\Require Query $Q$ ($T$ tokens), candidate set $\calD = \{d_1,\dots,d_N\}$, target $K$, margin $M$, batch $B$, calibration $\alpha_{\mathrm{ef}}$, confidence $\delta$
\Ensure Estimated top-$K$ set $\widehat{\calT}_K \subseteq \calD$
\State $\calA_1 \gets [N]$,\quad $\Omega \gets \emptyset$,\quad $r \gets 1$ \Comment{Active set; observed entries; round counter}
\State $[\LCB_i, \UCB_i] \gets [a\,T,\, b\,T]$ for all $i \in [N]$ \Comment{Trivial initial interval}
\State $\pi \gets$ random permutation of $[T]$ (uniform, fixed seed; Appendix~\ref{app:uniform_pac})
\While{$|\calA_r| > K + M$ \textbf{and} $(r{-}1)B < T$}
    \State $\calB_r \gets \{\pi[(r{-}1)B{+}1],\dots,\pi[\min(rB, T)]\}$ \Comment{Next $B$ tokens of $\pi$}
    \ForAll{$i \in \calA_r$ \textbf{in parallel} (one SIMD pass)}
        \State Reveal $\{H_{i,t} : t \in \calB_r\}$;\quad $\Omega \gets \Omega \cup (\{i\}\times \calB_r)$
        \State Update $\widehat{S}_i$ and $[\LCB_i, \UCB_i]$ via Eq.~\ref{eq:LCB_UCB} (radius $\propto \alpha_{\mathrm{ef}}$, $\delta$)
    \EndFor
    \State $\tau_r \gets$ $K$-th largest $\LCB_i$ over $i \in \calA_r$ \Comment{Elimination threshold}
    \State $\calA_{r+1} \gets \{i \in \calA_r : \UCB_i \ge \tau_r\}$ \Comment{Permanent elimination}
    \State $r \gets r + 1$
\EndWhile
\ForAll{$i \in \calA_r$} \Comment{$K$-margin rescore on survivors}
    \State $S_i \gets \sum_{t=1}^{T} H_{i,t}$ using high-precision kernel
\EndFor
\State \Return $\widehat{\calT}_K \gets \arg\mathrm{topK}_{i \in \calA_r} S_i$
\end{algorithmic}
\end{algorithm}}

\paragraph{Walkthrough.}
\emph{Initialisation (lines 1--3).} All $N$ candidates start in the active set $\calA_1$; the observed set $\Omega$ is empty and every interval is the trivial support bound $[aT, bT]$ (no cells revealed yet). A single per-query permutation $\pi$ fixes the reveal order of the $T$ query tokens; it is drawn uniformly with a fixed seed and shared across all documents, which is the structural condition under which Theorem~\ref{thm:pac} applies (Appendix~\ref{app:uniform_pac}).

\emph{Elimination loop (lines 4--13).} Each round selects the next $B$ tokens of $\pi$ (line 5) and, in a single vectorised pass over the surviving set (lines 6--9), reveals those $B$ cells for every active document, appends them to $\Omega$, and refreshes the partial-sum estimate $\widehat{S}_i$ and the hybrid interval $[\LCB_i,\UCB_i]$ via Eq.~\ref{eq:LCB_UCB}. The elimination threshold $\tau_r$ is the $K$-th largest lower bound over the active set (line 10); any document whose \emph{upper} bound is below $\tau_r$ cannot belong to the top-$K$ and is permanently dropped (line 11). The loop exits once the active set is small enough for exact completion ($|\calA_r|\le K{+}M$) or all $T$ tokens have been revealed.

\emph{Exact rescore (lines 14--16).} The $\le K{+}M$ survivors are rescored on all $T$ query tokens with the high-precision kernel (the same fused call as Full-MaxSim, identical float-add order), so the returned scores are bit-identical to the exhaustive baseline and the final top-$K$ is exact over the survivor set. The margin $M$ gives the elimination loop slack: borderline documents that the loop would otherwise have to disambiguate with more reveals are instead deferred to this cheap exact rescore.

\paragraph{Complexity.}
The loop reveals $B|\calA_r|$ cells per round and the active set shrinks monotonically, so the total cells touched are $\sum_r B|\calA_r| + T\,|\calA_{\mathrm{final}}|$, which is far below the $NT$ of exhaustive scoring whenever the score distribution has a clear top-$K$ separation (the regime confirmed empirically in \S\ref{sec:experiments}). The bookkeeping (interval updates, threshold, elimination) is $O(|\calA_r|)$ per round and is dominated by the kernel cost (Appendix~\ref{app:impl_details}, Table~\ref{tab:overhead_breakdown}).

\subsection{The \textsc{numkong} C Extension}
\label{app:impl_details}

The Full-MaxSim baseline and \method{} share a single C extension (referred to throughout as our \textsc{numkong} extension), so that the wall-clock comparisons in Table~\ref{tab:wallclock_cpu} and Table~\ref{tab:m1_neon} are apples-to-apples by construction: only the reveal schedule differs, not the kernel quality.
Our \textsc{numkong}-based \method{} kernel (CB-NK) will be released as open-source software upon publication. It is not publicly available at submission time, to preserve review anonymity.

\paragraph{Target instruction sets.}
The same C source supports three SIMD back-ends, selected at compile time:
(i) x86-64 AVX2 ($256$-bit, $8$ fp32 lanes), used on $\textsc{Cpu-S}$ (AMD EPYC 7763) for Table~\ref{tab:wallclock_cpu};
(ii) ARMv8 NEON ($128$-bit, $4$ fp32 lanes), used on $\textsc{Cpu-M1}$ (Apple M1 Max) for the edge-deployment study (Table~\ref{tab:m1_neon});
(iii) x86-64 AVX-512 ($512$-bit, $16$ fp32 lanes), which compiles cleanly but is not benchmarked in this submission.
The dot-product micro-kernel is the only platform-specific piece of code; the algorithmic skeleton (bound updates, elimination, $K$-margin rescore) is shared across all three back-ends.

\paragraph{Memory layout.}
Documents are stored as a packed flat array in tile-friendly stride order, so that per-document MaxSim accumulation reads sequential cache lines.
Padding rows are zero-norm, so they cannot dominate any per-query-token max and do not bias the score.

\paragraph{Compute path.}
Full-MaxSim runs in fp32 throughout.
\method{} (CB-NK) uses an int8-quantised coarse pass for the cheap LCB/UCB updates inside the elimination loop, followed by an aligned fp32 rescore for the $K{+}M$ surviving documents.
The $K{+}M$ rescore swaps to Full-MaxSim's exact kernel call (same float-add order, same packed input layout), so survivor scores are bit-identical to the exhaustive baseline; this property is what allows Overlap@$K$ to be a clean fidelity measure rather than a quantisation-noise measure.

\paragraph{Multi-threading model.}
Parallelism is across queries via OpenMP: each query is processed by a single thread, so the $1$-thread vs.\ $8$-thread columns in Table~\ref{tab:wallclock_cpu} (and the corresponding M1 figures) represent a fixed query batch where threads partition the query set.
This avoids intra-query synchronisation overhead and matches the deployment regime where queries arrive concurrently.

\paragraph{Default parameter setting.}
Unless stated otherwise, all \method{} experiments use the following deployed configuration:
$\alpha_{\mathrm{ef}}{=}0.20$ (the Pareto figures additionally sweep $\alpha_{\mathrm{ef}}\in\{0.01,0.02,0.05,0.10,0.20,0.30,0.40,0.50,1.0\}$);
round size $B{=}4$ (number of query tokens revealed per elimination round);
rescore margin $M{=}5$ (extra survivors carried into the final exhaustive-aligned rescore pass);
PAC failure probability $\delta{=}0.01$;
empirical Bernstein--Serfling bounds~\citep{bardenet2015concentration} (\S\ref{sec:bounds}).
The reveal order is a Fisher--Yates random permutation of $[T]$ with a fixed seed for reproducibility; at $\alpha_{\mathrm{ef}}{=}1$ this matches the conditions of Theorem~\ref{thm:pac}.

\paragraph{Hardware platforms.}
The hardware platforms used in the experiments are:
$\textsc{Cpu-S}$ (AMD EPYC 7763, AVX2) for Table~\ref{tab:wallclock_cpu};
$\textsc{Cpu-M1}$ (Apple M1 Max, NEON) for Table~\ref{tab:m1_neon};
and $\textsc{Gpu}$ (NVIDIA A100 80\,GB) for all GPU sweeps and the Pareto figure (Figure~\ref{fig:results_grid_}).

\paragraph{Hardware realization.}\label{app:hw_realization}
Algorithm~\ref{alg:progressive}'s batched structure aligns naturally with modern SIMD execution: each round reveals $B{=}4$ tokens for all surviving documents in a single vectorized pass. We implement \method{} as a fused C kernel parameterized by the underlying instruction set (AVX2/AVX-512 on x86, NEON on Apple Silicon) using a cache-separated quantized memory layout. The wall-clock results in Section~\ref{sec:main_results} use this fused realization on both Intel/AMD x86 and Apple Silicon.

\paragraph{Cache-separated quantized layout.}
Document token embeddings are stored in two parallel buffers per document: an int8 buffer used for fast partial-sum updates inside the elimination loop, and an aligned fp32 buffer used by the final $K$-margin rescore.
The int8 layout uses percentile-clip quantization (clip to the 99.9th percentile of absolute values, then linear quantize to int8) so that AVX-512 \texttt{VPDPBUSD} / NEON \texttt{sdot} dot-products run at full throughput while the rescore uses bit-faithful float arithmetic.
Because elimination decisions use the int8 estimates while the $K{+}M$ survivors are rescored in exact fp32, the survivor \emph{scores} are always exact, but quantization can change \emph{which} borderline documents survive to the rescore. Any resulting top-$K$ error is therefore already reflected in our reported Overlap@$K$, which is measured end-to-end against exact Full-MaxSim, and the margin $M$ absorbs most such borderline cases. Because int8 dot products round differently across SIMD back-ends, the surviving set, and hence Overlap@$K$, can differ slightly between \textsc{Cpu-S} (AVX2) and \textsc{Cpu-M1} (NEON) on the same data.

\paragraph{Register tiling.}
The inner loop processes a $B{=}4$ batch of query tokens against $4$ document token positions per cycle, which fits in 4 AVX-512 / NEON vector registers and keeps the hot data in L1.
The four query tokens of the current $\calB_r$ are kept in registers across all surviving documents in a round so that document tokens are streamed in once.

\paragraph{Cross-platform.}
The same C source compiles unchanged for x86 (AVX2/AVX-512) and ARM (NEON, including Apple Silicon).
The only platform-specific code is the dot-product micro-kernel, which is selected at compile time via preprocessor macros.
Empirically the NEON port on \textsc{Cpu-M1} matches the AVX2 wall-clock on \textsc{Cpu-S} at single-thread on small corpora and exceeds it at larger $K$.

\paragraph{Cross-ISA comparison against \emph{maxsim-cpu}.}
The published \emph{maxsim-cpu}~\citep{maxsimcpu2025mxbai} Rust SIMD MaxSim is tuned primarily for Apple Silicon, so its standing relative to our \textsc{numkong} kernel depends on the instruction set. On \textsc{Cpu-M1} (NEON) it is faster than our Full-MaxSim, yet \method{} is still $4.8\times$ faster than it at single thread (Table~\ref{tab:m1_neon}). On x86 the ordering reverses: on an AMD EPYC Zen-3 server (AVX2, the same instruction set and microarchitecture family as \textsc{Cpu-S}), our \textsc{numkong} Full-MaxSim is already ${\sim}2.3\times$ faster than \emph{maxsim-cpu}, so \method{} widens to ${\sim}11\times$ over it at single thread (ArguAna $11.2\times$, SciDocs $11.5\times$; $K{=}5$, $\alpha_{\mathrm{ef}}{=}0.2$, Overlap@$5\geq 0.91$). The kernel-level acceleration thus holds against a production third-party SIMD baseline on both ISAs; in particular, the AVX2 result shows that \method{}'s speedup is not an artifact of a slow Full-MaxSim baseline, since that baseline itself outpaces \emph{maxsim-cpu} on x86. This x86 measurement was taken on an EPYC Zen-3 (AVX2) node rather than the exact \textsc{Cpu-S} 7763; our Full-MaxSim there tracked the \textsc{Cpu-S} Full timings of Table~\ref{tab:wallclock_cpu} (ArguAna $180$ vs.\ $211$\,ms; SciDocs $509$ vs.\ $513$\,ms), supporting it as an ISA-matched proxy. Table~\ref{tab:mcpu_x86} reports the full breakdown.

\begin{table}[t]
\centering
\footnotesize
\setlength{\tabcolsep}{4pt}
\renewcommand{\arraystretch}{1.1}
\caption{\textbf{x86 cross-ISA check vs.\ \emph{maxsim-cpu}} (single thread, $K{=}5$, $\alpha_{\mathrm{ef}}{=}0.2$, $M{=}5$, $\delta{=}0.01$) on an AMD EPYC Zen-3 (AVX2) node, an ISA-matched proxy for \textsc{Cpu-S}. \method{}~(CB-NK) is ${\sim}11\times$ faster than \emph{maxsim-cpu}, and our \textsc{numkong} Full-MaxSim is itself ${\sim}2.3\times$ faster than \emph{maxsim-cpu} on this ISA.}
\label{tab:mcpu_x86}
\resizebox{\columnwidth}{!}{%
\begin{tabular}{lrccrrrr}
\toprule
\textbf{Dataset} & \textbf{$N$} & \textbf{Cov} & \textbf{Ov@$5$} & \textbf{Full (ms)} & \textbf{m-cpu (ms)} & \textbf{CB (ms)} & \textbf{CB vs m-cpu} \\
\midrule
ArguAna & $8.7$\,K  & $14\%$ & $0.93$ & $180$  & $404$   & \textbf{$36$}  & \textbf{$11.2\times$} \\
SciDocs & $25.7$\,K & $14\%$ & $0.92$ & $509$  & $1{,}174$ & \textbf{$102$} & \textbf{$11.5\times$} \\
\bottomrule
\end{tabular}%
}
\end{table}

\subsection{Parameter Selection}
\label{app:parameter_selection}

\method{} exposes two practical knobs: the calibration parameter $\alpha_{\mathrm{ef}}$ and the margin $M$.
In practice, $\alpha_{\mathrm{ef}}$ governs the aggressiveness of elimination: smaller values tighten the decision radius and reduce coverage, while larger values are more conservative.
The margin $M$ has a small, monotone effect: increasing $M$ from $0$ to $5$ adds roughly $2$ percentage points of Overlap@$K$ at negligible runtime cost, since rescoring $K{+}M$ documents on all $T$ tokens is a fixed small overhead. We use $M{=}5$ by default and select $\alpha_{\mathrm{ef}}$ based on a desired quality--coverage trade-off. Table~\ref{tab:ablations_M} sweeps $M\!\in\!\{0,5,10\}$ on two representative text corpora (SciDocs, HotpotQA-$500$\,K) at $\alpha_{\mathrm{ef}}{=}0.2$, $K{=}5$, $1$ thread.

\begin{table}[t]
\centering
\centering
\footnotesize
\setlength{\tabcolsep}{4pt}
\renewcommand{\arraystretch}{1.1}
\caption{\textbf{$K$-margin rescore ablation} ($\alpha_{\mathrm{ef}}{=}0.2$, $K{=}5$, $1$ thread, $50$ queries per corpus). Without the rescore ($M{=}0$) borderline-eliminated documents leak through; $M{=}5$ restores Ov@$5\geq 0.98$ on both corpora at a small latency cost. Numbers are not directly comparable to the full-benchmark Ov@$5$ in Table~\ref{tab:wallclock_cpu}, which uses the complete per-corpus query set.}
\label{tab:ablations_M}
\resizebox{\ifdim\width>\linewidth \linewidth\else \width\fi}{!}{%
\begin{tabular}{ll|rrr}
\toprule
Dataset & $M$ & Cov & Ov@$5$ & Lat.\ (ms) \\
\midrule
SciDocs   & 0  & 18\% & 0.94 & 107 \\
          & 5  & 18\% & 0.98 & 129 \\
          & 10 & 19\% & 0.98 & 129 \\
\addlinespace[1pt]
HotpotQA  & 0  & 14\% & 0.93 & 790 \\
          & 5  & 14\% & 0.98 & 838 \\
          & 10 & 14\% & 0.98 & 853 \\
\bottomrule
\end{tabular}
}

\end{table}

\paragraph{Calibrated relaxation.}
The relaxation parameter $\alpha_{\mathrm{ef}}\in(0,1]$ rescales the Bernstein--Serfling radius of Eq.~\ref{eq:effective_radius} (the same Bardenet--Maillard radius that appears in the proof of Theorem~\ref{thm:pac}), applied at every elimination step.
At $\alpha_{\mathrm{ef}}{=}1$ the bound is the unshrunk simplified Bernstein--Serfling-style radius and Theorem~\ref{thm:pac} delivers its conditional $\delta$-PAC guarantee.
For $\alpha_{\mathrm{ef}}{<}1$ the radii shrink, more aggressive eliminations are permitted, and the formal guarantee weakens to a calibrated relaxation: empirically, the rank-recovery error grows smoothly and predictably with the coverage saving.
Crucially, $\alpha_{\mathrm{ef}}$ is a single continuous knob with an interpretable $\delta$-PAC endpoint at $\alpha_{\mathrm{ef}}{=}1$, so practitioners can dial in a desired quality--coverage operating point along a one-dimensional Pareto.


\FloatBarrier
\section{Taxonomy of Efficient Late-Interaction Retrieval}
\label{app:taxonomy}

Figure~\ref{fig:taxonomy} situates \method{} relative to existing late-interaction acceleration methods along two axes: \emph{when} pruning happens (index-time vs.\ query-time) and \emph{what} is pruned (candidates, tokens, embedding dimensions, or atomic MaxSim cells). Index-time methods (PLAID, MUVERA, token pruning) commit before seeing the query and therefore prune conservatively; query-time methods (Ball-carving, \method{}) can use the actual query, with \method{}, to our knowledge, the first to prune at the atomic MaxSim-cell level.

\begin{figure}[h]
\centering
\resizebox{0.5\columnwidth}{!}{%
\begin{tikzpicture}[
    >=Stealth,
    main/.style={rectangle, rounded corners=3pt, draw=black!70, line width=1pt, fill=gray!15, font=\normalsize\bfseries, align=center, minimum width=3.8cm, minimum height=1.4cm},
    cluster/.style={rectangle, rounded corners=2pt, draw=black!60, line width=0.8pt, fill=gray!10, font=\normalsize, align=center, minimum width=3.2cm, minimum height=1.2cm},
    leaf/.style={rectangle, rounded corners=2pt, draw=black!50, line width=0.6pt, fill=white, font=\small, align=center, minimum width=3.6cm, minimum height=1.8cm, text width=3.4cm},
    leafHighlight/.style={rectangle, rounded corners=2pt, draw=orange!80!black, line width=1.5pt, fill=yellow!15, font=\small, align=center, minimum width=3.6cm, minimum height=1.8cm, text width=3.4cm},
    arrow/.style={->, thick, black!70},
]

\node[main] (root) at (0,0) {Efficient Late-Interaction\\Retrieval};

\node[cluster] (index) at (3,2.2) {\textbf{Index-Time}\\{\footnotesize(Static / Offline)}};
\node[cluster] (query) at (3,-2.2) {\textbf{Query-Time}\\{\footnotesize(Dynamic / Online)}};

\node[leaf] (compression) at (8,3.8) {\textbf{Compression \& Approx.}\\[2pt]{\footnotesize(Quantization, Sketches)}\\{\footnotesize e.g., PLAID, ColBERTv2,}\\{\footnotesize MUVERA, WARP}};

\node[leaf] (staticprune) at (8,1.2) {\textbf{Doc Token Pruning}\\[2pt]{\footnotesize(Remove tokens)}\\{\footnotesize e.g., Static Pruning}};

\node[leaf] (candprune) at (8,-1.2) {\textbf{Candidate Reduction}\\[2pt]{\footnotesize(Reduce $N$ docs)}\\{\footnotesize e.g., HNSW,}\\{\footnotesize Query Token Pruning}};

\node[leafHighlight] (interprune) at (8,-3.8) {\textbf{Interaction Pruning}\\[2pt]{\footnotesize(Reduce MaxSims)}\\{\footnotesize e.g., \textbf{\method{} (Ours)}}};

\draw[arrow] (root) -- (index);
\draw[arrow] (root) -- (query);
\draw[arrow] (index) -- (compression);
\draw[arrow] (index) -- (staticprune);
\draw[arrow] (query) -- (candprune);
\draw[arrow] (query) -- (interprune);

\end{tikzpicture}%
}
\caption{\textbf{Taxonomy of efficient late-interaction retrieval.}
Methods are classified by \emph{when} they prune (index-time vs.\ query-time)
and \emph{what} they prune. \method{} is the first to dynamically prune
the atomic interaction matrix $H$ during query-time scoring.}
\label{fig:taxonomy}
\end{figure}

\subsection{Extended Related Work}
\label{app:related_work}
This expands the condensed discussion in \S\ref{sec:related_work}.

\paragraph{Index-Time Compression \& Token Pruning.}
Approaches like PLAID \citep{plaid}, ColBERTv2 \citep{colbertv2}, and MUVERA \citep{muvera} accelerate retrieval via centroid-based compression, quantization, or fixed-dimensional encodings, improving the practicality of late-interaction methods that were initially constrained by considerable storage requirements. More recently, LEMUR~\citep{jaasaari2026lemur} reduces multi-vector retrieval to single-vector ANN through a learned per-corpus MLP, exact-MaxSim reranking the top-$k'$ ANN candidates. Additional system and indexing advances such as WARP \citep{scheerer2025warp} further improve scalability and usability.
Similarly, token pruning methods \citep{lassance2021study, tonellotto2021query} permanently discard non-informative tokens to reduce the index size ($N$) or query length ($T$), including near-lossless vector count reduction \citep{clavie2024reducing} and approaches that use a fixed number of representative tokens \citep{macavaney2025efficient}.
While effective, these methods are fixed at index-time and typically require offline modifications. \method{} is \emph{orthogonal} to them: it operates purely at query-time on standard indices, dynamically pruning the atomic interaction matrix $H$ during scoring. The two layers can be combined: \emph{any} of the above index-time systems can produce the candidate set $\mathcal{D}$ that \method{} reranks.

\paragraph{Efficient Systems \& Bound-Based Pruning.}
System-level optimizations like DESSERT \citep{dessert} use approximate retrieval to speed up candidate generation. In sparse retrieval, algorithms like WAND \citep{broder2003efficient_wand} and BMW~\citep{ding2011faster_bmw} use score upper bounds to skip documents. \method{} bridges these concepts, applying bound-based early stopping to \emph{dense late-interaction}. Unlike WAND, which prunes inverted list pointers, we prune atomic MaxSim operations $h(d,t)$ to certify the Top-$K$ set with statistical guarantees.

\paragraph{MaxSim-Level Pruning (Our Approach).}
To our knowledge, no prior work adaptively \emph{estimates} the exhaustive top-$K$ identity from a partially-revealed MaxSim matrix \emph{within} the exact scoring loop, using the actual query to decide which cells to compute. Existing methods reduce the number of candidates ($N$) or tokens ($T$) \emph{before} scoring, and therefore before seeing the query. \method{} frames the scoring process itself as a finite-population top-$K$ identification problem, using the query to drive the elimination, and progressively revealing only the subset of MaxSim entries needed to certify the ranking.

\paragraph{Finite-Population Bandits and Top-$k$ Arm Identification.}
Our method is inspired by fixed-confidence Top-$K$ Arm Identification \citep{kalyanakrishnan2012pac,chen2014combinatorial}, with one structural twist: standard best-arm identification (BAI) estimates the unknown means of \emph{stochastic} arms, whereas a row of our score table $H$ is a \emph{finite} population of $T$ deterministic values sampled \emph{without replacement}; the randomness is only in the reveal order. Fixed-confidence BAI machinery (UCB \citep{auer2002finite}, UCB-E \citep{audibert2010best}, LUCB \citep{kalyanakrishnan2012pac}) gives us the interval-driven reveal policy and stopping criterion, but standard infinite-population bounds are conservative here. We therefore replace the stochastic-arm radius with the empirical Bernstein--Serfling concentration of \citet{bardenet2015concentration}, which collapses deterministically as $n_i \to T$ and yields tighter intervals (\S\ref{sec:bounds}). MAB applications include prompt learning \citep{shi2024best}, LLM evaluation \citep{zhou2024speedingup}, $k$-NN search \citep{indyk2019adaptive}.

\clearpage
\section{Table of Notations}
\label{app:notation}

The notations used in the paper are described below.
\begin{table}[H]
\centering
\footnotesize
\setlength{\tabcolsep}{4pt}
\renewcommand{\arraystretch}{1.1}
\caption{Notations used in the paper.}
\label{tab:notation}
\begin{tabular}{cl}
\toprule
Symbol & Description \\
\midrule
\multicolumn{2}{l}{\textit{Input}} \\
\midrule
$Q = \{\mathbf{q}_1, \ldots, \mathbf{q}_T\}$ & A query represented as a set of $T$ token embeddings \\
$d$ & A document from the collection $\mathcal{D}$ \\
$\mathcal{D} = \{d_1, \ldots, d_N\}$ & The candidate document set with $N$ documents \\
$T$ & The number of query tokens \\
$L_d$ & The number of tokens in document $d$ \\
$l$ & The embedding dimension ($\mathbf{q}_t, \mathbf{e}_{d,j} \in \mathbb{R}^l$) \\
$K$ & The number of top documents to identify \\
\midrule
\multicolumn{2}{l}{\textit{Scoring}} \\
\midrule
$\mathrm{sim}(\cdot, \cdot)$ & Similarity function (e.g., cosine similarity); bounded support $[a,b]$ \\
$h(d, t)$ & MaxSim score: $\max_{j \in [L_d]} \mathrm{sim}(\mathbf{e}_{d,j}, \mathbf{q}_t)$ \\
$S(d; Q)$ & Total late-interaction score: $\sum_{t=1}^{T} h(d, t)$ \\
$\mathcal{T}^{\star}_K$ & The true (exhaustive) Top-$K$ document set \\
$\widehat{\mathcal{T}}_K$ & The returned (estimated) Top-$K$ document set \\
$\mathrm{Overlap}@K$ & Headline fidelity metric: $|\mathcal{T}^\star_K \cap \widehat{\mathcal{T}}_K|/K$ (Eq.~\ref{eq:overlap_def}) \\
\midrule
\multicolumn{2}{l}{\textit{Matrix \& Observation}} \\
\midrule
$\mathbf{H} \in \mathbb{R}^{N \times T}$ & The MaxSim matrix with entries $H_{i,t} = h(d_i, t)$ \\
$S_i$ & Total score for document $i$: $\sum_{t=1}^{T} H_{i,t}$ \\
$\Omega \subseteq [N] \times [T]$ & The set of observed (revealed) matrix entries \\
$\mathcal{O}_i,\ \mathcal{U}_i$ & Observed and unobserved token indices for $i$; $\mathcal{U}_i = [T] \setminus \mathcal{O}_i$ \\
$n_i$ & Number of revealed tokens for document $i$: $|\mathcal{O}_i|$ \\
$\gamma(\Omega)$ & Coverage: fraction of matrix revealed, $|\Omega| / (N \times T)$ \\
$\pi$ & Per-query reveal permutation of $[T]$ (uniform, fixed seed) \\
\midrule
\multicolumn{2}{l}{\textit{Bounds \& Algorithm State}} \\
\midrule
$[a, b]$ & Global support of $H_{i,t}$ (e.g., $[-1, 1]$ for cosine similarity) \\
$\hat{\mu}_i$ & Empirical mean over revealed entries: $\frac{1}{n_i} \sum_{t \in \mathcal{O}_i} H_{i,t}$ \\
$\hat{S}_i$ & Partial-sum estimator: $T \cdot \hat{\mu}_i$ \\
$\hat{\sigma}_i$ & Empirical standard deviation over $\{H_{i,t}\}_{t \in \mathcal{O}_i}$ \\
$\mathrm{LB}^{\mathrm{hard}}_i, \mathrm{UB}^{\mathrm{hard}}_i$ & Deterministic hard bounds (\S\ref{sec:bounds}) \\
$r^{\mathrm{eff}}_i$ & Variance-adaptive decision radius (Eq.~\ref{eq:effective_radius}) \\
$\mathrm{LCB}_i, \mathrm{UCB}_i$ & Hybrid lower/upper confidence bounds (Eq.~\ref{eq:LCB_UCB}) \\
$\mathcal{A}_r$ & Active set of surviving documents at round $r$ \\
$\mathcal{B}_r$ & Set of $B$ token indices revealed in round $r$ \\
$\tau_r$ & Round-$r$ elimination threshold: $K$-th largest $\mathrm{LCB}_i$ over $\mathcal{A}_r$ \\
\midrule
\multicolumn{2}{l}{\textit{Algorithm Parameters}} \\
\midrule
$\alpha_{\mathrm{ef}} \in (0,1]$ & Calibration knob; $\alpha_{\mathrm{ef}}{=}1$ is the $\delta$-PAC corner (Theorem~\ref{thm:pac}) \\
$M$ & Rescore margin: $K{+}M$ survivors rescored exactly on all $T$ tokens \\
$B$ & Round (batch) size: tokens revealed per round \\
$\delta \in (0,1)$ & PAC error tolerance \\
$\rho_n$ & Finite-population correction factor (Eq.~\ref{eq:effective_radius}) \\
\bottomrule
\end{tabular}
\end{table}

\end{document}